\documentclass[12pt,twoside,letterpaper]{report}
\usepackage{appendix}
\usepackage{graphicx}
\usepackage{bm}
\usepackage{subcaption}
\usepackage{soul}
\usepackage{xcolor}

\usepackage[byname]{smartref}
\usepackage{mathrsfs}
\usepackage{siunitx}
\usepackage{amsmath}
\usepackage{amsfonts}
\usepackage{algorithmic}
\usepackage{filecontents}
\usepackage[normalem]{ulem}
\usepackage{nomencl}
\makenomenclature

\usepackage{txfonts}
\usepackage{tocloft}
\makeatletter
\numberwithin{figure}{chapter}
{\clearemptydoublepage
 \begin{center}
  \section*{Acknowledgements}
 \end{center}
 \begingroup
}{\newpage\endgroup}

{\clearemptydoublepage 
 \begin{center}
  \section*{Dedication}
 \end{center}
 \begingroup
}{\newpage\endgroup}

\newenvironment{preliminary}%
{\pagestyle{plain}\pagenumbering{roman}}%
{\pagenumbering{arabic}}

\addtoreflist{chapter}

\newcommand{\qed}{\nobreak \ifvmode \relax \else
      \ifdim\lastskip<1.5em \hskip-\lastskip
      \hskip1.5em plus0em minus0.5em \fi \nobreak
      \vrule height0.75em width0.5em depth0.25em\fi}

\if@twoside 
\def\ps@thesis{\let\@mkboth\markboth
   \def\@oddfoot{}
   \let\@evenfoot\@oddfoot
   \def\@oddhead{
      {\sc\rightmark} \hfil \rm\thepage
      }
   \def\@evenhead{
      \rm\thepage \hfil {\sc\leftmark}
      }
   \def\chaptermark##1{\markboth{\ifnum \c@secnumdepth >\m@ne
      Chapter\ \thechapter. \ \fi ##1}{}}
   \def\sectionmark##1{\markright{\ifnum \c@secnumdepth >\z@
      \thesection. \ \fi ##1}}}
\else 
\def\ps@thesis{\let\@mkboth\markboth
   \def\@oddfoot{}
   \def\@oddhead{
      {\sc\rightmark} \hfil \rm\thepage
      }
   \def\chaptermark##1{\markright{\ifnum \c@secnumdepth >\m@ne
      Chapter\ \thechapter. \ \fi ##1}}}
\fi

\newcommand\isco[1]{%
  \edef\@tempa{#1}%
  \def\@tempb{}%
  \ifx\@tempa\@tempb
	\else \\\underline{Co-Supervisor:}\vspace{0.35in}\\\dots\dots\dots\dots\dots\dots\dots\\{#1}\\
  \fi
}

\newcommand\isjoint[1]{%
  \edef\@tempa{#1}%
  \def\@tempb{}%
  \ifx\@tempa\@tempb
	\else \\\underline{Joint Supervisor:}\vspace{0.35in}\\\dots\dots\dots\dots\dots\dots\dots\\{#1}\\
  \fi
}

\newcommand\isalt[1]{%
  \edef\@tempa{#1}%
  \def\@tempb{}%
  \ifx\@tempa\@tempb
	\else \\\underline{Alternate Supervisor:}\vspace{0.35in}\\\dots\dots\dots\dots\dots\dots\dots\\{#1}\\
  \fi
}

\newcommand\isdefinedsig[1]{%
  \edef\@tempa{#1}%
  \def\@tempb{}%
  \ifx\@tempa\@tempb
	\else \\ \dots\dots\dots\dots\dots\dots\dots\\{#1}\\
  \fi
}
\newcommand\isdefinedspinetitle[1]{%
  \edef\@tempa{#1}%
  \def\@tempb{}%
  \ifx\@tempa\@tempb
	\else (Spine title: #1)\\
  \fi
}
\newcommand\coauthor[1]{%
  \edef\@tempa{#1}%
  \def\@tempb{}%
  \ifx\@tempa\@tempb
	\else \newpage \Large Co-Authorship Statement\normalsize\\\indent\\#1\\
  \fi
}

\newcommand\acknowlege[1]{%
  \edef\@tempa{#1}%
  \def\@tempb{}%
  \ifx\@tempa\@tempb
	\else \newpage \Large \begin{center} \textbf{Acknowledgments} \end{center} \normalsize\indent\\#1\newpage
  \fi
}

\newcommand{\department}{Electrical and Computer Engineering}
\newcommand{\degree}{Doctor of Philosophy}
\newcommand{\firstname}{Raffi}
\newcommand{\middlename}{Garabet}
\newcommand{\lastname}{Toukhtarian}
\newcommand{\authorname}{{\firstname} {\middlename} {\lastname}}
\newcommand{\titl}{Modeling and Simulation for Varying Die Gap Fluid Extrusion}
\newcommand{\gyear}{\number\year}

\newcommand{\makeacknowlege} {

First, I would like to express sincere gratitude and appreciation to my supervisor Prof. Benoit Boulet for his great guidance and the intellectual freedom I was granted throughout this research. This work would not have been possible without his great dedication, constructive criticism and support.

A special thanks to my co-supervisor Prof. Roni Khazaka for his mentorship and guidance. I also gratefully acknowledge Prof. Savvas Hatzikiriakos and Mr. Haile Atsbha for sharing with me their profound knowledge and expertise. 

I gratefully acknowledge the Natural Sciences and Engineering Research Council of Canada (Automotive Partnership Canada) for the financial support and giving me the opportunity to pursue my passion in science and engineering.

Finally, I would like to thank my loving friends and family for their encouragement and support along my journey over the past five years. 

}

\newcommand{\listappendixname}{List of Appendices}
\newlistof{myappendices}{app}{\listappendixname}
\newcommand{\myappendices}[1]{%
\addcontentsline{app}{myappendices}{#1}\par}

\renewcommand{\maketitle}
{\begin{titlepage}
   \setcounter{page}{1}
   \begin{large}
   \begin{center}
      \mbox{}
      \vfill
      {\MakeUppercase{\titl}}\\
      \vfill
      by \\
      \vfill
      {\firstname} {\middlename} {\lastname}\\
      \vfill
      Department of {\department}\\
      \vfill
		A thesis submitted to McGill University in partial fulfillment\\
		of the requirements for the degree of\\
		\degree\\
		\vfill
		McGill University\\
		Montr\'eal, Qu\'ebec, Canada\\
		\vfill
      {\copyright} {\authorname} {\gyear}  \\
      \vspace*{.2in}
   \end{center}
   \end{large}
   \end{titlepage}

}

\makeatother
\begin{document}


\begin{preliminary}

\maketitle
\newpage
\addcontentsline{toc}{chapter}{Abstract}
\Large\begin{center}\textbf{Abstract}\end{center}\normalsize

The analysis and the simulation of fluid extruded out of a die opening are of great importance in different industrial applications. Mathematical modeling and experimental studies of extrusion out of a constant die gap has been the subject of numerous publications in the literature. Extrusion Blow Molding is a polymer forming technique used to manufacture a wide range of large hollow parts such as fuel tanks used in the automotive industry. The die gap is varied during extrusion to adjust the manufactured product's final thickness and to compensate for uneven stretching during molding. In this work, mathematical modeling of varying die gap extrusion is introduced. Two models using different approaches are presented to model the time dependent effects of varying the die gap on the extrudate.

In the first approach, physics based partial differential equations are used to simulate extrusion. The Navier Stokes equations are used as governing equations to simulate molten polymers. The governing equations are solved using Arbitrary Lagrangian Eulerian based Finite Element Method. A novel numerical method for defining the free surface of the extrudate is proposed. In addition, the effects of varying the die gap on the material velocity, pressure and extrudate shape are reported.

In the second approach, a parameter identification scheme is used to obtain the parameters of a novel model with a predefined structure. The purpose of the proposed model is to replicate time dependent extrudate thickness of the physics based model. The parameters of the model are identified by minimizing the response error between the two models due to step inputs. The parameter identification based model provides a computationally low cost alternative to the physics based model. Moreover, the parameter identification based model has a transport Partial Differential Equation/Ordinary Differential Equation cascade structure that is suitable for designing feedback controllers for Extrusion Blow Molding. 

\vfill
\newpage
\addcontentsline{toc}{chapter}{Abr\'{e}g\'{e}}
\Large\begin{center}\textbf{Abr\'{e}g\'{e}}\end{center}\normalsize

L'analyse et la simulation d'un fluide extrud\'{e} \`{a} partir d'un orifice de matrice rev\^{e}tent une grande importance dans diff\'{e}rentes applications industrielles. La mod\'{e}lisation math\'{e}matique et les \'{e}tudes exp\'{e}rimentales sur l'extrusion \`{a} partir d'un orifice de section constante ont fait l'objet de nombreuses publications dans la litt\'{e}rature. Le moulage par extrusion-soufflage est une technique de formage de polym\`{e}re utilis\'{e}e pour fabriquer une vaste gamme de grandes pi\`{e}ces creuses telles que les r\'{e}servoirs de carburant utilis\'{e}s dans l'industrie automobile. L'espace libre entre les fili\`{e}res varie pendant l'extrusion pour ajuster l'\'{e}paisseur finale du produit fabriqu\'{e} et pour compenser l'\'{e}tirement irr\'{e}gulier pendant le moulage. Dans ce travail, nous introduisons la mod\'{e}lisation math\'{e}matique de diff\'{e}rents types d'extrusion. Deux mod\`{e}les utilisant des approches diff\'{e}rentes sont pr\'{e}sent\'{e}s pour mod\'{e}liser les effets d'une variation de l'intervalle de fili\`{e}re sur l'extrudat en fonction du temps.
 
Dans la premi\`{e}re approche, des \'{e}quations aux d\'{e}riv\'{e}es partielles bas\'{e}es sur la physique sont utilis\'{e}es pour simuler l'extrusion. Les \'{e}quations de Navier Stokes sont utilis\'{e}es comme \'{e}quations de base pour simuler des polym\`{e}res fondus. Les \'{e}quations qui gouvernent cette dynamique sont r\'{e}solues \`{a} l'aide de la m\'{e}thode des \'{e}l\'{e}ments finis bas\'{e}e sur la m\'{e}thode arbitraire lagrangienne eul\'{e}rienne. Une nouvelle m\'{e}thode num\'{e}rique pour d\'{e}finir la surface libre de l'extrudat est propos\'{e}e. De plus, les effets de la variation de l'intervalle de fili\`{e}re sur la vitesse du mat\'{e}riau, la pression et la forme de l'extrudat sont rapport\'{e}s.
 
Dans la seconde approche, un sch\'{e}ma d'identification de param\`{e}tre est utilis\'{e} pour obtenir les param\`{e}tres d'un nouveau mod\`{e}le avec une structure pr\'{e}d\'{e}finie. Le mod\`{e}le propos\'{e} a pour objectif de reproduire l'\'{e}paisseur d'extrudat du mod\`{e}le bas\'{e} sur la physique en fonction du temps. Les param\`{e}tres du mod\`{e}le sont identifi\'{e}s en minimisant l'erreur de r\'{e}ponse \`{a} l'\'{e}chelon entre les deux mod\`{e}les. Le mod\`{e}le bas\'{e} sur l'identification des param\`{e}tres fournit une alternative au mod\`{e}le bas\'{e} sur la physique, \`{a} faible co\^{u}t en calcul. De plus, le mod\`{e}le bas\'{e} sur l'identification de param\`{e}tres a une structure cascade d'\'{e}quation aux d\'{e}riv\'{e}es partielles de transport de mati\`{e}re et d'\'{e}quation diff\'{e}rentielle ordinaire qui convient \`{a} la conception de contr\^{o}leurs de r\'{e}troaction pour le moulage par extrusion-soufflage.

\vfill
\newpage

\addcontentsline{toc}{chapter}{Claims of Originality}
\Large\begin{center}\textbf{Claims of Originality}\end{center}\normalsize

The thesis includes the following contributions.

\begin{enumerate}
\item Most work related to fluid extrusion in the literature address constant die gap extrusion. In this work, Finite Element Method is used to simulate a configuration where the die gap varies during extrusion. The effects of varying the die gap on the material velocity, pressure and extrudate shape is presented and discussed. The extrudate shape exhibits newly reported phenomena such as bulging and necking. Different factors affecting bulging and necking are studied and analyzed. To the author's knowledge, a Finite Element simulator with a varying die gap has not been published in the literature. 

\item Varying the die gap during extrusion results in large extrudate thickness variations compared to constant die gap extrusion. A novel scheme is presented to define the free surface of the extrudate. The scheme introduces extra nodes that represent the free surface. The larger number of extra nodes defines the free surface with higher resolution, which is especially important to represent the large free surface variations. The new scheme is shown to eliminate free surface oscillations that occur when coarse meshes are used for higher Reynolds numbers. 

\item In addition to the Finite Element simulator, a novel model for modeling extrudate thickness of a varying die gap extruder is proposed. Instead of understanding the physics behind a very complex process, the intended output is replicated using a novel parameter identification based approach. The proposed model addresses some of the limitations of the FEM simulator such as complexity and large computational cost.

\item The thesis introduces readers with Control Systems background to different Computational Fluid Dynamics concepts used for simulating extrusion. In addition, the reader is presented with discussions and explanations for the different phenomena observed during extrusion. Moreover, readers with Computational Fluid Dynamics background are introduced to modeling and identification techniques usually used in Control Systems.
\end{enumerate}

This thesis contributes to the following peer reviewed publication:
\begin{itemize}
\item R. Toukhtarian, S. G. Hatzikiriakos, H. Atsbha, B. Boulet, Modeling Polymer Extrusion With Varying Die Gap Using Arbitrary Lagrangian Eulerian (ALE) Method, Physics of Fluids 30 (9) (2018) 093103.
\begin{itemize}
\item R. Toukhtarian: Developing the proposed simulator. 
\item S. G. Hatzikiriakos: Sharing expertise in Rheology and Computational Fluid Dynamics. The coauthor also participated in editing the manuscript.
\item H. Atsbha: Sharing practical manufacturing expertise and editing the manuscript.
\item B. Boulet: Supervision of the work and editing the manuscript.
\end{itemize}
\end{itemize}

An invention disclosure is submitted to the Office of Innovation at McGill University. 
\begin{itemize}
\item R. Toukhtarian, B. Boulet, Parameter Identification Based Model with Transport PDE/Nonlinear ODE Cascade Structure for the Prediction of Polymer Extrusion with Varying Die Gap.
\begin{itemize}
\item R. Toukhtarian: Developing the proposed model. 
\item B. Boulet: Supervision of model development.
\end{itemize}
\end{itemize}

After reporting the invention, the following manuscript will be submitted to peer-reviewed journal:
\begin{itemize}
\item R. Toukhtarian, M. Darabi, S. G. Hatzikiriakos, H. Atsbha, B. Boulet, Parameter Identification of Transport PDE/Nonlinear ODE Cascade Model for Polymer Extrusion with Varying Die Gap.
\begin{itemize}
\item R. Toukhtarian: Developing the proposed model. 
\item M. Darabi: Aid in development of the proposed model. 
\item S. G. Hatzikiriakos: Sharing expertise in Rheology and Computational Fluid Dynamics. The coauthor also participated in editing the manuscript.
\item H. Atsbha: Sharing practical manufacturing expertise and editing the manuscript.
\item B. Boulet: Supervision of the work and editing the manuscript.
\end{itemize}
\end{itemize}

\vfill
\newpage

\addcontentsline{toc}{chapter}{Acknowledgments}
\acknowlege{\makeacknowlege}

\tableofcontents\newpage
\newpage
\addcontentsline{toc}{chapter}{List of Figures}
\listoffigures
\newpage
\addcontentsline{toc}{chapter}{List of Tables}
\listoftables\newpage
\renewcommand{\nomname}{List of Symbols and Abbreviation}
\printnomenclature 
\hfill
\newpage
\end{preliminary}


\chapter{Introduction}
\label{ch:intro}

\section{Background}
\label{back}
The thesis addresses the modeling of extrusion with varying die gap, which is especially important for Extrusion Blow Molding (EBM). In this section, an overview and background of the thesis is presented.
\subsection{Project Overview}
This work is part of a larger project funded through an Automotive Partnership Canada program. The different teams participating in the project are Kautex Textron Inc, McGill University, University of British Colombia (UBC), and the National Research Council of Canada (NRC). Kautex Inc. is one of the leading companies manufacturing fuel tanks used in the automotive industry. One of the novel products of Kautex Inc. is the Next Generation Fuel System (NGFS), which is produced by EBM. The main purpose of the project is to find new techniques that would improve the manufacturing process of the NGFS. The NGFS is produced by a twin sheet extrusion process, which allows the integration of components within the tank. This manufacturing technique distinguishes the NGFS from other gas tank manufacturing processes. The new technology also reduces tank weight and emissions while meeting stringent specifications such as the ability to handle high pressures \cite{NGFS}. 

Every team in the project has a specific task. The role of the Kautex team is to share practical experience and highlight manufacturing challenges for the other research teams to consider. The role of the UBC team is to study the rheological properties of the materials used and the development of models that can replicate the material behavior. The role of the NRC team is the development of simulators that simulate the overall EBM process. The role of the McGill team is to investigate ways to automate the process by applying feedback techniques.

\subsection{Extrusion Blow Molding}
\subsubsection{Overview}
Blow Molding is a plastic forming technique used to manufacture hollow parts. Blow molding is used to produce a wide range of products such as sail boats, toys, bumpers, and different types of containers. The main advantages of EBM over other molding techniques such as injection blow molding is the ability to manufacture larger parts with low tooling cost. The main drawback of EBM is the limited control over the final product's thickness, which is an essential feature of the product \cite{Rosato}. 

\begin{figure}
\centering
	\includegraphics[width=0.6\linewidth]{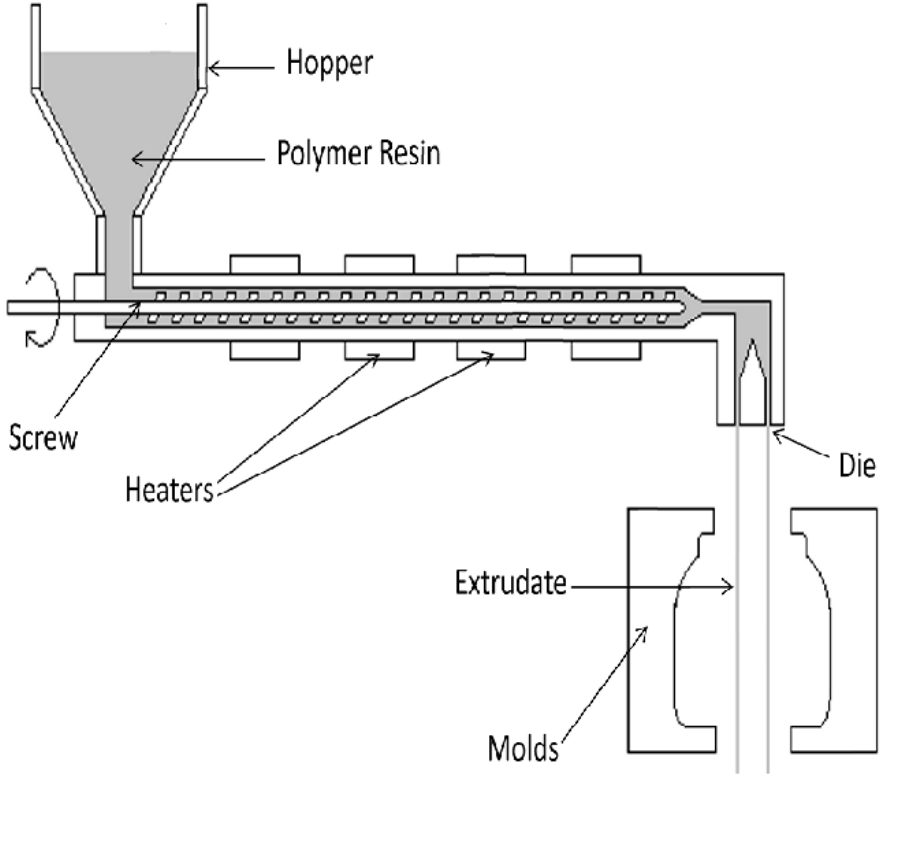}
	\caption{Extruder diagram.}
	\label{fig:Ext}
\end{figure}
 
The three main steps of EBM are extrusion, molding, and cooling. In the extrusion process, polymer resin is fed into a hopper. Then, the resin is fed into an extruder that grinds and transports it using a rotating screw. During grinding, the polymer temperature increases due to friction causing the polymer to melt. Some extruders are enhanced with extra heaters, which aid in increasing and regulating the polymer temperature. The molten polymer is then forced out of a die, forming a vertically suspended extrudate. Different types of dies produce different extrudate shapes. Slit dies produce sheet shaped extrudates while annular dies produce cylindrical extrudates. During molding, the suspended extrudate is forced into a mold by suction or blowing causing the extrudate to acquire the inner shape of the mold. After molding, the mold is cooled by air or water and the molded part is taken out of the mold. Finally, the extra plastic at the edges is trimmed. The trimmings may go through regrind to be reused in the hopper. Figure \ref{fig:Ext} shows the different parts of the EBM machine.

\subsubsection{Die Gap Programming}
\begin{figure}
\centering
	\includegraphics[width=0.4\linewidth]{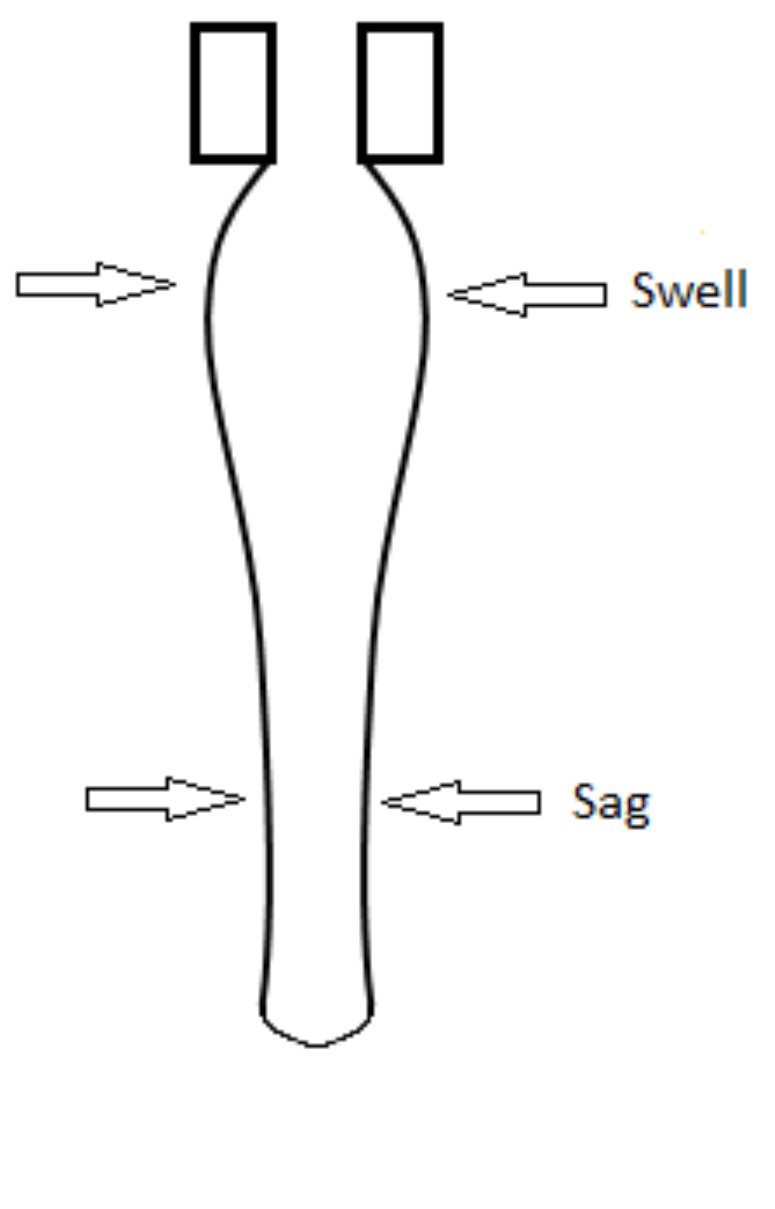}
	\caption{Thickness profile of a sheet extrudate.}
	\label{fig:sheet}
\end{figure}

The ability to adjust the final polymer product thickness is crucial to obtain a product with desired physical characteristics. In some cases, it is desirable to have an even thickness along the walls of the final product. During extrusion, molten polymer exiting the die opening may show the swelling effect: The thickness of the polymer extrudate becomes larger than the die opening due to the sudden absence of the die walls. In addition, the extrudate may show the sagging effect due to stretching caused by its own weight under gravity. Figure \ref{fig:sheet} shows the swelling and sagging effects on the thickness profile of the cross section of a sheet shaped extrudate. 

During molding, different parts of the extrudate are subjected to different degrees of stretching. Stretching of the extrudate is dependent on the mold shape. The varying degree of stretching at different areas will cause non-uniform thinning along the extrudate. 

In order to control the final part thickness and compensate for the sag, swell and uneven stretching, the die opening is varied during extrusion. The machine operator inputs a reference die opening and a controller forces the die opening to be equal to the reference. The die gap of the extruder is measured by a sensor. The sensor measurement is used by the controller to adjust the die gap by an actuator such as a hydraulic actuator.

\begin{figure}
\centering
	\includegraphics[width=0.7\linewidth]{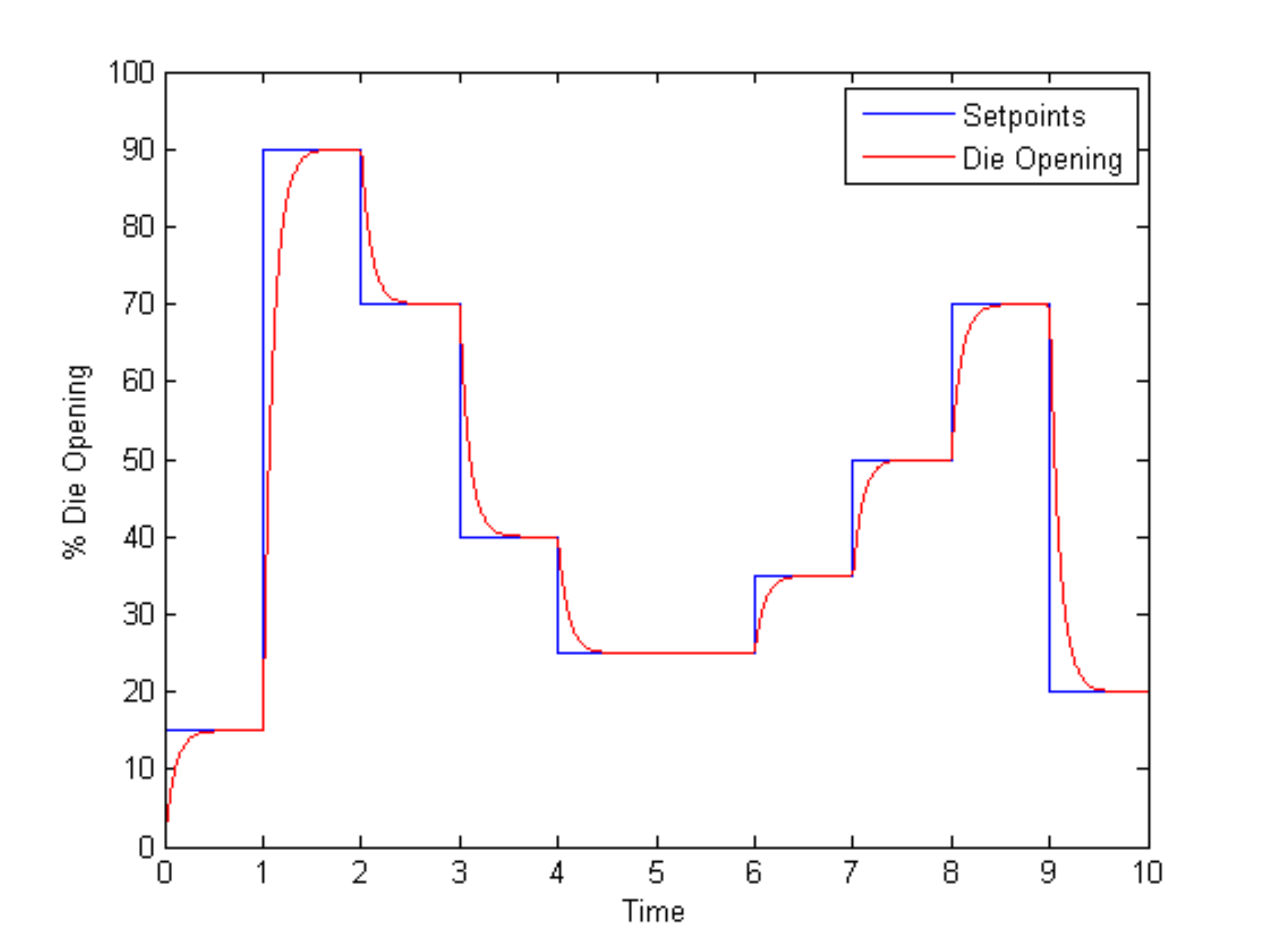}
	\caption{Setpoints and die gap change with respect to time.}
	\label{fig:Set}
\end{figure}

The time needed for the extrusion process (extrusion cycle) is divided into equal intervals in which the die gap reference is set to a constant value called setpoint. Figure \ref{fig:Set} shows an example of the variation of the setpoints and the die opening with respect to time. The figure shows how the die gap tends to be equal to the reference setpoints assigned by the operator. In Figure \ref{fig:Set}, the extrusion cycle is divided into 10 equal intervals. The task of tuning the setpoints in order to obtain the desired thickness profile of the molded part is known as die gap programming (parison programming for annular dies) \cite{Haile1,Haile2,Opti,yu,huang2015}.

\subsubsection{Challenges}
In principle, die gap programming is done when a new product is introduced. Die gap programming is a very demanding task usually done by trial and error. Computer aided engineering (CAE) tools such as BlowView\textsuperscript{\textregistered} and Polyflow\textsuperscript{\textregistered} are used to gain an insight into the process. However, CAE tools are not typically used in the industry for prediction due to their limited accuracy. Machine operators spend days trying to find proper setpoints that produce an acceptable thickness profile for a new product. In addition to being time consuming, die gap programming is expensive due to the large number of scrapped parts generated during trial and error.

EBM is sensitive to numerous factors that make EBM susceptible to machine drift. In some cases, introducing a new batch of polymer resin of the same kind will cause variations in the product thickness. In addition, a portion of the used raw material may be from regrind of scrapped parts. Therefore, the machine operators should stay cautious to identify any machine drift in order to retune the setpoints when needed.  

\section{Modeling and Simulation of Polymer Extrusion}
The main objective of this thesis is the modeling and the simulation of the extrusion process, which is a challenging problem in the literature. Computer modeling of fluid extrusion is not just used for EBM, it is also valuable for different industrial applications such as ink-jet printing and cable coating \cite{spectral2015}. Fluid forced out of a die undergoes complex behavior that is dependent on a wide variety of factors \cite{Mitsoulis1,Mitsoulis2,CES_Extrusion1,CES_Extrusion4}. The extrusion modeling and simulation is usually referred to as the die swell or extrusion swell problem. Extrusion is usually simulated using a variety of physics based Partial Differential Equations (PDEs), which have moving boundaries due to the free surface. Furthermore, different types of numerical methods are proposed to solve the governing PDEs. 

\subsection{Governing Equations}
The governing equations modeling fluid motion consist of the conservation and constitutive equations.

\subsubsection{Conservation Equations}
The conservation equations are derived from conservation of mass, momentum and energy laws applied to fluid motion. The conservation of momentum equation is a statement of Newton's second law of motion. It relates the sum of the forces acting on the fluid to the fluid acceleration or rate of change of momentum. The conservation of momentum is given by
\begin{equation}
   \label{eq:Momentum}
   \rho \frac{D\vec{u}}{D t}=\nabla\cdot\bm{\sigma}+\rho \vec{f}.
\end{equation}
$\rho$ is the density, $\vec{u}$ is the velocity vector, $\bm{\sigma}$ is the Cauchy stress tensor and $\vec{f}$ is the body force per unit mass of fluid. $D(\cdot)/D t$ is the material time derivative. 

The fluid moves in way such that mass is neither created nor destroyed. The conservation of mass equation is given by 
\begin{equation}
   \label{eq:Mass}
   \frac{\partial\rho}{\partial t} + \nabla\cdot(\rho\vec{u})=0.
\end{equation}
In the literature, many papers address the extrusion of compressible fluids \cite{compress1,compress2, compress3, compress4}. Polymer melts are only compressible under very high pressures \cite{mitsoulisbook}. Therefore, polymer melts are usually assumed to be incompressible, thus, the time derivative of the density is set to zero in (\ref{eq:Mass}). 

The conservation of energy equation is considered when extrusion  is studied under non-isothermal conditions. In this work, we assume that the extrusion is performed under isothermal conditions. Temperature effects on extrudate swell is studied and reported in different papers \cite{VinodPoF,CES_Extrusion1,noniso1,noniso2}.

\subsubsection{Constitutive equations}
In addition to the conservation equations, a constitutive equation is required to relate the stress to the fluid deformation. The constitutive equation for Newtonian fluids is given by
\begin{equation}
   \label{eq:Constitutive}
   \bm{\sigma} = -p\bm{I} + \underbrace{\mu\left(\nabla\vec{u}+\nabla\vec{u}^T \right)}_{\bm{\tau}}.
\end{equation}
$\bm{I}$, $\mu$ and $\bm{\tau}$ are the unit tensor, the viscosity and the viscous stress tensor, respectively. The constitutive equation for Newtonian fluids is linear. Substituting (\ref{eq:Constitutive}) in (\ref{eq:Momentum}) and assuming incompressible flow in (\ref{eq:Mass}) result in the Navier Stokes equations. Navier Stokes equations are extensively used to model fluids in various scientific and engineering applications. 

Polymer melts exhibit non-Newtonian fluid behavior. Non-Newtonian visco-elastic models are proposed to better simulate phenomena observed in real setups. An appropriate constitutive equation depends on the polymer type, flow problem configuration, and precision required \cite{bird}. Various non-Newtonian constitutive equations including integral, differential and molecular models are proposed to simulate polymer extrusion. The extrudate thickness over the die gap is known as the die swell ratio. The die swell ratio is used as a main criterion to validate the results of different models with experimental setups \cite{ViscoSwell1,ViscoSwell2,ViscoSwell3,ViscoSwell4,Vinod1,Vinod3}. 

\subsection{Numerical Methods}
The governing equations coupled with moving boundaries are very difficult to solve analytically. Different numerical methods are proposed to overcome various challenges encountered while solving the governing equations.

\subsubsection{Spatial Discretization and Temporal Integration}
Finite Element methods (FEM) are the most commonly used methods for spatial discretization of the governing equations in extrusion simulation. After discretization, the equation variables are solved at a finite number of mesh nodes in space. A wide variety of FEM schemes are used for the die swell problem \cite{ALE_DS2,SFEM,elastic,STGFEM1,STGFEM2}. Other methods such as finite difference and finite volume are also used \cite{FDM1,MAC1,FDM3,FVM1}.

Generic implicit and explicit time integration methods such as Backward Euler, Forward Euler and Trapezoidal Rule are used to compute the time dependent solution \cite{Hughes3,HughesBook}. Implicit methods for nonlinear equations are solved using iterative methods such as the Predictor Multi-Corrector method \cite{HughesBook, PoFPMC1, PoFPMC2}. Moreover, solving the incompressibility condition alongside the rest of the governing equations requires appropriate numerical treatment \cite{Hughes1,Hughes2,inco1,inco2}.

\subsubsection{Mesh Movement}
A fundamental consideration when developing a simulator with moving boundaries is the scheme in which the mesh moves. The mesh should cover the moving computational area without being distorted. In Lagrangian methods, the mesh flows with the material. As a result, the mesh will inherently be covering the entire domain while the boundaries are moving. Therefore, Lagrangian methods are effective in defining different material interfaces and free surfaces. Moreover, Lagrangian methods facilitate the use of history dependent constitutive equations since a mesh node trajectory resembles the trajectory of a fluid particle. However, large material deformations result in severe mesh distortion presenting difficulties in Lagrangian methods. Thus, Lagrangian methods are usually used in the simulation of solid mechanics where the deformation is small \cite{Lag1,Lag2,Lag3,Lag4}. 

On the other hand, classical Eulerian methods have a fixed mesh. Such methods are suitable for large material deformation in a defined domain but ineffective in defining free surfaces and material interfaces. Eulerian methods are usually used to simulate fluid flows in predetermined computational domains \cite{PoFLDC, Euler1, Euler2, Euler3}. Arbitrary Lagrangian Eulerian (ALE) methods have been proposed to include the benefits of Eulerian and Lagrangian methods \cite{Sung, Duarte, Donea,kronsteiner, PoFALE1,PoFALE2}. In ALE methods, the mesh moves according to a scheme that aims to cover a moving domain while reducing mesh deformation. Therefore, if a suitable scheme is applied, ALE methods could effectively handle large material deformation while defining domain boundaries accurately.

ALE methods are widely used in extrusion simulations \cite{Ganvir2,spectral2015,ALE_DS,ALE_DS2,ALE_DS3}. In order to use ALE methods, ALE form of the governing equations should be used. The variables in the ALE form of the equations represent physical quantities at nodes moving with arbitrary velocities. Consider a scalar $f$ describing a physical quantity at a given mesh node with a given position. The value of $f$ at a given instant is independent of the node velocity. However, the time derivative of $f$ at a given instant is dependent on the node velocity, since the velocity determines the node position after an infinitesimal amount of time. The relationship between the time derivative of $f$ for a mesh node moving with the material and a mesh node moving with an arbitrary velocity is given by
\begin{equation}
   \label{eq:ALE}
   \frac{Df}{Dt}=\frac{\partial{f}}{\partial{t}}+(\vec{u}-\vec{u}_{m})\cdot\nabla f.
\end{equation}
$\vec{u}$ and $\vec{u}_m$ are the material velocity and mesh node velocity respectively. The material time derivative $Df/Dt$ is equal to the node time derivative $\partial{f}/\partial{t}$ if the node moves with the material ($\vec{u}=\vec{u}_m$). If $\vec{u}_m=0$, the right hand side will represent the material time derivative in an Eulerian framework. Similarly, the relationship between the time derivative of  vector $\vec{u}$ for a mesh node moving with the material and a mesh node moving with an arbitrary velocity is given by
\begin{equation}
   \label{eq:ALEu}
   \frac{D\vec{u}}{Dt}=\frac{\partial{\vec{u}}}{\partial{t}}+\left((\vec{u}-\vec{u}_{m})\cdot\nabla\right) \vec{u}.
\end{equation}
Replacing the material time derivative in (\ref{eq:Momentum}) with the right hand side of (\ref{eq:ALEu}) results in the ALE form of the conservation of momentum equation. 

\begin{equation}
   \label{eq:NS1}
\rho \frac{\partial{\vec{u}}}{\partial{t}}+\rho\left((\vec{u}-\vec{u}_{m})\cdot\nabla\right) \vec{u}=\nabla\cdot\bm{\sigma}+\rho \vec{f}.
\end{equation}

\subsubsection{Free Surface}

\begin{figure}
\centering
	\includegraphics[width=0.7\linewidth]{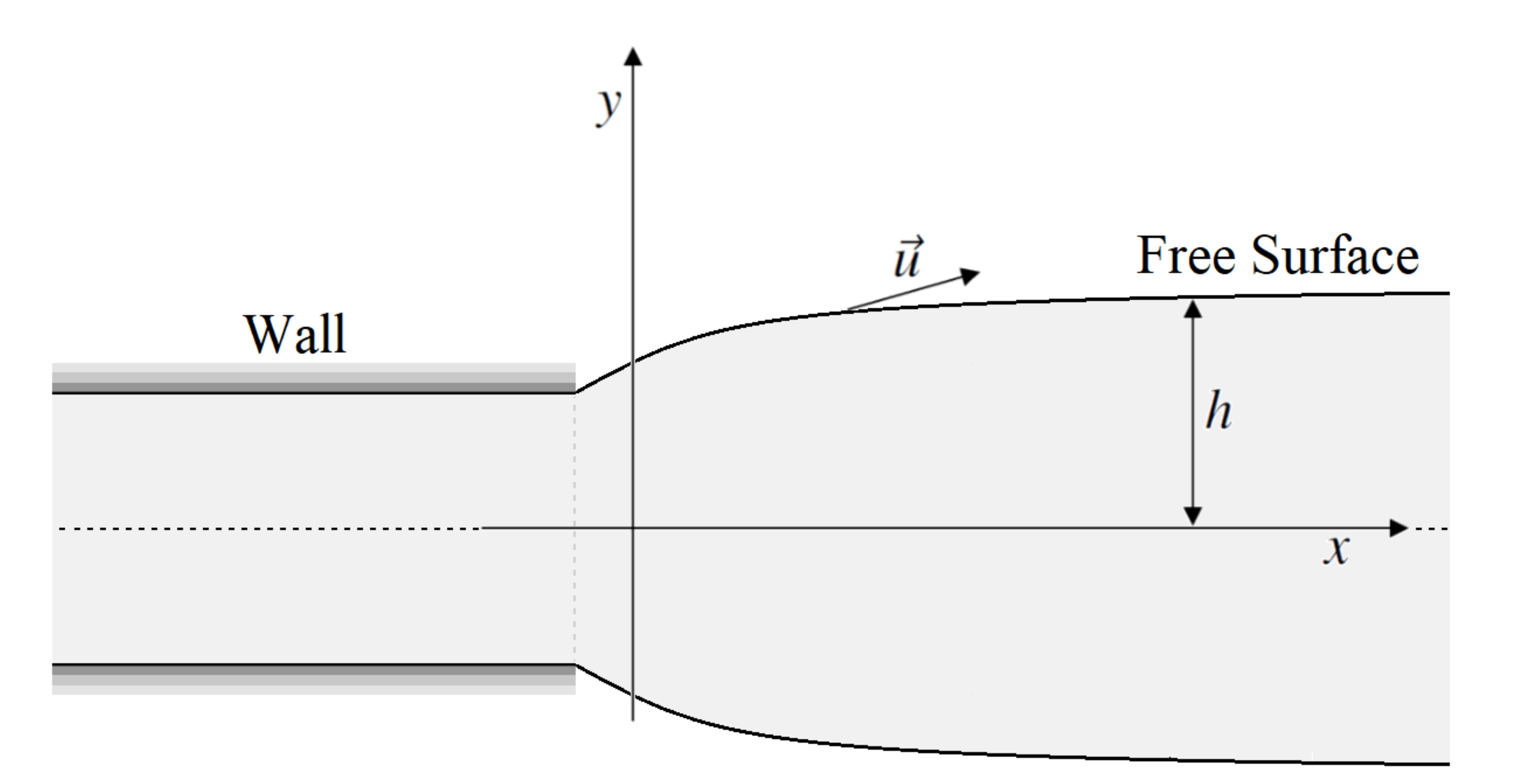}
	\caption{Cross section of slit extruder.}
	\label{fig:FS}
\end{figure}

Simulating the free surface is one of the main challenges in swell prediction. In Figure \ref{fig:FS}, the cross section of a slit die extruder that forms sheet shaped extrudates is depicted. Fluid is forced out between two walls that form the die. The fluid swells once it exits the die. The curves at the boundaries between the fluid and the air represent the free surface. The kinematic equation is used for the extrudate swell problem in different numerical methods. The kinematic equation is given by
\begin{equation}
   \label{eq:kinematic}
   \frac{\partial h}{\partial t}+ u_x \frac{\partial h}{\partial x} = u_y.
\end{equation}
$h$ is the height of the free surface. $u_x$ and $u_y$ are the $x$ and $y$ component of $\vec{u}$, where $\vec{u}$ is the material velocity at the free surface as shown in Figure \ref{fig:FS}. In \cite{Nickell, Omodei1979, Omodei1980}, the kinematic equation in the steady state $(\partial h/\partial t=0)$ is used to simulate the steady die swell using an Eulerian framework. Although classical Eulerian methods have difficulty in defining free surface, the free surface and mesh positions are computed iteratively. These methods have convergence problems for surfaces with large surface tension, where the ratio of the surface tension to viscous force is high. A numerical method for die swell with higher surface tension is described in \cite{Georgiou1988,Silliman}. Molten polymers have high viscosity, thus, surface tension is usually neglected in polymer extrusion. The kinematic equation is also used in ALE based numerical methods for die swell prediction \cite{ALE_DS2,ALE_DS3}. 

Lagrangian methods do not require the kinematic equation since the computational domain moves with the material and determines the free surface. In some time dependent simulations, every time step is divided into two sub-steps. In the first sub-step, the equations in the Lagrangian form are solved predicting the free surface shape. Afterwards, the velocity of the mesh is calculated according to the calculated surface location. Finally, in the second sub-step, the equations in the ALE form are solved \cite{Ganvir,Ramaswamy}. In addition to defining the free surface, the Lagrangian sub-step facilitates solving history dependent constitutive equations for polymers exhibiting memory phenomena. Other generic methods for solving the free surface such as Volume of Fluid (VOF) \cite{VOF1,VOF2} and Marker and Cell (MAC) \cite{MAC1,MAC2,MAC3} are also used for simulating die swell. VOF and MAC solve the Eulerian form of the equations using the Finite Difference Method.

\section{Problem Statement and Scope}
Simulations using physics based governing equations have numerous limitations. Molten plastics exhibit very complex fluid behavior. Finding constitutive equations that model the complex behavior is still an open problem. Moreover, spatial discretization of the governing equations results in a very large number of algebraic equations (steady simulations) or Ordinary Differential Equations (ODEs) (time dependent simulations). The number of equations is usually in the order of thousands and is dependent on the number of discretization nodes. Most of the governing equations include non-linearities, thus, the equations are solved by iterative numerical methods. For time dependent simulations, the equations are solved at every time step, which makes the simulations very computationally demanding. 

When simulating practical setups, simplifications are required due to the high computational cost. Major factors such as temperature and gravity may not be taken into consideration. Moreover, simpler configurations such as simpler die shapes and constant die gap extrusion are considered. In addition, two dimensional simulations are usually proposed instead of three dimensional simulations. Due to modeling and computational difficulties, current simulation methods have a long way to go to adequately represent practical industrial process conditions. 

As mentioned in  subsection \ref{back}, the major drawbacks of EBM are having limited control over the final product thickness and machine drift. Automatic control systems are used to address machine drift and process control in different industrial applications \cite{Benoit1,Benoit2,CES_Feedback1,CES_Feedback2,CES_Feedback3}. The initial objective of this work was to design intelligent controllers to address the limitations of EBM. Different controller were suggested to improve EBM. An In-Cycle controller that adjusts the die gap if the extrudate thickness drifts from an original desired thickness was suggested. For example, if a new material were introduced, a sensor would measure the change in thickness of extrudate and compensate for the errors during the extrusion cycle automatically. 

In addition, a Cycle-to-Cycle controller was suggested to address die gap programming. Instead of implementing die gap programming by trial and error, the controller calculates the setpoints automatically in an iterative process. Extrudate or final product thickness is measured and the error is obtained by comparing the obtained thickness to a desired thickness. The setpoints for the subsequent trial are calculated according to a mathematical model, a control algorithm and the error obtained in the current trial. The Cycle-to-Cycle controller could be implemented only during die gap programming or it can stay operational during manufacturing to automatically compensate for any machine drift. 

Designing the suggested controllers requires a mathematical model of the controlled system. The controllers change the die gap to control the extrudate or final product thickness. Therefore, a dynamical model that has the die gap as an input and the extrudate or product thickness as an output is required. In the literature, models or experimental data addressing the effects of changing the die gap on the extrudate thickness are limited. To our team's knowledge, extrusion with varying die gap has not been addressed in the open literature. To address this issue, the UBC team designed an experimental slit die extruder with an adjustable die gap. Experimental reportings and models based on the experimental extruder are presented in a recent publication \cite{Vinod4}. The McGill team collaborated with the UBC team in an experiment to obtain data from the setup at the UBC campus. Obtaining the required data was largely unsuccessful due to difficulties in implementation, difficulties in measurement and logistical reasons. 

Lack of suitable models and experimental data in the literature shifted the objectives from designing controllers to developing control oriented models for EBM. An FEM simulator using the conservation and constitutive equations of Newtonian fluids is developed. The simulator is developed such that the die closes and opens. The die gap is varied during the simulation in order to predict the effects of die gap change on the extrudate. The FEM simulator is too complicated for controller design, hence, the results of the FEM simulator are used to propose a model suitable for control.

\subsection{FEM Model}
\begin{figure}
\centering
	\includegraphics[width=0.32\linewidth]{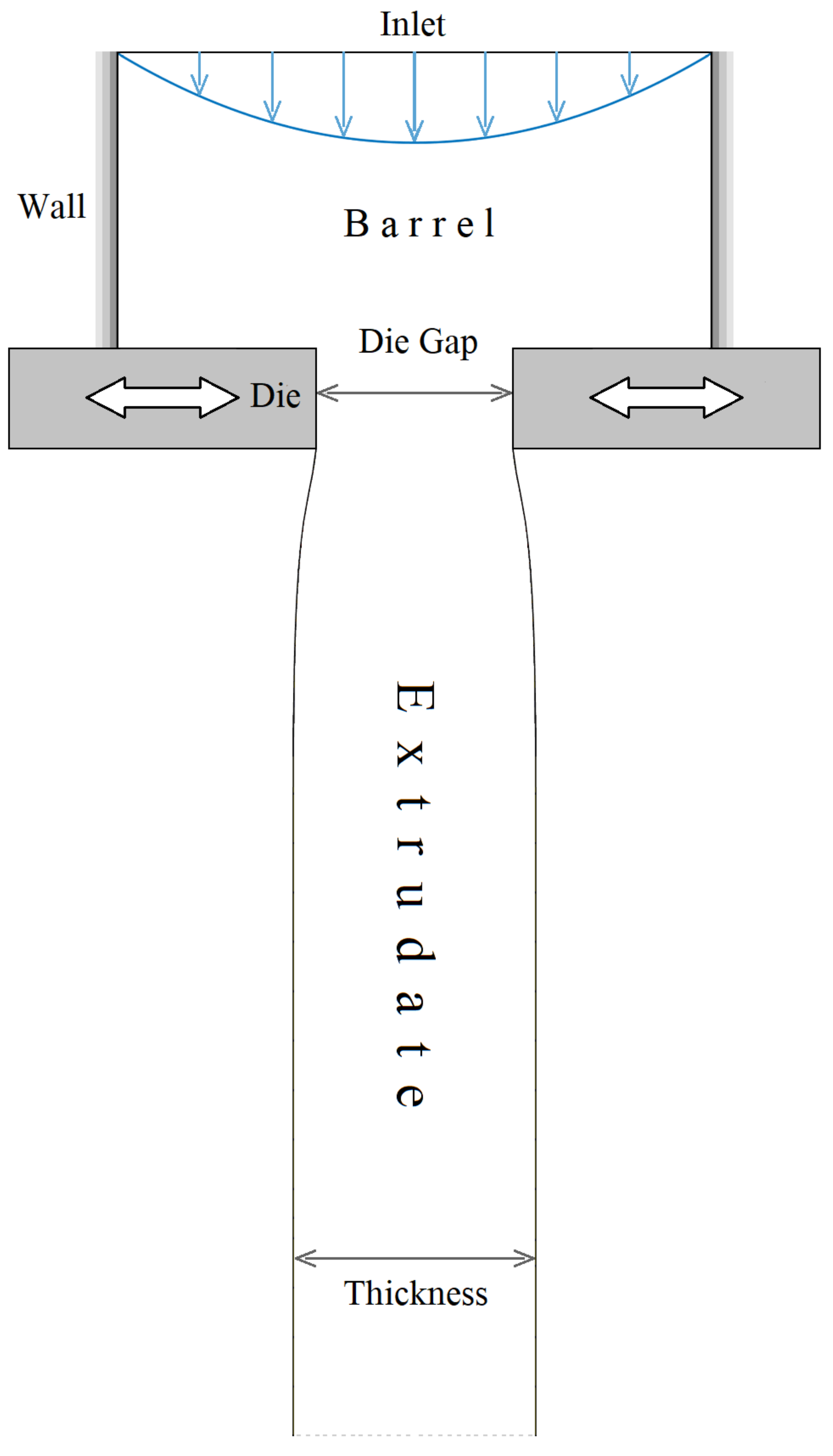}
	\caption{Extruder configuration of FEM simulator.}
	\label{fig:Ext_C}
\end{figure}

A time dependent simulator is developed using the Navier Stokes equations. The Navier Stokes governing equations model isothermal incompressible Newtonian fluids. A Finite Element Method combined with an ALE mesh moving scheme is used to solve the governing equations. Figure \ref{fig:Ext_C} shows the cross section of the slit die extruder configuration used in the simulator. The fluid enters though an inlet into a barrel. Afterwards, the fluid is squeezed out of the die and forms a sheet shaped extrudate. The die walls move toward and away form each other symmetrically to change the die gap. 

The steady die swell ratio results are compared to die swell ratios obtained in the literature for validation. Although the main focus of this work is the die gap change effects on the extrudate thickness, the fluid velocity and pressure are presented and studied given that such results are not reported in the literature. The results show new transient phenomena (bulging and necking) that form on the extrudate due to the die gap change. Different factors affecting bulging and necking such as die length, die gap change and duration of die gap change are examined.

The die gap change causes large variations on the free surface compared to the variations obtained when constant die gap is considered. It is observed that the large surface variations cause numerical oscillations if the mesh is not fine enough. A new method for defining the free surface is introduced to avoid free surface oscillations. The new method enables the use of coarse mesh, which makes it possible to run simulations with shorter computational time. 

\subsection{Parameter Identification based Model}
The simulated physics based model is composed of thousands of nonlinear ODEs that are not suitable for controller design. Good control oriented models are characterized by being simple enough to design controllers, requiring small computational cost to run online, and having the ability to be identified from the process using simple tests. 

A parameter identification based approach is proposed to model the extrusion process. The data used for parameter identification is obtained from an input step based test. The proposed model has a transport PDE/nonlinear ODE cascade structure. Parameters of different blocks of the system are identified by minimizing the error between the response of the FEM simulator and the response of the proposed model. PDE/ODE cascade models are used to simulate and control different applications with complex dynamics \cite{PDEODE1,PDEODE2,PDEODE3,PDEODE4,PDEODE5,PDEODE6,PDEODE7,PDEODE8}. The complex dynamics in these applications can be due to different phenomena such as material transport and wave propagation. 

The parameter identification based model has several advantages over physics based FEM models. First, the parameter identification based model is suitable for controller design. Second, the proposed model is computationally less expensive. Having a computationally cheap model facilitates the implementation of optimization methods that use a large number of trials such as machine learning. Third, the proposed model can model different setup configurations and fluid types. The same identification procedure can be used to simulate a wide range of conditions. Instead of attempting to understand and replicate the physics behind a very complex process, data from a setup could be collected and used to identify the proposed model.

\section{Thesis Outline}

This thesis includes five chapters and three appendices. In Chapter 2, the governing equations and numerical methods used to develop the FEM simulator are discussed. The simulation results of the FEM simulator are presented and analyzed in Chapter 3. In Chapter 4, the parameter identification based model is proposed. Chapter 4 ends with a comparison of the extrudate thickness simulated using the two models presented in this thesis. Finally, in Chapter 5, the thesis is summarized and directions for future research work are proposed. 

\chapter{Modeling and Implementation of the FEM Simulator}

\section{Introduction}
The mathematical models and numerical methods used to simulate the configuration described in Figure \ref{fig:Ext_C} are presented in this chapter. The fluid is assumed to be Newtonian and the Navier Stokes equations are discretized using FEM. An ALE scheme is used to accommodate for the moving die walls and free surface. Every time step is divided into two sub-steps where the governing equations are solved in the Lagrangian and ALE form. The free surface is defined during the Lagrangian sub-step as the nodes move with the material. The ALE sub-step is implemented to enable the mesh to track the free surface while preventing the mesh nodes from flowing downstream with the material. Between the two sub-steps an interpolation based scheme for defining the free surface and computing the ALE mesh velocity is used. When the die gap is varied, thickness variations of the extrudate are larger compared to a constant die gap extrusion. A new interpolation scheme is proposed to define the free surface with higher resolution using the same computational mesh. Defining the free surface with higher resolution enables more accurate thickness variation prediction.

\section{Governing Equations and Boundary Conditions}
The Navier Stokes equations are considered to model isothermal, incompressible, and Newtonian fluid flow. To obtain the non-dimensionalized Navier Stokes equations, lengths are divided by length of the die $L_s$, velocities by the maximum velocity at the inlet $U_{s}$ and time $t$ by $t_s=L_s/U_s$. The pressure and viscous stress tensor are scaled by $\mu U_s/L_s$. Using the dimensionless quantities and neglecting gravity and other external forces, the momentum equation can be written in the ALE form as

\begin{equation}
   \label{eq:NS}
   Re\left(\frac{\partial{\vec{u}}}{\partial{t}}+\left((\vec{u}-\vec{u}_{m})\cdot\nabla\right)\vec{u}\right)=-\nabla{p}+\nabla\cdot\bm{\tau},
\end{equation}
where $Re$ is the Reynolds number defined by $\rho U_s L_s/ \mu$. The conservation of momentum equation used in this work (\ref{eq:NS}) is obtained by replacing (\ref{eq:Constitutive}) in (\ref{eq:NS1}). The conservation of momentum equation is coupled with the conservation of mass equation given by
\begin{equation}
   \label{eq:Continuity}
   \nabla\cdot\vec{u}=0.
\end{equation}
The conservation of mass equation (\ref{eq:Continuity}) is obtained by setting $\partial{\rho}/\partial{t}$ in (\ref{eq:Mass}) to zero to model incompressible flows.

\begin{figure}
    \centering
	\includegraphics[width=\linewidth]{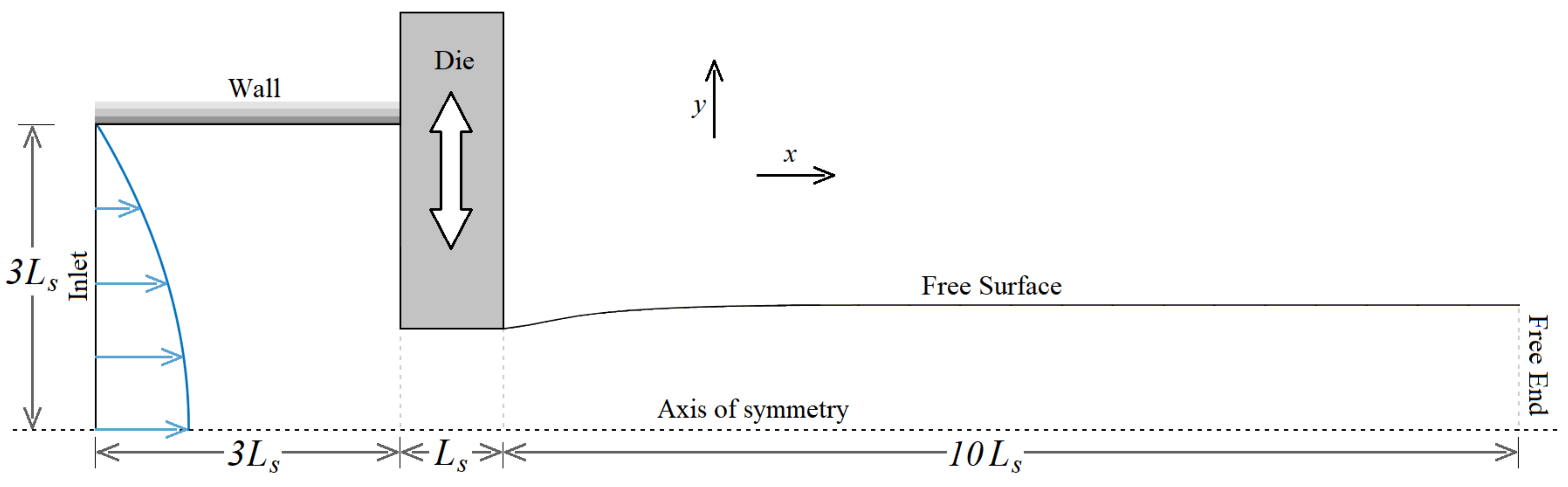}
	\caption{Computational domain.}
	\label{fig:Ext2}
\end{figure}

The configuration in Figure \ref{fig:Ext_C} is symmetrical, hence, the flow field is symmetrical as well. Therefore, only half of the fluid domain is considered as shown in Figure \ref{fig:Ext2}.
Figure \ref{fig:Ext2} shows the computational domain's dimensions. The flow is assumed to be a fully developed Poiseuille flow at the inlet. A no slip condition is assumed at the wall and the die. The wall is stationary, therefore, the velocity and acceleration of the fluid at the wall are equal to zero. The die moves in a direction perpendicular to the axis of symmetry thus changing the exit gap. Therefore, the fluid in contact with the die has the same velocity and acceleration as that of the die, which are always in the $y$-direction. The point of intersection of the wall and the die is assumed to be part of the wall. At the axis of symmetry, the non-diagonal elements of the viscous stress tensor and the vertical velocities are equal to zero $(\tau_{xy}=\tau_{yx}=u_y=0)$. At the free surface and free end, $\vec{n}\cdot\boldsymbol{\sigma}\cdot\vec{t}=\vec{n}\cdot\boldsymbol{\sigma}\cdot\vec{n}=0$ where $\vec{n}$ and $\vec{t}$ are the normal and tangential vectors to the boundary. 

It is assumed that the die gap has been constant for a long time before $t=0$. Therefore, the initial node coordinates, velocities and pressures are computed using a steady-state solver. The solver uses a Picard iteration that is classified as a kinematic iteration scheme \cite{Georgiou1988,Omodei1979,Omodei1980}. The steady state variables must satisfy (\ref{eq:NS}) and (\ref{eq:Continuity}) where $\vec{u}_m$ and $\partial{\vec{u}}/\partial{t}$ are equal to zero. In addition, the node velocities and positions at the free surface must satisfy the steady kinematic condition given by
\begin{equation}
   \label{eq:Kinematic}
   \frac{\partial{h}}{\partial{x}}=\frac{u_y}{u_x}.
\end{equation}
The steady kinematic equation (\ref{eq:Kinematic}) is obtained by setting $\partial{h}/\partial{t}$ in (\ref{eq:kinematic}) to zero since the variables are computed at steady state. The steady kinematic condition ensures that the free surface is a streamline where a particle on the surface has a trajectory that remains on the free surface. Therefore, the velocity of the particle should always remain tangent to the trajectory and the free surface. 

\section{Numerical Implementation}
Equations (\ref{eq:NS}) and (\ref{eq:Continuity}) are spatially discretized using the Galerkin Finite Element Method (GFEM). The trapezoidal rule is used for temporal discretization. At every time step, two consecutive sub-steps are performed \cite{Ganvir,Ramaswamy}. In the first sub-step, fluid particle displacement is computed by solving the Lagrangian form of (\ref{eq:NS}) and (\ref{eq:Continuity}) where $\vec{u}_m$ is set to $\vec{u}$. The mesh node positions are part of the unknowns since the mesh moves with the material. The new node positions define the free surface in the next time step. In the second sub-step, the new free surface position is used to compute $\vec{u}_m$ which has components only in the $y$-direction to prevent the mesh from moving downstream with the flow. Once $\vec{u}_m$ is computed, the ALE form of (\ref{eq:NS}) and (\ref{eq:Continuity}) is computed in the second sub-step.
\subsection{Spatial and Temporal Discretization}
\begin{figure}
    \centering
	\includegraphics[width=.5\linewidth]{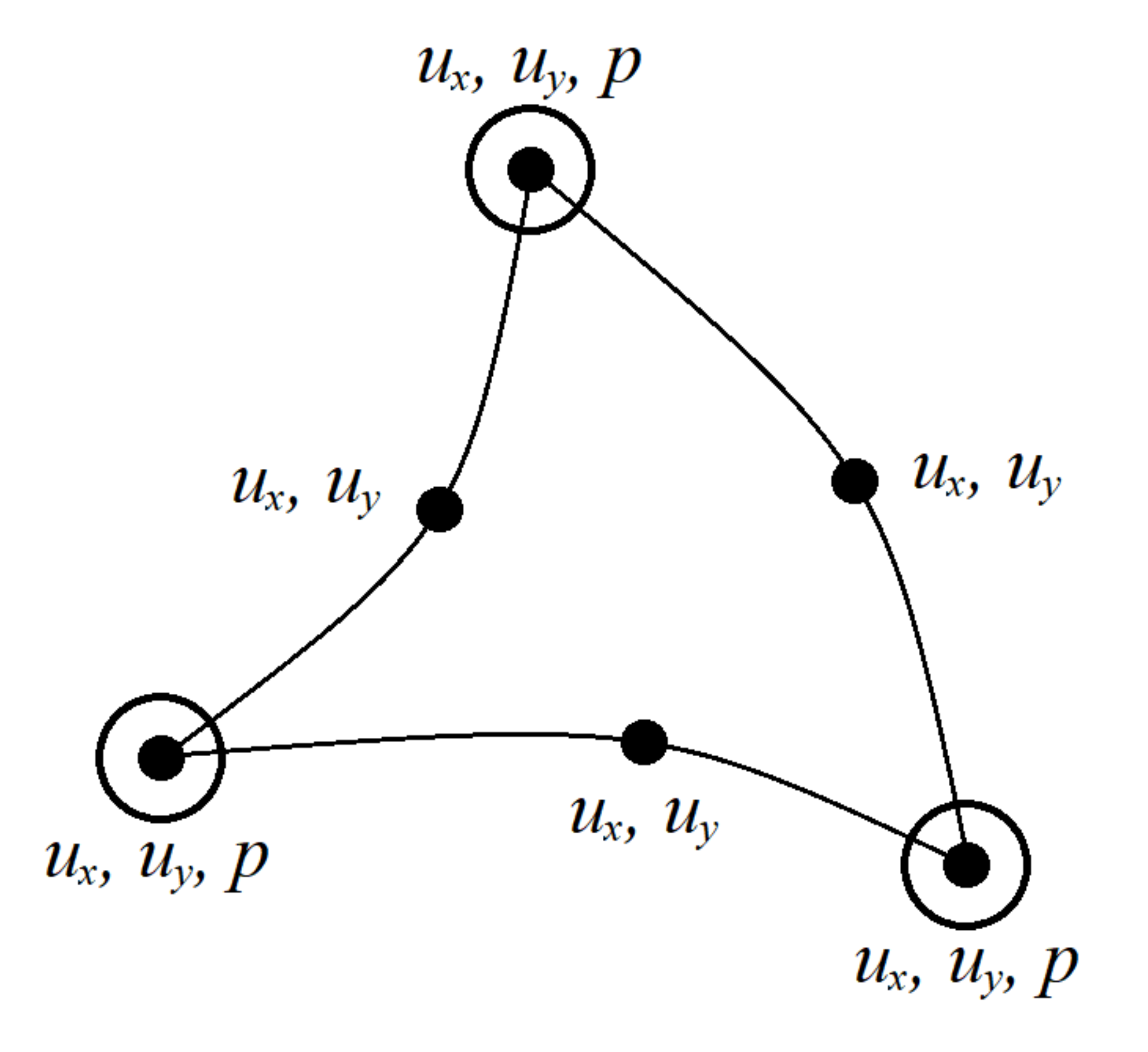}
	\caption{Taylor-Hood $P_1-P_2$ isoparametic triangular element.}
	\label{fig:element}
\end{figure}

Standard GFEM is used to spatially discretize the governing equations. Taylor-Hood $P_1-P_2$ isoparametic triangular elements are used. Each element is composed of 6 velocity and 3 pressure nodes as shown in Figure \ref{fig:element}. The element satisfies the Ladyzhenskaya, Babuska and Brezzi (LBB) condition. Quadratic basis functions are used to interpolate the velocity, while linear basis functions are used for the pressure \cite{Gresho}. After spatial discretization, the trapezoidal rule is used to discretize in time. It is second order accurate in time and considered as part of the Newmark family of time integration schemes \cite{Chung1993,Jansen2000}.

\subsection{Predictor Multi-Corrector Method}
The discretized equations are solved using the Predictor Multi-Corrector method for incompressible fluids  \cite{HughesBook, Hughes1, Hughes2, PoFPMC1, PoFPMC2}. Predictor Multi-Corrector method is an iterative method where the user specifies the number of iterations or a desired error tolerance. In this work, the solution is reached when the error $e_{PMC}$ is smaller than the error tolerance $e_{tol}$.
\begin{equation}
   \label{eq:Error}
   e_{PMC} := \frac{\left\Vert V_{i+1}^{n+1}-V_{i}^{n+1}\right\Vert_{2}}{\left\Vert V_{i+1}^{n+1}\right\Vert_{2}},
\end{equation}
where $V_{i}^{n}$ is a vector containing node velocities at time step $n$ and iteration $i$. $e_{tol}=10^{-5}$ is used in this work.

\section{Free Surface Estimation}
After the Lagrangian sub-step, the new free surface position is estimated at the original $x$-location of the node before its drift. The estimated free surface node position is set to be the position of the node at the end of the time step causing the mesh to move only in the $y$-direction. The positions of the inner nodes are calculated according to the nodes at the boundaries. The mesh velocity, $\vec{u}_m$, is calculated by dividing the difference in the node positions by the time step $\Delta t$. $\vec{u}_m$ is used in the second sub-step where the ALE form of the equations is solved.
In this section, a method for estimating the free surface position is revisited and a new method is proposed.
\begin{figure}
	\includegraphics[width=0.7\linewidth]{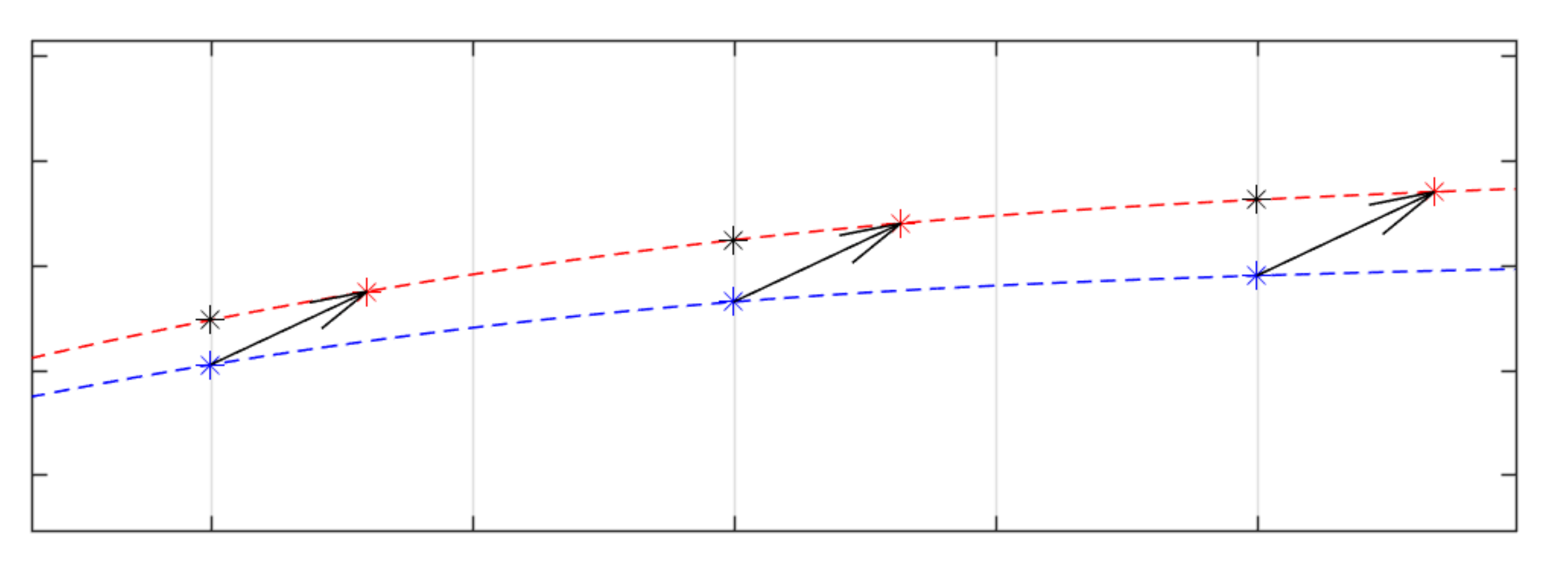}
	\centering
	\caption{Free surface interpolation.}
	\label{fig:Surf1}
\end{figure}
\subsection{Free Surface Interpolation}
In \cite{Ganvir}, the location of the free surface is estimated using cubic spline interpolation derived based on the position of the nodes at the end of the Lagrangian sub-step. In \cite{Ganvir}, the die gap is constant therefore the free surface forms a smooth curve with small variations between consecutive nodes. The blue asterisks in Figure \ref{fig:Surf1} represent the nodes at the beginning of the time step. After the Lagrangian sub-step, the nodes drift with flow to the new position represented with the red asterisk. The new position of the nodes is used to derive a cubic spline function. The black asterisks represent the interpolation of the new free surface node position at the original $x$-location of the nodes using the cubic spline function. 

\subsection{Free Surface Interpolation with Extra Nodes}
When varying die gap is considered, larger thickness variations of the extrudate are expected. The large variation may necessitate a finer mesh to represent the free surface accurately. Finer mesh will require smaller time steps causing the simulation time to increase drastically. To address this issue, a new method for estimating the free surface position is proposed. The proposed method increases accuracy and stability.
\begin{figure}
	\includegraphics[width=.7\linewidth]{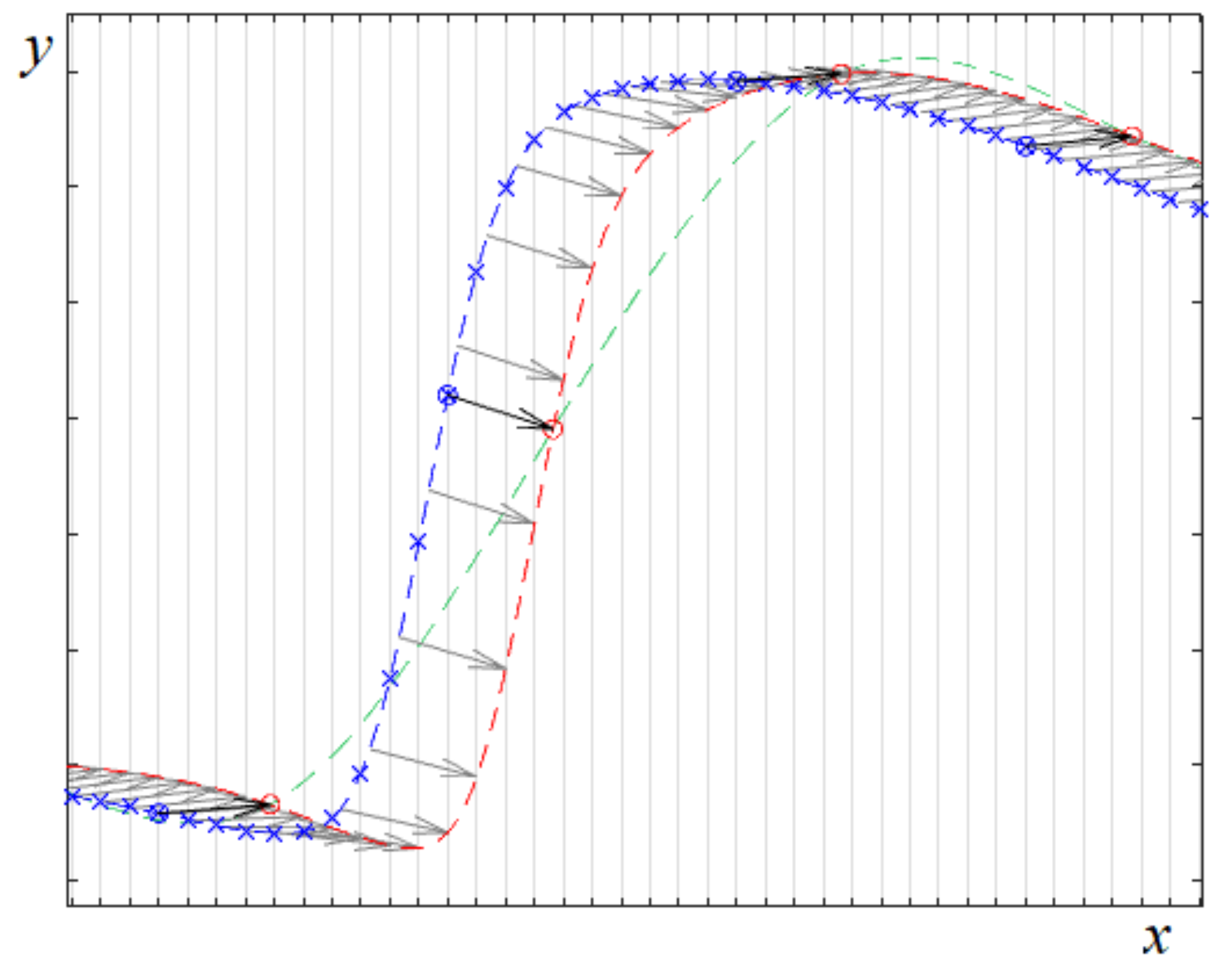}
	\centering
	\caption{Free surface interpolation with extra nodes.}
	\label{fig:Surf2}
\end{figure}

In this method, a user defined number of extra nodes are added on the free surface. The extra nodes are represented as blue crosses in Figure \ref{fig:Surf2} while the main nodes are represented by circles. The extra nodes are used to derive a cubic spline function that will estimate the free surface position with higher precision for the next time step. The velocities and pressures are only computed for the main nodes. The black vectors represent the main nodes' drift during the Lagrangian calculation. Since the mesh nodes move with the material, the vectors represent fluid particle displacements at the nodes. The vertical gray lines represent the $x$-axis location of the nodes at the end of the time step. The gray vectors represent the estimated drift of the nodes that end up at the vertical gray lines. The $x$-axis location of the particles at the beginning of the time step, that will end up at the vertical gray lines are known ($x$-axis location of vertical grey lines), while the $y$-axis locations are the intended unknowns. The gray vectors are estimated by interpolating the black vectors. This method is divided into four main steps. 

\begin{enumerate}
  \item Estimate the $x$-component of the gray vectors by interpolating the $x$-component of the black vectors to obtain the $x$-axis location of the nodes at the beginning of the Lagrangian sub-step. Subtract the $x$-component of each gray vector from the $x$-axis location of the corresponding vertical line to obtain the $x$-location of the tail of each gray vector.
  \item Using the $x$-axis location of the tails of the gray vectors, compute the $y$-axis location of the particles at the beginning of the time step using the cubic spline function of the free surface derived from the previous time step.
  \item Estimate the $y$-component of the gray vectors by interpolating the $y$-component of the black vectors. Add the $y$-component of the gray vectors to the $y$-axis location computed in the second part. This will yield in the $y$-axis location of the tips of the gray vectors.
  \item Derive a cubic spline function of the free surface using the gray vectors' tips. The derived cubic spline function will be used in Step 2 of the next time step. 
\end{enumerate}

The dashed red curve represents the free surface estimation using the proposed method while the dashed green curve represent the free surface estimation if the previous method is used. It can be observed that a cubic spline function using only the main nodes is inadequate to represent curves with large variations. The new method computes two extra cubic spline functions for the $x$ and $y$ components of the gray vector. The computational time needed to evaluate these functions is much shorter than the extra time needed to run the simulation with a finer mesh (larger number of variables), which requires a smaller $\Delta t$ (larger number of time steps).

\section{Mesh}
\begin{figure}
	\includegraphics[width=\linewidth]{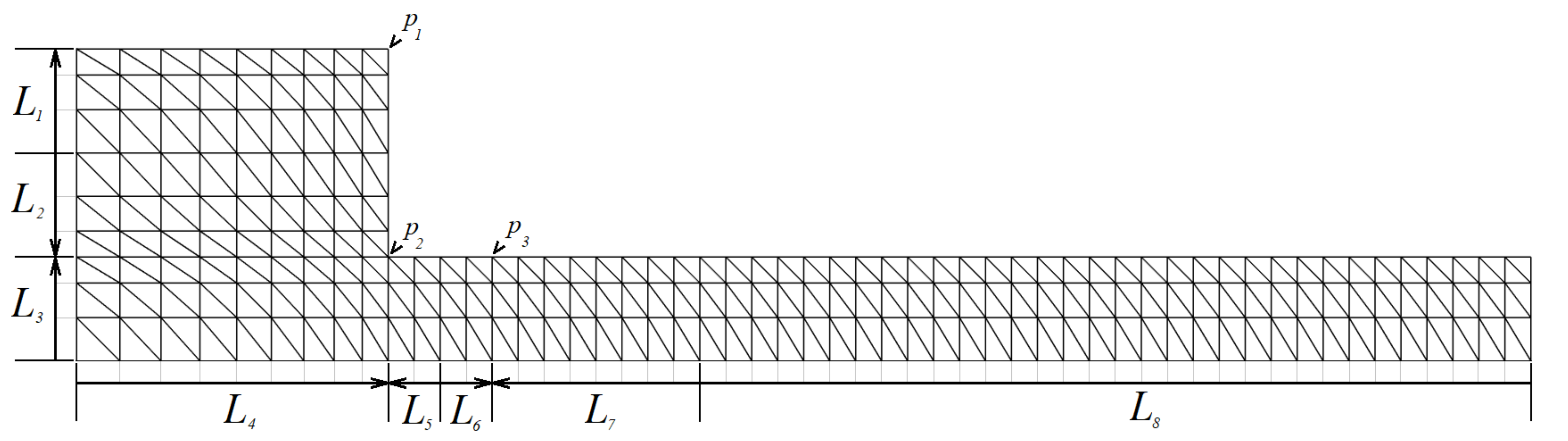}
	\centering
	\caption{Segments of Mesh2.}
	\label{fig:Seg}
\end{figure}

A custom mesh generator is developed for this work. Every mesh is generated according to 8 segments that are divided into several sub-segments as seen in Figure \ref{fig:Seg}. The elements of the mesh are right triangles before the simulation starts and the sub-segments are equal to the lengths of adjacent and opposite sides of the triangles. The user defines the length $L_i$, length of minimum sub-segment $L_{min(i)}$ and the number of sub-segments $n_i$ for each segment $i$. Segment 8 is divided equally, so only $L_8$ and $n_8$ is defined by the user. For Segments 1 to 7, the length difference between consecutive sub-segments is equal to $\left(L_i-(n_iL_{min(i)})\right)/\left(\sum_{k=1}^{k=n_i-1}k\right)$. $L_{min(i)}$ should be chosen greater or equal to $L_i/n_i$. In Figure \ref{fig:Seg}, the arrows point to the location of the smallest sub-segment of a segment. The location of the smallest sub-segments are chosen in order to obtain the densest mesh at points $p_1$, $p_2$, and $p_3$. Points $p_1$, $p_2$, and $p_3$ are located at the positions where singularity points are obtained during simulations.

In this paper, all the meshes used have the same lengths of segments $L_1=1$, $L_2=1$, $L_3=1$, $L_4=3$, $L_5=0.5$, $L_6=0.5$, $L_7=2$ and $L_8=8$. Except for Mesh3L where the extruder has a die with double the length ($L_5=1$ and $L_6=1$) and Mesh2.5 where the extrudate length is 4 ($L_7=L_8=2$).  $L_{min(i)}$ for all segments are chosen to be equal to $L_{min}$ for a given mesh. Moreover, all the sub-segments of segments 7 and 8 are equal to $L_{min}$ to maximize the number of nodes representing the free surface. The parameters of the different meshes used in this work are shown in Table \ref{table:1}. 

\begin{table}
  \resizebox{\linewidth}{!}
  {\begin{tabular}{ | c || c | c | c | c | c | c | c | c | c | c | c | c | }
  
    \hline
          & $L_{min}$ & $n_1$ & $n_2$ & $n_3$ & $n_4$ & $n_5$ & $n_6$ & $n_7$ & $n_8$ & $n_f$ & Nodes & Corner Nodes\\ \hline    \hline
    Mesh1 & 0.5 & 2 & 2 & 2 & 6 & 1 & 1 & 4 & 16 & 4 & 389 & 115\\ \hline
    Mesh2 & 0.25 & 3 & 3 & 3 & 9 & 2 & 2 & 8 & 32 & 6 & 977 & 276\\ \hline
    Mesh2.5 & 0.125 & 3 & 3 & 4 & 9 & 3 & 3& 12 & 10 & 5 & 903 & 250\\ \hline
    Mesh3 & 0.125 & 5 & 5 & 6 & 15 & 3 & 3 & 16 & 64 & 10 & 3259 & 874\\ \hline
    Mesh3L & 0.125 & 5 & 5 & 6 & 15 & 6 & 6 & 16 & 64 & 10 & 3415 & 916\\ \hline
    Mesh4 & 0.0625 & 6 & 6 & 7 & 18 & 5 & 5 & 32 & 128 & 12 & 6543 & 1740\\ \hline
    Mesh5 & 0.03125 & 12 & 12 & 14 & 36 & 10 & 10 & 64 & 256 & 24 & 25341 & 6543\\ \hline
    Mesh6 & 0.015625 & 24 & 24 & 28 & 72 & 20 & 20 & 128 & 512 & 48 & 99705 & 25341\\ \hline
  
  \end{tabular}}
\centering
\caption{Mesh parameters.}
\label{table:1}
\end{table}

As shown in Figure \ref{fig:element}, the elements are composed of 6 nodes including 3 corner nodes. During simulation, velocity in $x$ and $y$ direction is evaluated at every node while pressure is evaluated only at the corner nodes. For example, Mesh4 has 27,912 variables which include $x$ and $y$ position of the nodes $(2\times6543)$, velocity at the nodes in the $x$ and $y$ direction $(2\times 6543)$ and the pressure at the corner nodes $(1740)$. 

\section{Inner Nodes Movement}

\begin{figure}
	\includegraphics[width=0.8\linewidth]{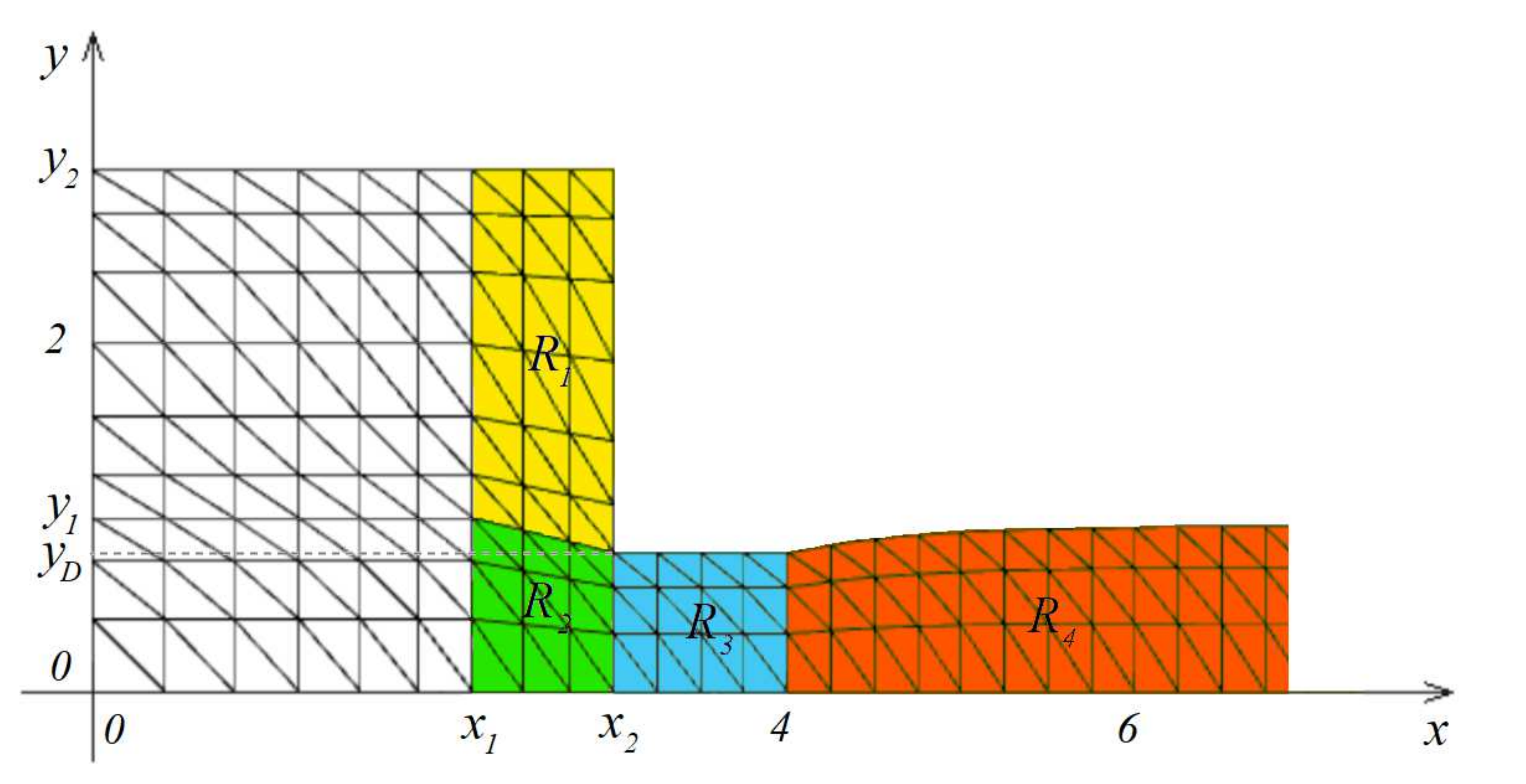}
	\centering
	\caption{Mesh2 deformed.}
	\label{fig:Close}
\end{figure}

Figure \ref{fig:Close} shows how Mesh2 may be deformed during a simulation. The inner nodes deform according to the outer nodes. Different regions of the inner nodes are shown in Figure \ref{fig:Close}. In the white region, the nodes are fixed. $n_f$ in Table \ref{table:1} is the number of sub-segments that constitute the width of the white region. Having a fixed mesh region in part of the barrel makes the assembly of the discretized matrices more efficient. The parts of the matrices associated with the white region do not need to be reassembled if the mesh is stationary. In regions $R_1$ (yellow), $R_2$ (green), and $R_3$ (blue) the inner nodes position is dependent on the bottom $y$-position of the die $y_D$. While in region $R_4$ (orange), the position of inner nodes is dependent on the $y$-position of the free surface node $y_s$ on top of the nodes. In Table \ref{table:2}, the expression of the shift in the $y$-direction of the inner node $\Delta y_n$ in terms of the original $x$-position of node $x$, original $y$-position of node $y$, $y_D$, and $y_s$ is given. $x_1$, $x_2$, $y_1$, and $y_2$ are shown in Figure \ref{fig:Close}. Note that, the $x$-position of the nodes does not change. The coordinate system used in Figure \ref{fig:Close} is chosen as the standard coordinate system in Chapters 2 and 3.
\begin{table}

  \begin{tabular}{ | c || c | }
  
    \hline
    $R_1$ & $\Delta y_n =\left(y_D-L_3\right)\left(\frac{x-x_1}{x_2-x_1}\right) \left(\frac{y_2-y}{y_2-y_1}\right)$ \\ \hline 
    $R_2$ & $\Delta y_n =\left(y_D-L_3\right)\left(\frac{x-x_1}{x_2-x_1}\right) \left(\frac{y}{y_1}\right)$ \\ \hline 
    $R_3$ & $\Delta y_n =\left(y_D-L_3\right)\left(\frac{y}{y_1}\right)$ \\ \hline 
    $R_4$ & $\Delta y_n =\left(y_s-L_3\right)\left(\frac{y}{y_1}\right)$ \\ \hline 
  
  \end{tabular}
  \centering
  \caption{Inner node position.}
  \label{table:2}
\end{table}

\section{Conclusion}
The governing equations and boundary conditions used to model extrusion are described in this chapter. Different numerical methods to solve the equations are presented. GFEM with Taylor-Hood $P_1-P_2$ isoparametric triangular elements is used for spatial discretization. Different parameters of the mesh generator used in the spatial discretization are explained. Trapezoidal Rule is used for temporal discretization. Predictor Multi-Corrector is used to solve the discretized equations in the Lagrangian and Eulerian form. A new method with extra nodes for defining the free surface is proposed to model the large free surface variations obtained during extrusion with varying die gap. Using the proposed method the free surface is defined with a larger number of nodes for a given mesh. 

\chapter{FEM Model Results}

\section{Introduction}
Mathematical modeling and numerical implementation of the FEM simulator are presented in Chapter 2. In this chapter, the results of the FEM simulator are presented and discussed. The simulator is implemented in order to develop control oriented models in later chapters and to design controllers in future work. Therefore, Matlab\textsuperscript{\textregistered} is chosen as a platform for coding the simulator due to its flexibility and its Control System Toolbox. The code consists of thousands of lines of code and required about 3 years of coding. Simulation run time for each simulation required to obtain the results in this chapter vary from a few hours to more than a week on modern 3.4 GHz processors. 

Simulation parameters such as simulation time, extrudate length, mesh used and values of $Re$ considered in this chapter are constrained by the computational time and memory cost. Transient simulations are much more computationally expensive than steady state simulations. The largest possible $\Delta t$ in time dependent simulations is limited by $L_{min}$ and $Re$. Therefore, a much finer mesh (Mesh6) is used in steady simulation compared to the transient simulations (Mesh4). This is also true in the literature; where steady simulations have a very fine mesh compared to time dependent simulations \cite{Mitsoulis1, georgiou1999}. 

\begin{figure}
    \centering
    \begin{subfigure}[b]{0.4\linewidth}
        \includegraphics[width=\linewidth]{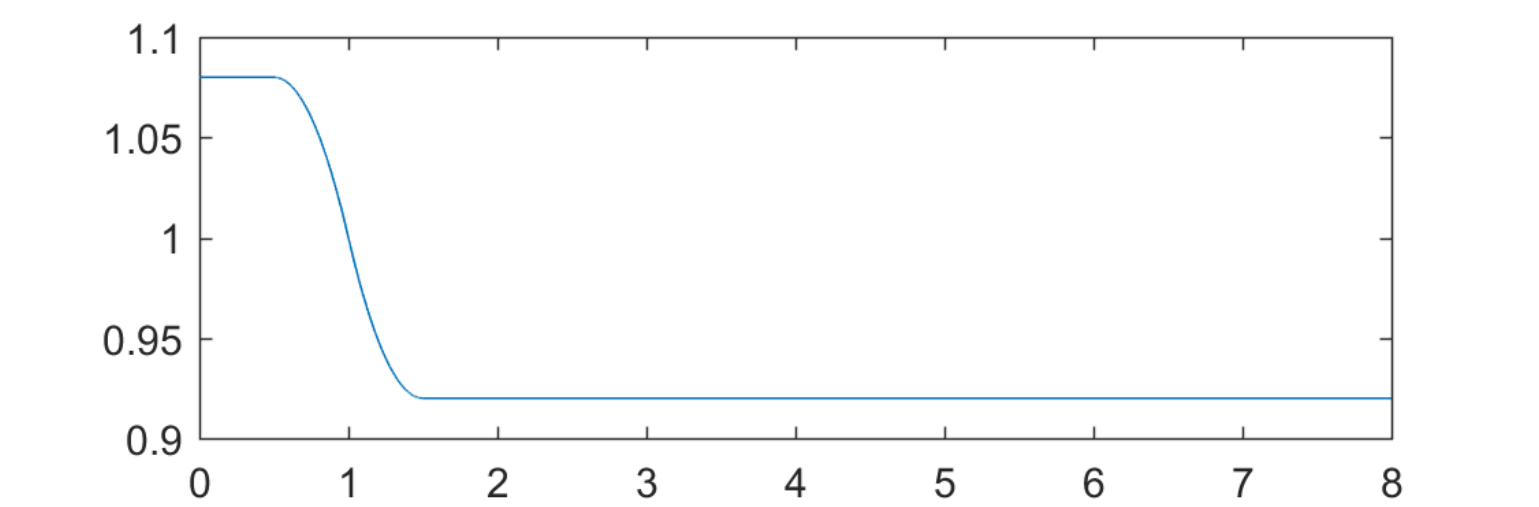}
        \caption{$y_D$ while closing.}
        \label{fig:xc}
    \end{subfigure}
    \begin{subfigure}[b]{0.4\linewidth}
        \includegraphics[width=\textwidth]{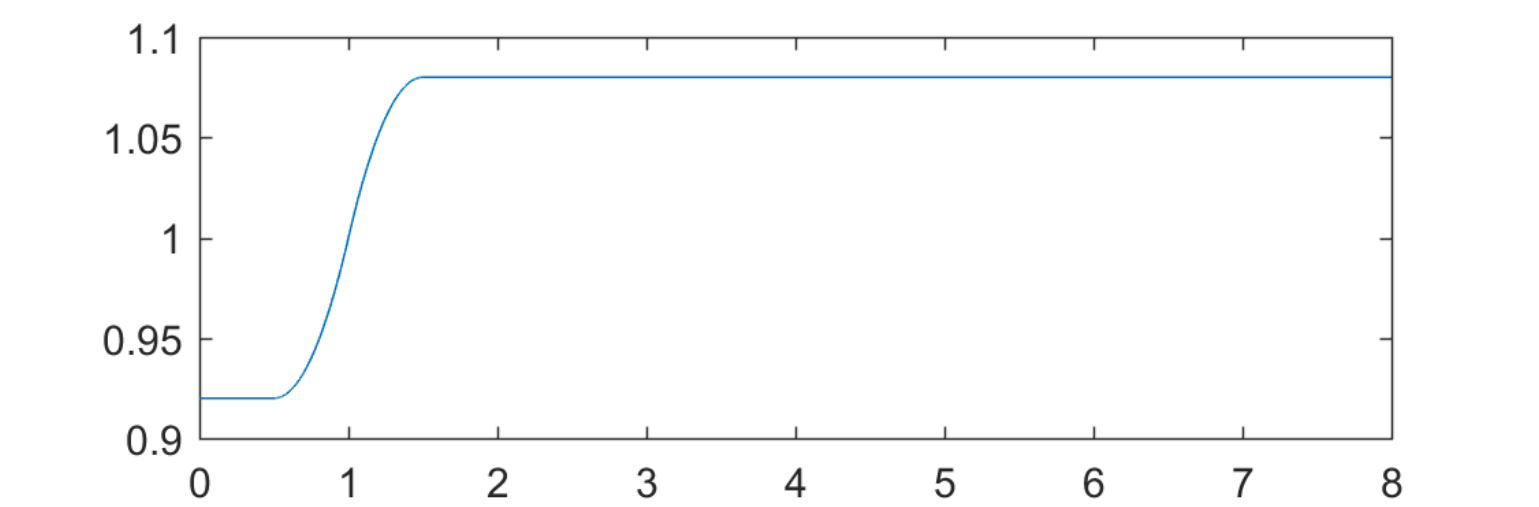}
        \caption{$y_D$ while opening.}
        \label{fig:xo}
    \end{subfigure}

    \begin{subfigure}[b]{0.4\linewidth}
        \includegraphics[width=\textwidth]{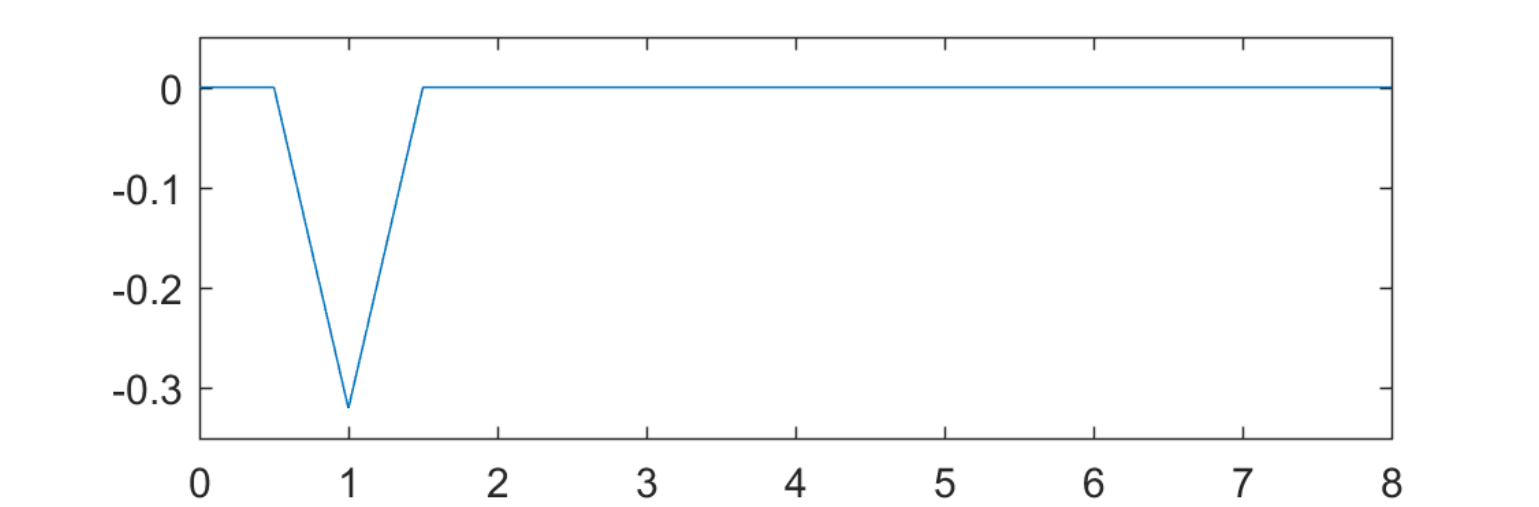}
        \caption{$u_D$ while closing.}
        \label{fig:vc}
    \end{subfigure}
    \begin{subfigure}[b]{0.4\linewidth}
        \includegraphics[width=\textwidth]{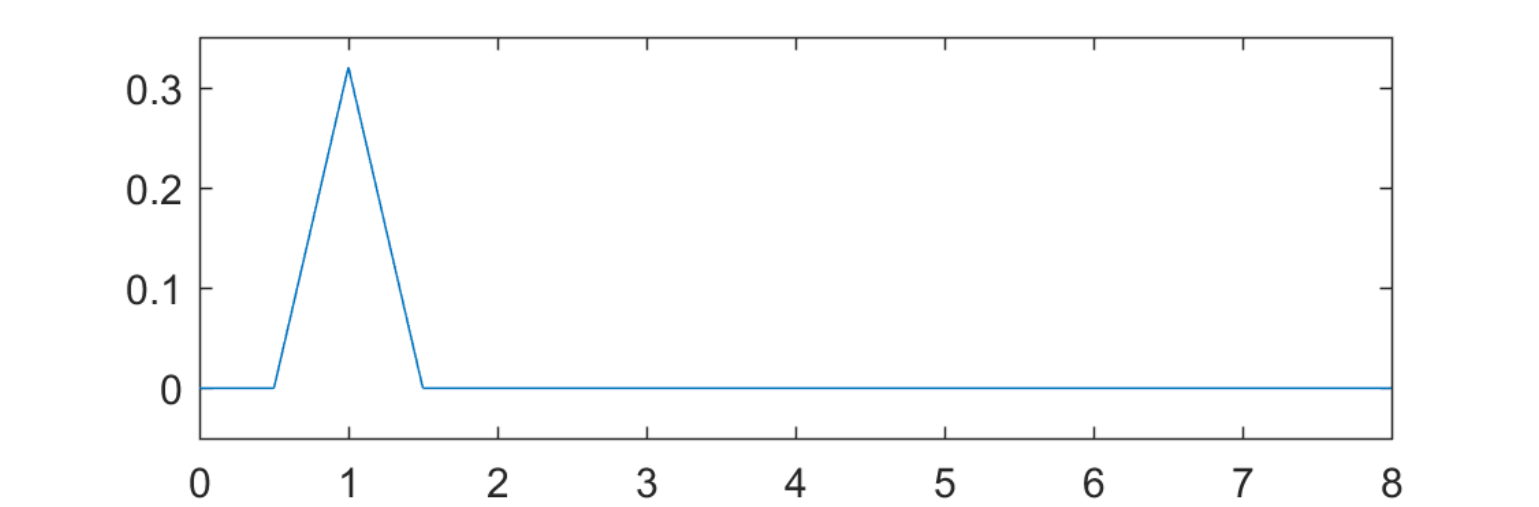}
        \caption{$u_D$ while opening.}
        \label{fig:vo}
    \end{subfigure}
    
    \begin{subfigure}[b]{0.4\linewidth}
        \includegraphics[width=\linewidth]{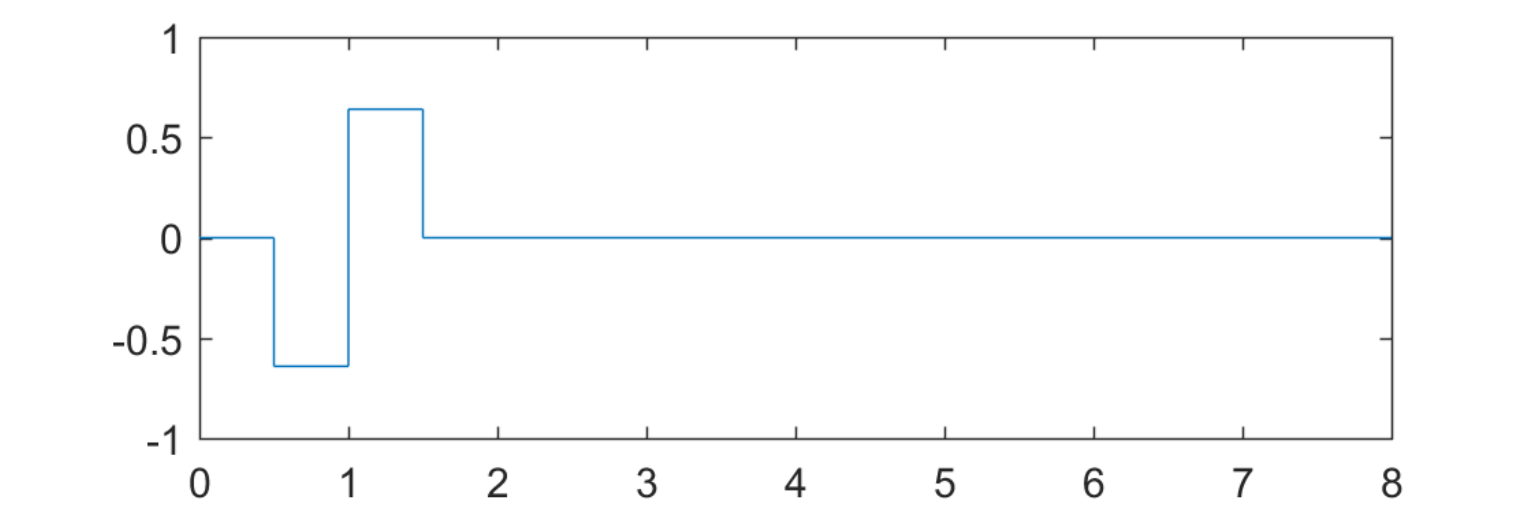}
        \caption{$a_D$ while closing.}
        \label{fig:ac}
    \end{subfigure}
    \begin{subfigure}[b]{0.4\linewidth}
        \includegraphics[width=\linewidth]{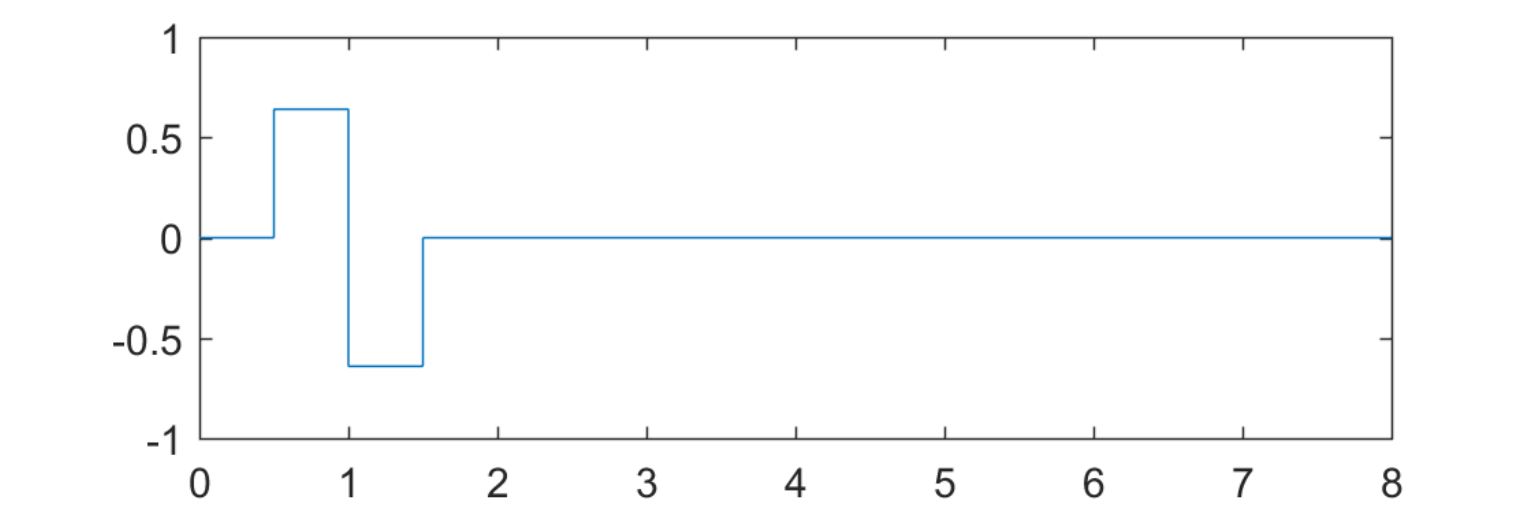}
        \caption{$a_D$ while opening.}
        \label{fig:ao}
    \end{subfigure}
    \caption{Die bottom kinematics while opening and closing.}\label{fig:wall}
\end{figure}

In Table \ref{table:3}, the parameters of the transient simulations presented in this paper are listed. A single change in the die gap is applied in each simulation. An example of the $y$-position $y_D$, velocity $u_D$, and acceleration $a_D$ of the bottom of the die is shown in Figure \ref{fig:wall}. $u_D$ and $a_D$ are applied as boundary conditions at the nodes that are located at the die. The die is accelerated by a positive value $a$ then decelerated by $-a$ while opening the die. Similarly, it is accelerated by a constant acceleration value $-a$ then decelerated by $a$ while closing the die. The minimum and maximum values of $y_D$ are $0.92$ and $1.08$, this implies the die gap change $2\Delta y_D$ be equal to $-0.32$. The time needed to change $y_D$ from one value to the other is $\Delta T$. $\Delta y_D$ and $\Delta T$ are selected by the user and $a$ is calculated accordingly. In all the simulations, the change is centered at $t=1$ and the median of $y_D$ is equal to $L_s/2$. As $\Delta t$ decreases, the number of iterations needed to converge in the Predictor Multi-Corrector method decreases for the two sub-steps. $\Delta t$ was chosen such that the maximum number of iterations is less than or equal to 5 for all simulations. It can be observed in Table \ref{table:3} that using a finer mesh, using longer dies, decreasing $Re$ or decreasing $\Delta T$ requires the use of a smaller $\Delta t$. Moreover, for simulations with negative values of die gap change, smaller or equal values of $\Delta t$ are needed compared to simulations with positive values of die gap change.

The chapter starts with mesh convergence analysis and comparing the die swell ratio of the steady simulator to results from the literature for validation. Then, the advantages of using the new free surface interpolation scheme are presented. In Section \ref{sec:Results}, the effects of varying the die gap on the pressure, velocity and extrudate shape is presented. Changing the die gap result in newly reported phenomena (bulging and necking). The different factors affecting bulging and necking are discussed in Section \ref{sec:BandN}. Finally, in Section \ref{sec:SmallRe}, simulation results for diminishingly small values of $Re$ are presented. 

\section{Mesh Convergence and Literature Validation at $t=0$}
Mesh convergence is studied using the steady simulator, which is used to compute the solution at $t=0$. Mesh convergence is investigated by comparing die swell ratio. The die swell ratio is the ratio of the maximum extrudate thickness over the die gap in steady state. 

\begin{table}

  \begin{tabular}{ | c | c | c | }
  
    \hline
           & $Re=1$ or $Re^*=2$ & $Re=10$ or $Re^*=20$ \\ \hline    \hline
    Mesh1  &  1.262 & 1.021 \\ \hline
    Mesh2  &  1.229 & 0.977 \\ \hline
    Mesh3  &  1.203 & 0.9535 \\ \hline
    Mesh4  &  1.186 & 0.9381 \\ \hline
    Mesh5  &  1.176 & 0.9282 \\ \hline
    Mesh6  &  1.171 & 0.9228 \\ \hline
    Mitsoulis \textit{et al.}\cite{Mitsoulis1}  &  $\simeq$1.14 & $\simeq$0.92 \\ \hline
  
  \end{tabular}
  \centering
  \caption{Die swell ratio.}
  \label{table:DS}
\end{table}

In Table \ref{table:DS}, the die swell ratio is shown using different meshes for $y_D=1$. At every second row, $L_{min}$ of the used mesh is decreased by half. The die swell error between two consecutive meshes continues to decrease as finer mesh is used until it reaches a value less than 0.43$\%$ for $Re=1$ and 0.56$\%$ for $Re=10$ between Mesh5 and Mesh6. The die swell ratio decreases as $Re$ increases as reported in the literature \cite{middleman1961,Mitsoulis1}.

\begin{figure}
	\includegraphics[width=\linewidth]{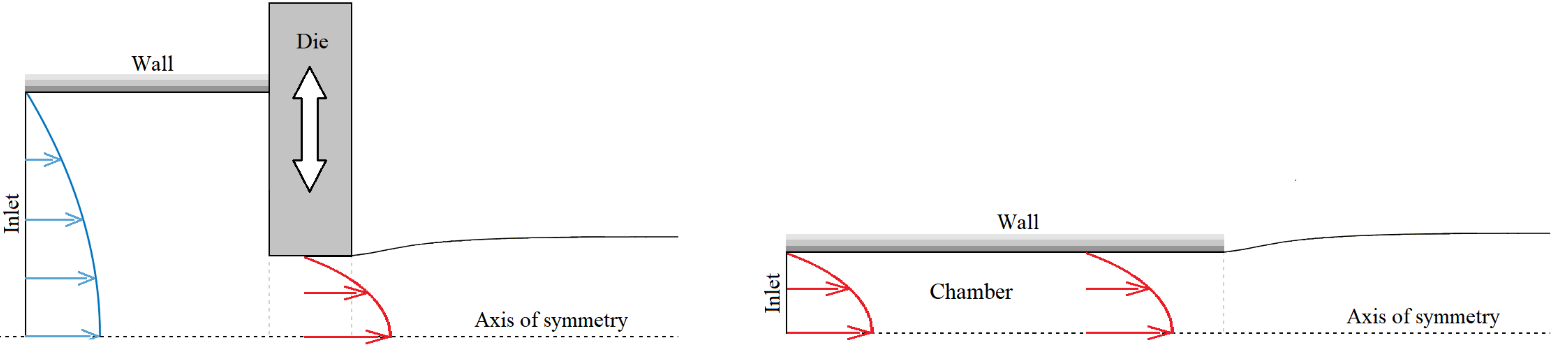}
	\centering
	\caption{Different extruder configurations.}
	\label{fig:2Config}
\end{figure}

In \cite{Mitsoulis1}, the fluid passes through a chamber without any contractions before exiting into free space. This configuration is typically used in literature addressing die swell where the fluid passes through a long chamber and forms a Poiseuille flow \cite{Tanner1,Russo,Mitsoulis2,Omodei1979,Omodei1980,spectral2015}. In this work, the die acts as a sudden contraction. The die may not be long enough for the fluid to form a fully developed Poiseuille flow. In Figure \ref{fig:2Config}, the different configurations are shown. In \cite{Mitsoulis1}, the characteristic velocity used to define $Re$ is the average velocity at the inflow, whereas, the maximum velocity at the inflow is used in this paper. Moreover, in this work, the velocity of the fluid in the die is dependent on the die gap. In order to compare the die swell of the two configurations, die swell results with Reynolds number $Re=(y_D/2)Re^*$ from this work, are compared to results with Reynolds number $Re^*$ in \cite{Mitsoulis1}. The relationship is derived by assuming that a Poiseuille flow is attained in the die. This assumption results in a velocity profile in the die similar to that of the chamber in \cite{Mitsoulis1}. Given that Poiseuille flow is assumed in the die, the average velocity is set to $2/3$ of the maximum velocity and the maximum velocity in the die is set to $3/y_D$ to obtain the relationship between $Re$ and $Re^*$.

The ratio is obtained at the maximum distance away from the die ($x=14$), although, it is recommended to obtain the ratio at larger distances specially for higher $Re$ values \cite{Mitsoulis1}. In Table \ref{table:DS}, the die swell ratio reported in \cite{Mitsoulis1} is obtained from Figure 4 of that work. The results of the two works for $Re=10$ are very close. The difference in the results may be decreased for $Re=1$ if a longer die length were used in this paper.

\section{Free Surface Interpolation Methods Comparison}
\label{sec:NewInter}
\begin{figure}
	\includegraphics[width=.9\linewidth]{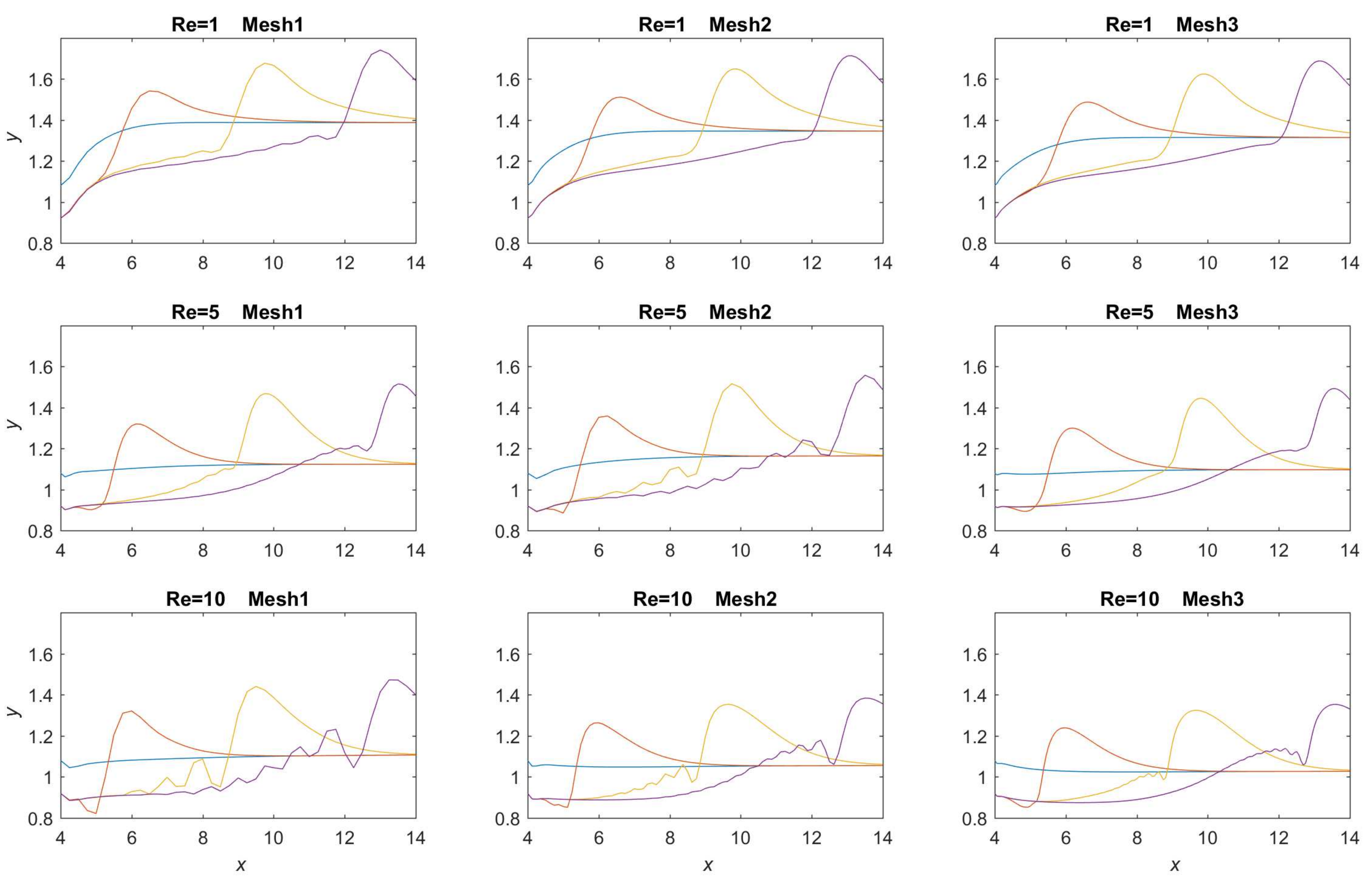}
	\centering
	\caption{Free surface plot using free surface interpolation.}
	\label{fig:ReMe}
\end{figure}

Figure \ref{fig:ReMe} shows the results of simulations 1, 6, 9, 2, 7, 12, 3, 8 and 21 listed in Table \ref{table:3}. In the figure, the free surface plot for $Re=1$, $Re=5$, and $Re=10$ using different meshes and no extra nodes are shown. The surface at $t=0.5$, $t=2$, $t=4$, and $t=6$ are shown in blue, red, yellow and purple, respectively. In this paper, the $x$-location difference between two consecutive discretization nodes on the free surface is equal to $L_{min}$ and is listed in Table \ref{table:1}. Oscillations are observed in Figure \ref{fig:ReMe} for larger values of $L_{min}$ and $Re$. As $Re$ increases, a finer mesh is required to prevent oscillations. The amplitude of the oscillations increases with time. 
\begin{figure}
	\includegraphics[width=0.7\linewidth]{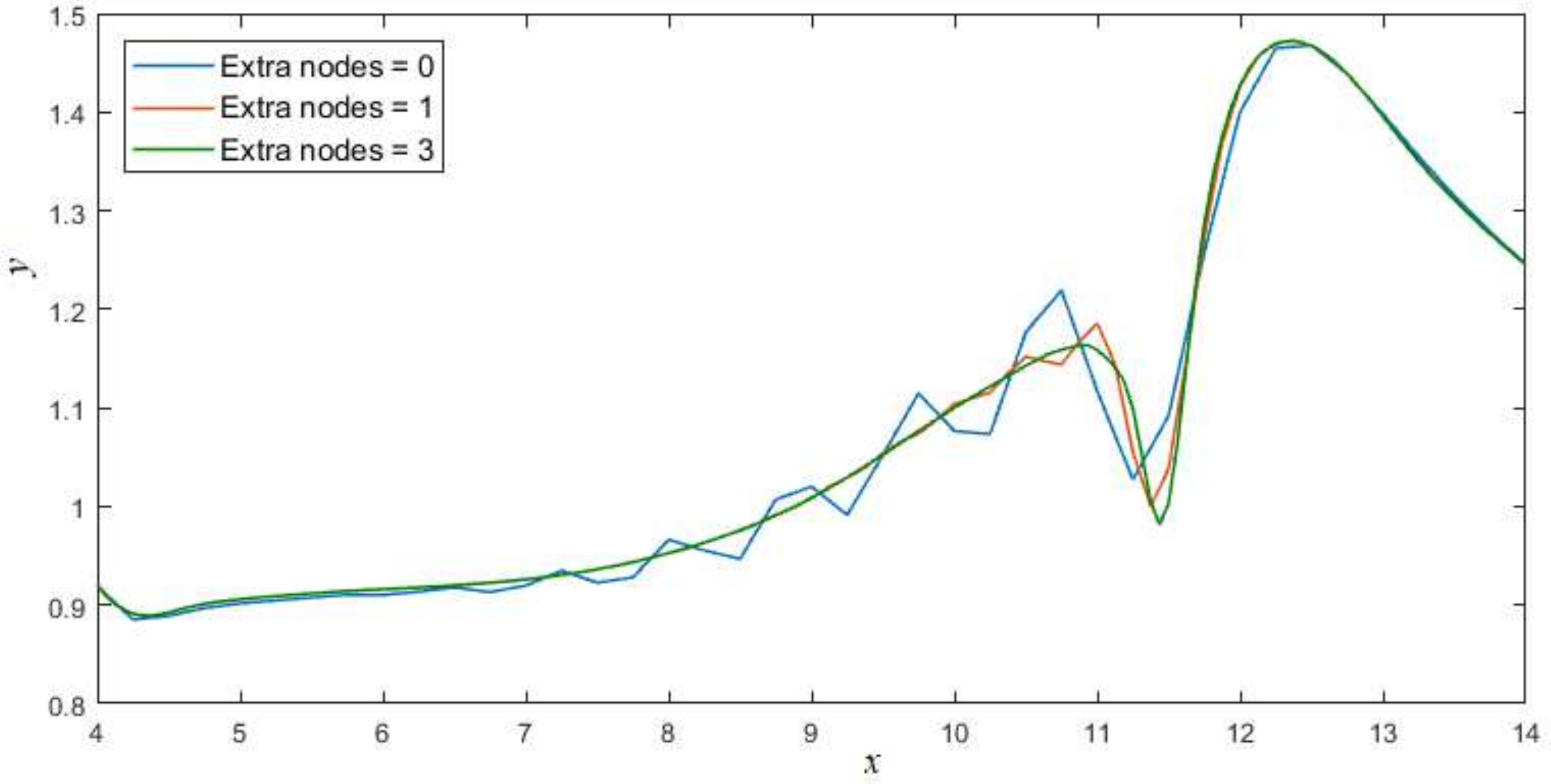}
	\centering
	\caption{Effect of extra nodes on free surface oscillations.}
	\label{fig:FSA4}
\end{figure}

Smaller values of $L_{min}$ not only require a finer mesh, but also require smaller $\Delta t$, which increases the simulation time drastically. Being able to simulate higher values of $Re$ using a coarse mesh decreases the simulation time cost. Therefore, the free surface interpolation method with extra nodes is proposed. The free surface with largest oscillations in Figure \ref{fig:ReMe} is simulated using the new method and shown in Figure \ref{fig:FSA4} where $Re=10$ and Mesh1 is used. Simulations 3, 4, and 5 were used to construct Figure \ref{fig:FSA4}. The new method with three extra nodes nearly eliminates oscillations. It can also be observed that a distance of $L_{min}$ between the nodes is not enough to capture the shape of the free surface. The new method makes it possible to use a coarse mesh (Mesh1) by eliminating free surface oscillations for the case with largest oscillations ($Re=10$). Note, higher number of extra nodes could be used with very little extra simulation time cost but three extra nodes are enough to give acceptable results in this case.
\begin{figure}
	\includegraphics[width=0.8\linewidth]{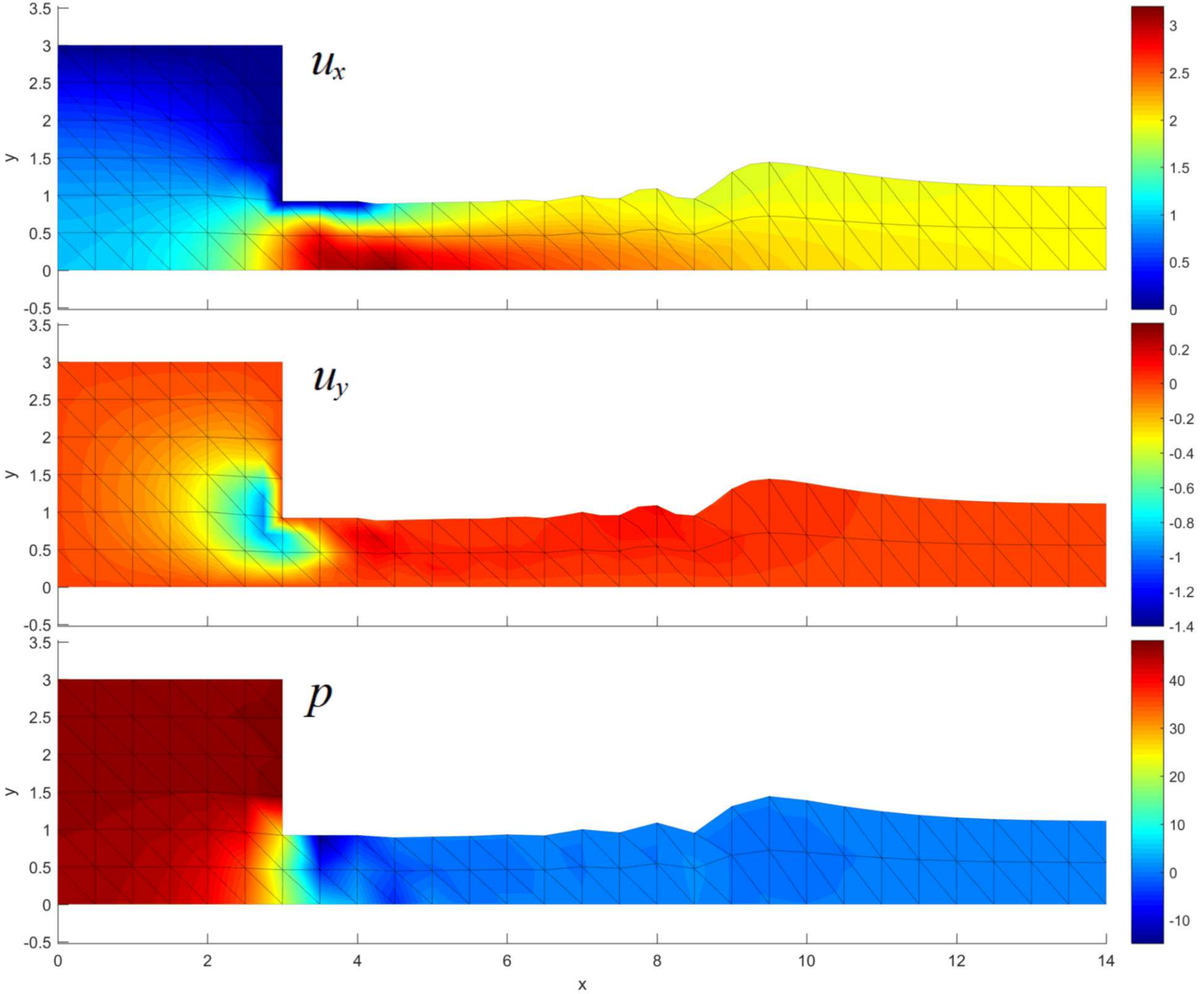}
	\centering
	\caption{$Re=10$ using Mesh1 at $t=4$.}
	\label{fig:Re10M1}
\end{figure}

The pressure and velocity for $Re=10$ and Mesh1 are presented in Figure \ref{fig:Re10M1}. It can be observed that the velocity and pressure do not show any oscillations in space even if a very coarse mesh is used. The oscillations are limited to the free surface. 

\section{Die Opening and Closing for $Re=1$ and $Re=10$}
\label{sec:Results}
In Appendix \ref{sec:Re1} and Appendix \ref{sec:Re10}, results for Simulations 10, 11, 22 and 23 are shown where the die gap is increased and decreased for $Re=1$ and $Re=10$. Mesh3 was used for $Re=1$ while Mesh4 was used for $Re=10$.
\begin{figure}
    \centering
    \begin{subfigure}[b]{0.7\linewidth}
        \includegraphics[width=\linewidth]{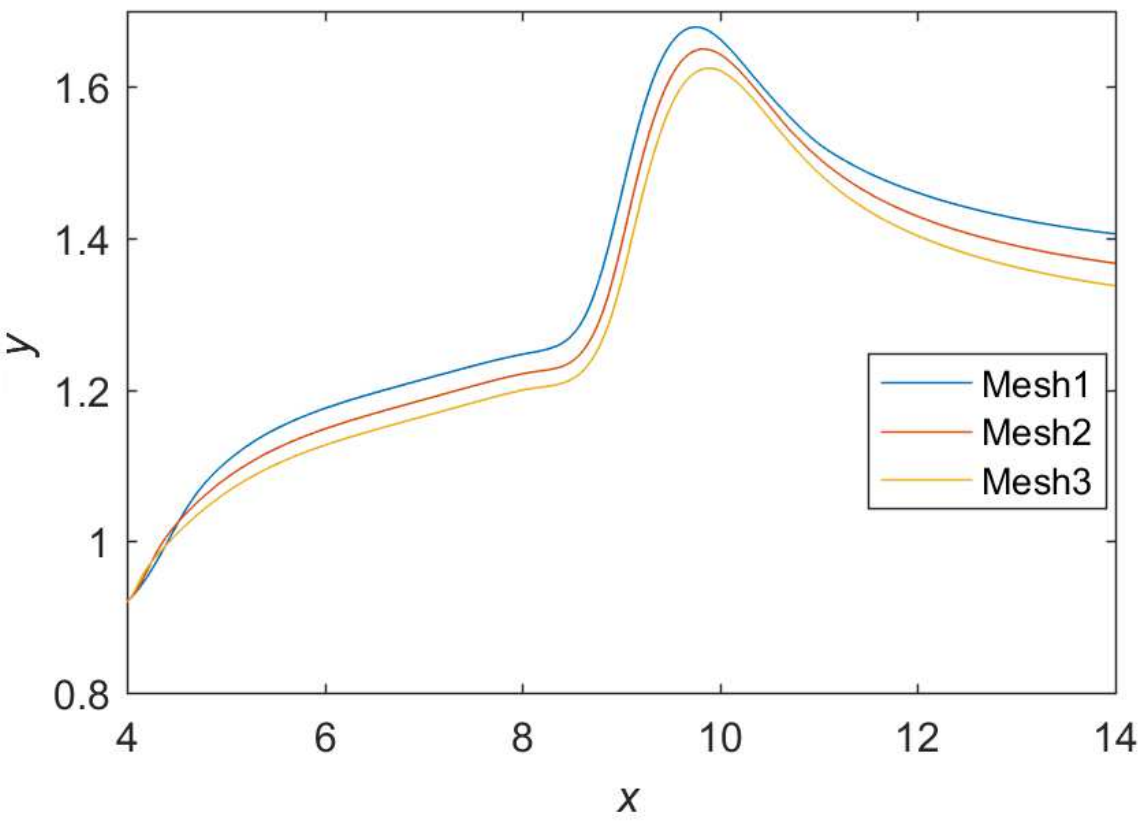}
        \caption{$Re=1$.}
        \label{fig:MRE1}
    \end{subfigure}
    
    \begin{subfigure}[b]{0.7\linewidth}
        \includegraphics[width=\textwidth]{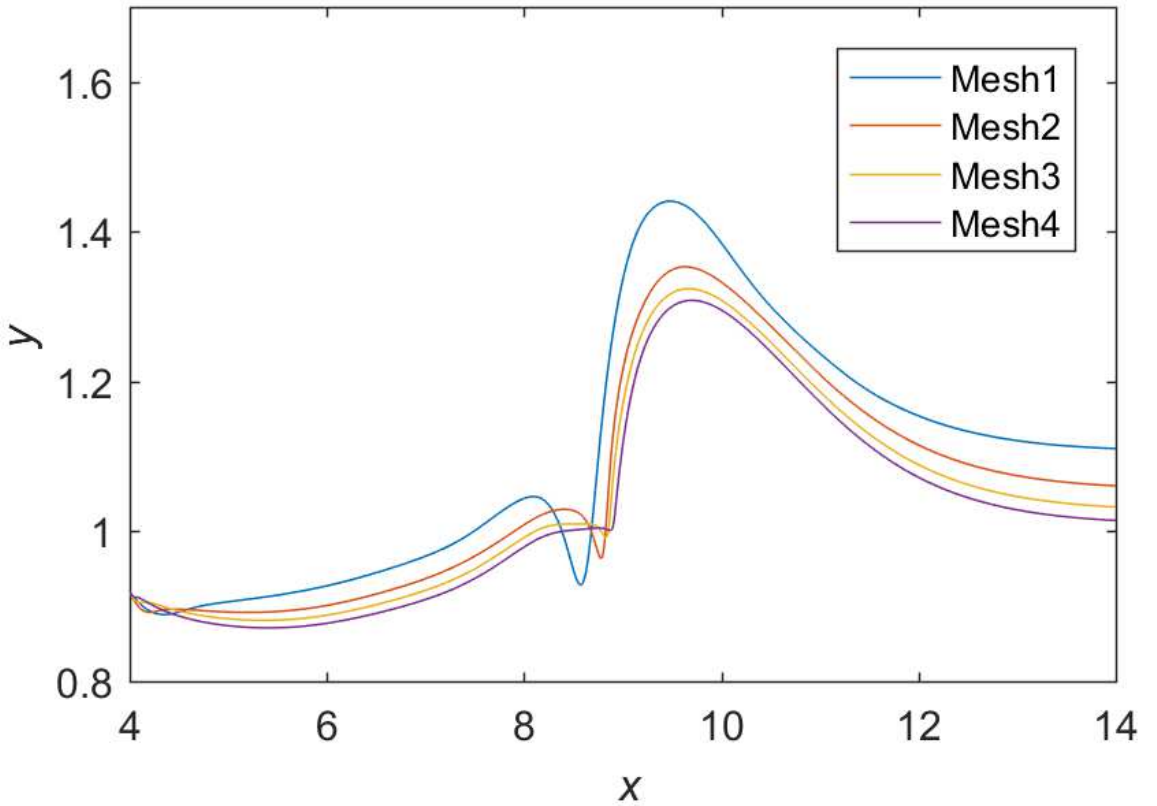}
        \caption{$Re=10$.}
        \label{fig:MRE2}
    \end{subfigure}
    \caption{Free surface at $t=4$ using different meshes.}
    \label{fig:MRE}
\end{figure}

In order to show the adequacy of the used meshes, the free surface using different meshes at $t=4$ is plotted in Figure \ref{fig:MRE}. The results of Simulations 10, 22, 24, 25, 26, 27 and 28 are used. It can be observed that extrudate thickness is overestimated when a coarser mesh is used. Using finer mesh specially near the exit of the die improves the free surface estimation \cite{Mitsoulis1, georgiou1999}. For $Re = 10$, the use of larger values of $\Delta t$ is possible, thus, a finer mesh (Mesh4) is used. For transient simulations, computational time increases drastically as $L_{min}$ decreases. A finer mesh was not possible due to excessive computational time required. To quantify the error, let the surface error be the integral with respect to $x$ of the absolute difference between the free surfaces obtained using two different meshes divided by the integral of the free surface obtained using the finer mesh. For $Re=10$, the error between Mesh3 and Mesh4 is less than 1.6$\%$ for all values of $0\leq t\leq 8$. For $Re=1$ the error between Mesh2 and Mesh3 is less than 2.3$\%$ for all values of $0\leq t\leq 8$. In addition, the error decreases as a finer mesh is used for $0\leq t\leq 8$ as observed in Figure \ref{fig:MRE}.

In Figure \ref{fig:MRE2}, a small valley before the peak is observed. The valley peters out as the mesh becomes finer. The valley becomes very small for Mesh4. The error could be further decreased if a finer mesh were possible. Nevertheless, the shape of the extrudate remains similar for the different meshes.

$u_x$, $u_y$ and $p$ are shown in Appendix \ref{sec:Re1} and Appendix \ref{sec:Re10} using Mesh3 and Mesh4 respectively as the results change with respect to time at $t=0$, $t=1$, $t=1.5$, $t=4$ and $t=8$. In this sub-section, observations and analyses of these results are discussed. 

\subsection{Velocity in the $x$-direction}
Figures \ref{fig:cRey1u} and \ref{fig:oRey1u} show $u_x$ for $Re=1$ where the die gap is decreased and increased respectively while figures \ref{fig:cRey10u} and \ref{fig:oRey10u} show $u_x$ for $Re=10$. At $x=0$, the flow forms a Poiseuille flow with max $u_x=1$ at $y=0$. The maximum values of $u_x$ are observed at $y=0$ and around the die due to the contraction. At the walls, the minimum value is observed and is equal to 0. As $x$ increases after the die exit, the absolute value of the derivative of $u_x$ in the $y-$direction decreases till $u_x$ reaches a constant value as $y$ changes. This constant value is greater for higher values of $Re$ since the extrudate is thinner. Moreover, for lower $Re$ values, the constant value of $u_x$ and final swell ratio is reached for smaller values of $x$ \cite{Mitsoulis1}.

\begin{figure}
    \centering
    \begin{subfigure}[b]{0.48\linewidth}
        \includegraphics[width=\linewidth]{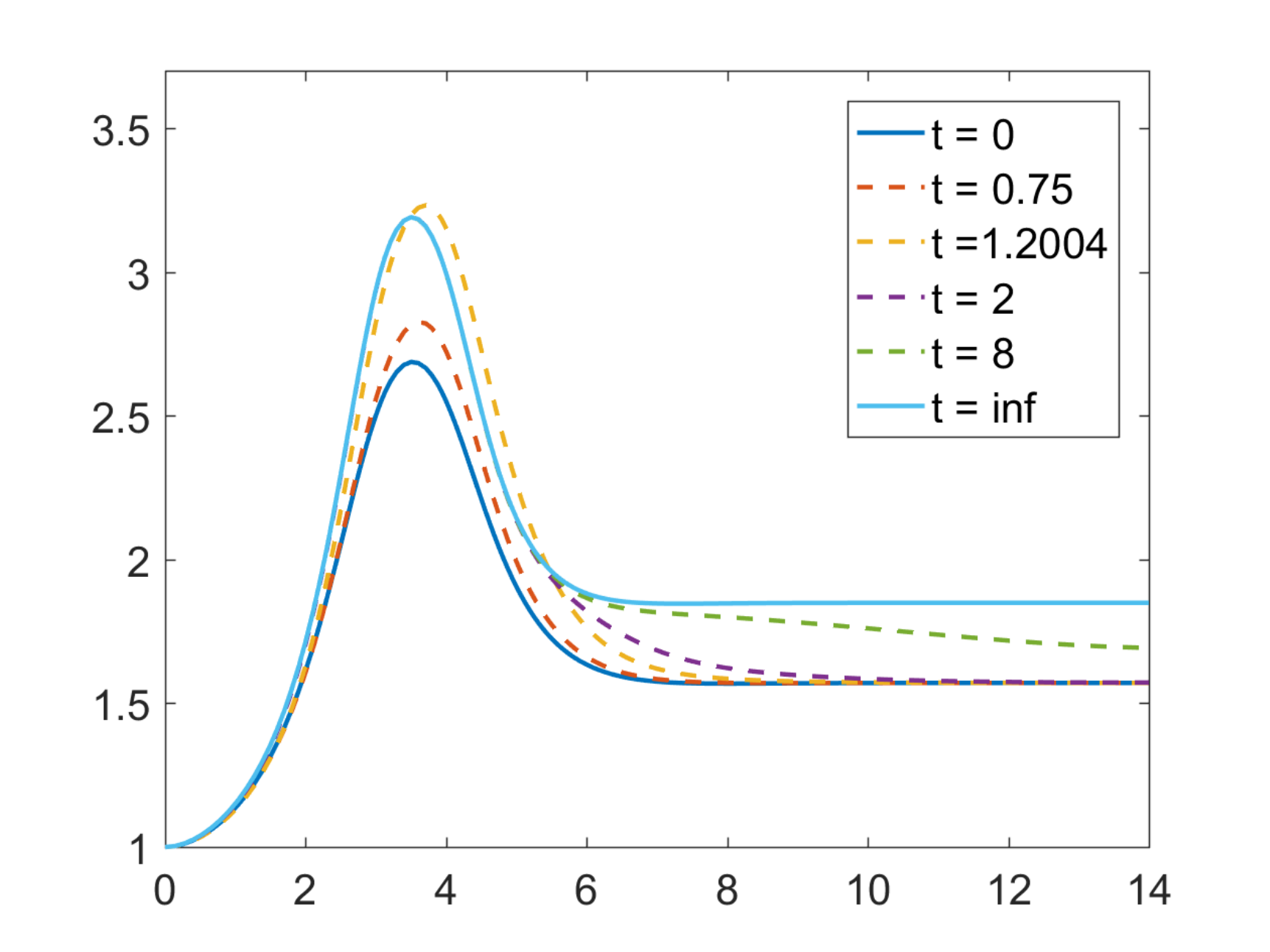}
        \caption{Die gap decrease for $Re=1$.}
        \label{fig:uxc1}
    \end{subfigure}
    \begin{subfigure}[b]{0.48\linewidth}
        \includegraphics[width=\linewidth]{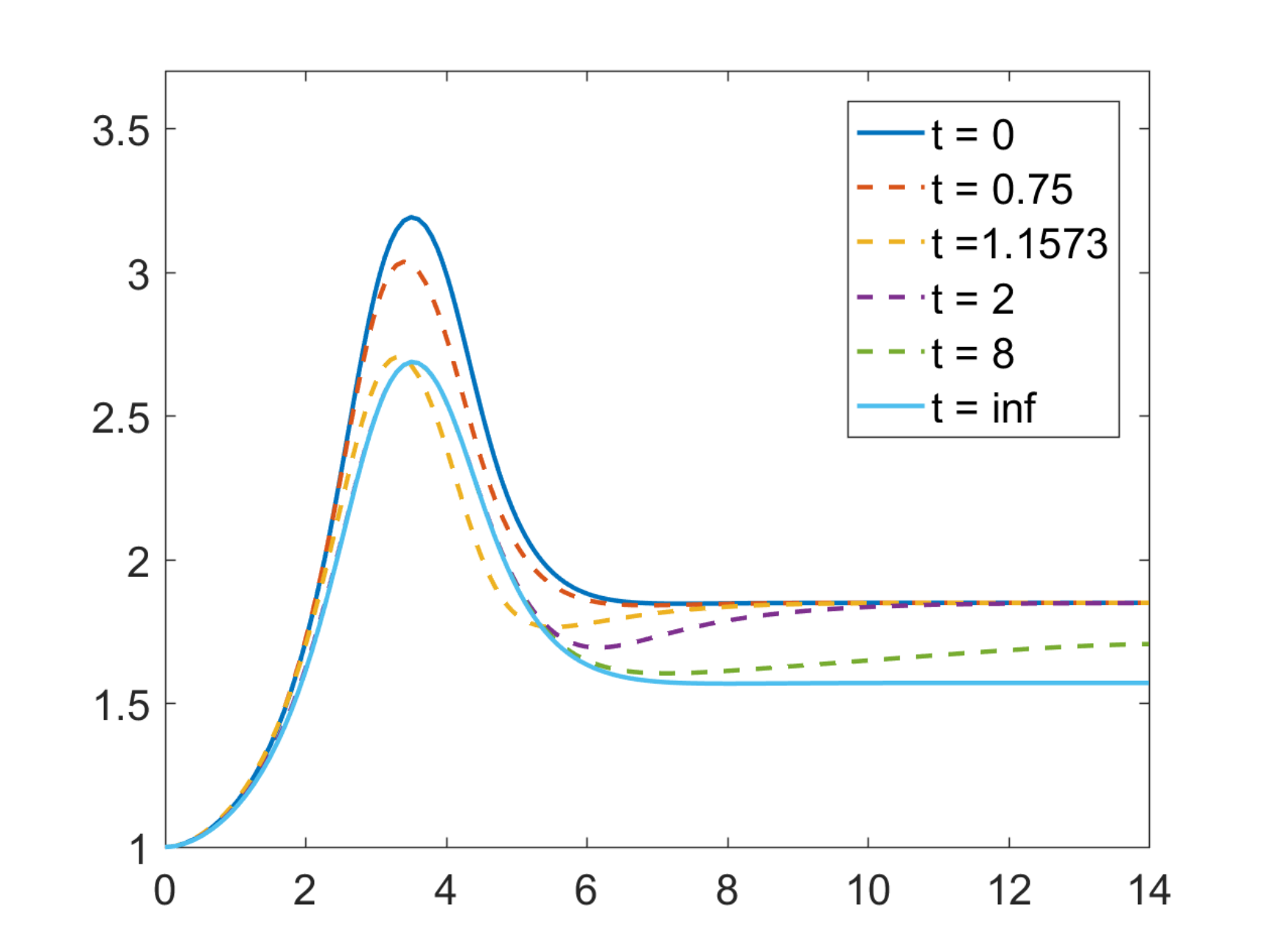}
        \caption{Die gap increase for $Re=1$.}
        \label{fig:uxo1}
    \end{subfigure}

    \begin{subfigure}[b]{0.48\linewidth}
        \includegraphics[width=\linewidth]{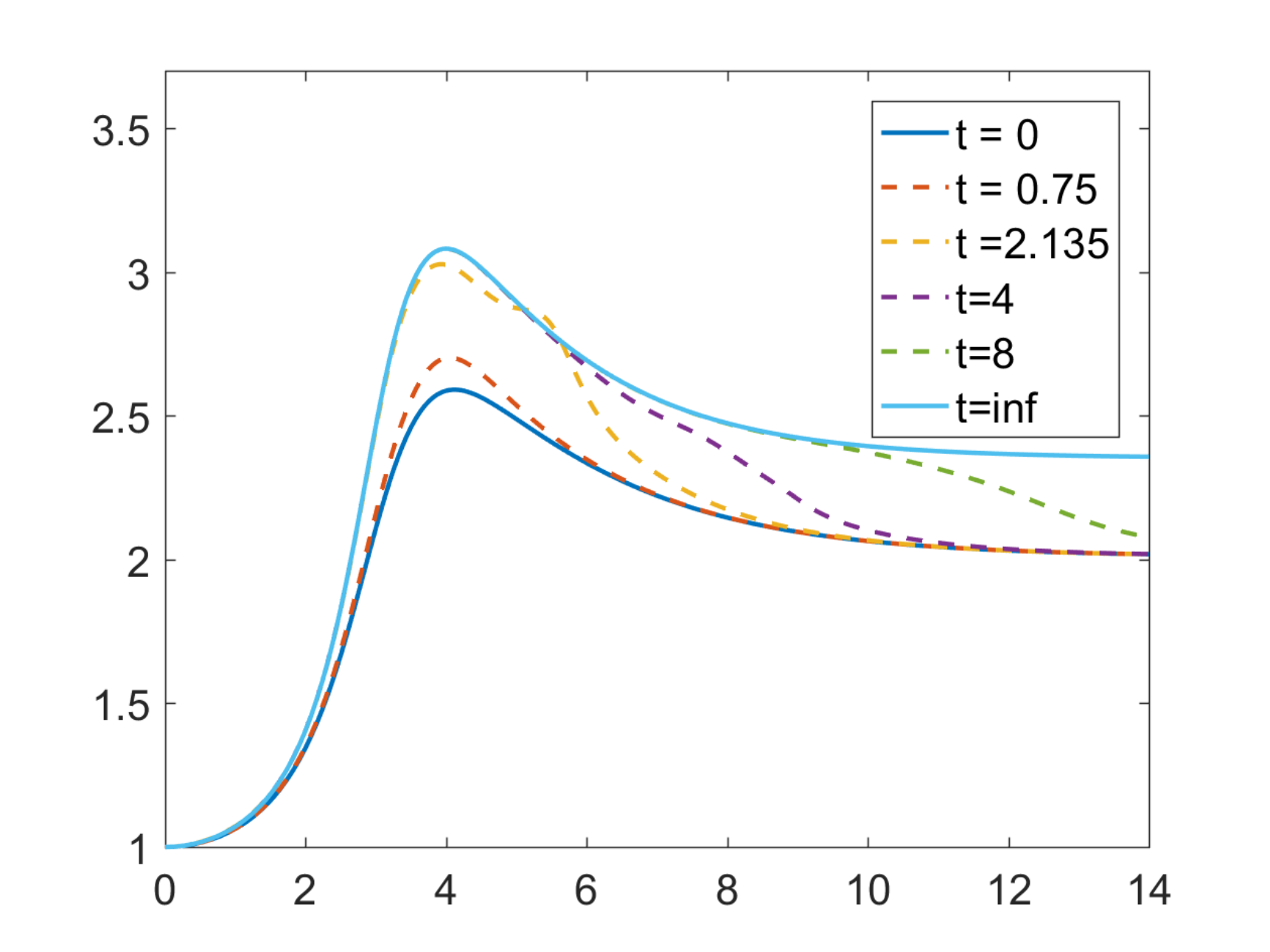}
        \caption{Die gap decrease for $Re=10$.}
        \label{fig:uxc10}
    \end{subfigure}
    \begin{subfigure}[b]{0.48\linewidth}
        \includegraphics[width=\linewidth]{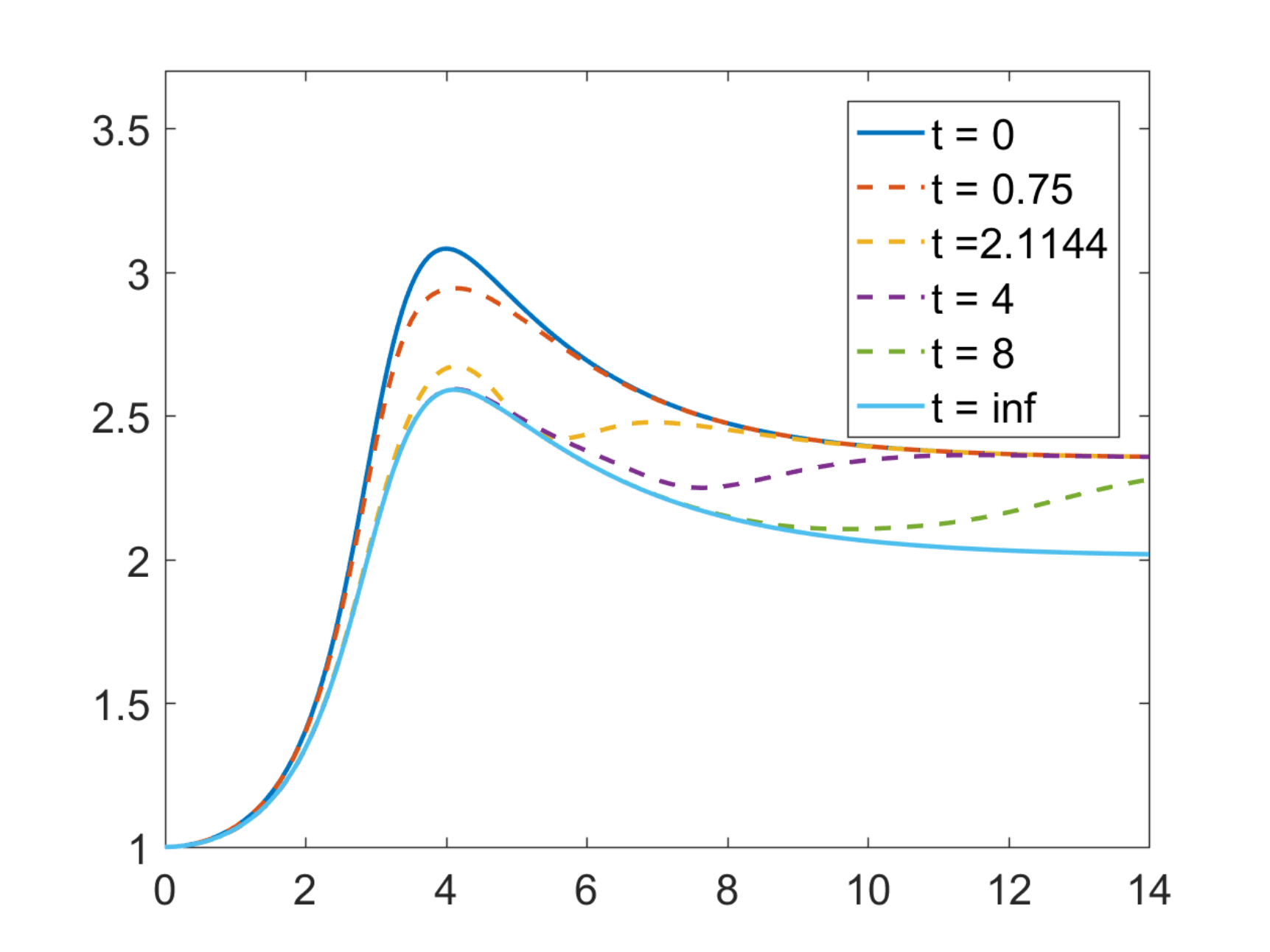}
        \caption{Die gap increase for $Re=10$.}
        \label{fig:uxo10}
    \end{subfigure}
    
    \caption{$u_{xc}$ as a function of $x$ for $Re=1$ and $Re=10$.}
    \label{fig:uxc}
\end{figure}
Let $u_x$ at $y=0$ be denoted by $u_{xc}$. In Figure \ref{fig:uxc}, $u_{xc}$ is presented with respect to $x$ at different instants of $t$. At $t=0$ or $t=\infty$, $u_{xc}$ increases as $x$ increases to reach a maximum value close to the exit of the die. After $x$ increases beyond the location of the maximum value, $u_{xc}$ decreases to a constant value. The simulation at $t=0$ and $t=\infty$ , are computed using the steady-state solver. Longer dies have $u_{xc}$ curves with wider peaks. Moreover, it can be observed that for $Re=1$ the peak and the constant value after exiting the die is reached faster as $x$ increases giving a narrower curve compared to $Re=10$. 

$u_{xc}$ has larger values at steady state ($t=0$ or $t=\infty$) for smaller die gap and for $x>0$ given that the volumetric flow is the same at the intake. Therefore, in the case of die gap decrease, $u_{xc}$ changes from a smaller value to a larger value except at $x=0$ which is equal to 1. The opposite is true for an increase in die gap. It can be observed in both cases, the change of $u_{xc}$ occurs first inside the die and then gradually progresses for higher values of $x$. For some values of $x$, $u_{xc}$ undergoes an overshoot or an undershoot during its transient change. Overshoot may occur when $u_{xc}$ is changing from a smaller to a larger value (die gap decrease) while undershoot may occur when $u_{xc}$ is changing from a larger to a smaller value (die gap increase). Overshoots and undershoots are turning points during the $u_{xc}$ transient change that show that $u_{xc}$ is not monotonically increasing or decreasing. Let the maximum overshoot be equal to the maximum of $u_{xc}(x,t)-u_{xc}(x,\infty)$ for $0 \leq x \leq 14$ and $0 \leq t \leq \infty$. Let the minimum undershoot be equal to the minimum of $u_{xc}(x,t)-u_{xc}(x,\infty)$ for $0 \leq x \leq 14$ and $0 \leq t \leq \infty$. For example, for $Re=1$, the maximum overshoot is equal to $0.2111$ which occurs at the node that has $x=4.4375$ at $t=1.2004$ and the minimum undershoot is equal to $-0.2059$ which occurs at the node that has $x=4.375$ at $t=1.1573$. Figure \ref{fig:uxc} shows $u_{xc}$ at $t=0$, $t=2$, $t=8$, $t=\infty$, and at the instant the minimum undershoot or maximum overshoot occurs.

\subsection{Velocity in the $y$-direction}
Figures \ref{fig:cRey1v} and \ref{fig:oRey1v} show $u_y$ for $Re=1$ and for decreasing and increasing die gap respectively while Figures \ref{fig:cRey10v} and \ref{fig:oRey10v} shows $u_y$ for $Re=10$. Minimum negative values of $u_y$ are located near the corner of the contraction where the fluid particles are heading downstream toward the contraction. Smaller minimum values with larger absolute gradients are observed for higher $Re$ values. The maximum values of $u_y$ are observed at the exit of the die due to the die swell ratio that is greater than 1 for $Re=1$. While for $Re=10$, negative values of $u_y$ at the exit of the die near the free surface due to the die swell ratio that is less than 1. During the opening and closing of the die, positive and negative values of $u_y$ are observed near the die walls, respectively. Consequently, a decrease or increase in $u_y$ is observed at the exit of the die during die gap increase or decrease, respectively. These variations in $u_y$ cause the formation of a bulge or neck like shape on the extrudate. The bulge and neck shape are observed more prominently when both sides of the extrudate profile are plotted as shown in Figure \ref{fig:ExtCO}.

\subsection{Pressure}
Figures \ref{fig:cRey1p} and \ref{fig:oRey1p} show $p$ for $Re=1$ and for decreasing and increasing die gap, respectively while Figures \ref{fig:cRey10p} and \ref{fig:oRey10p} shows $p$ for $Re=10$. The pressure at the free surface and end of the die is set to $0$ as a boundary condition. The maximum and minimum values of the pressure are observed at singularity points that are located at the corner of the contraction and at the exit of the die. It can be observed that the pressure builds up inside the barrel. Notably, a sudden increase in pressure is observed at the contraction. During the opening and closing of the die, extreme peaks and valleys of pressure are observed respectively at the die and wall contact point ($x=3$ and $y=3$). This phenomenon is also observed in lid driven cavities where a moving lid is in contact with a static wall \cite{bruneau, PoFLDC}. The contact point is a singularity that reaches its maximum absolute pressure values at $t=1$.

\subsection{Free Surface Shape}
It can be observed that when the die gap is decreased, a bulge on the free surface is formed. The bulge moves downstream as time progresses. On the other hand, when the die gap increases a neck like shape due to the thinning of the extrudate is formed. These phenomena make die gap programming more complicated. Consider that the extruder operator is trying to make a certain portion of the extrudate thicker. Intuitively, the operator will increase the die gap during the extrusion of that portion of the extrudate. The increase in the die gap will create a thinning (neck) in the extrudate before the extrudate becomes thicker. 

\section{Factors Affecting Bulging and Necking}
\label{sec:BandN}
\begin{figure}
	\includegraphics[width=.6\linewidth]{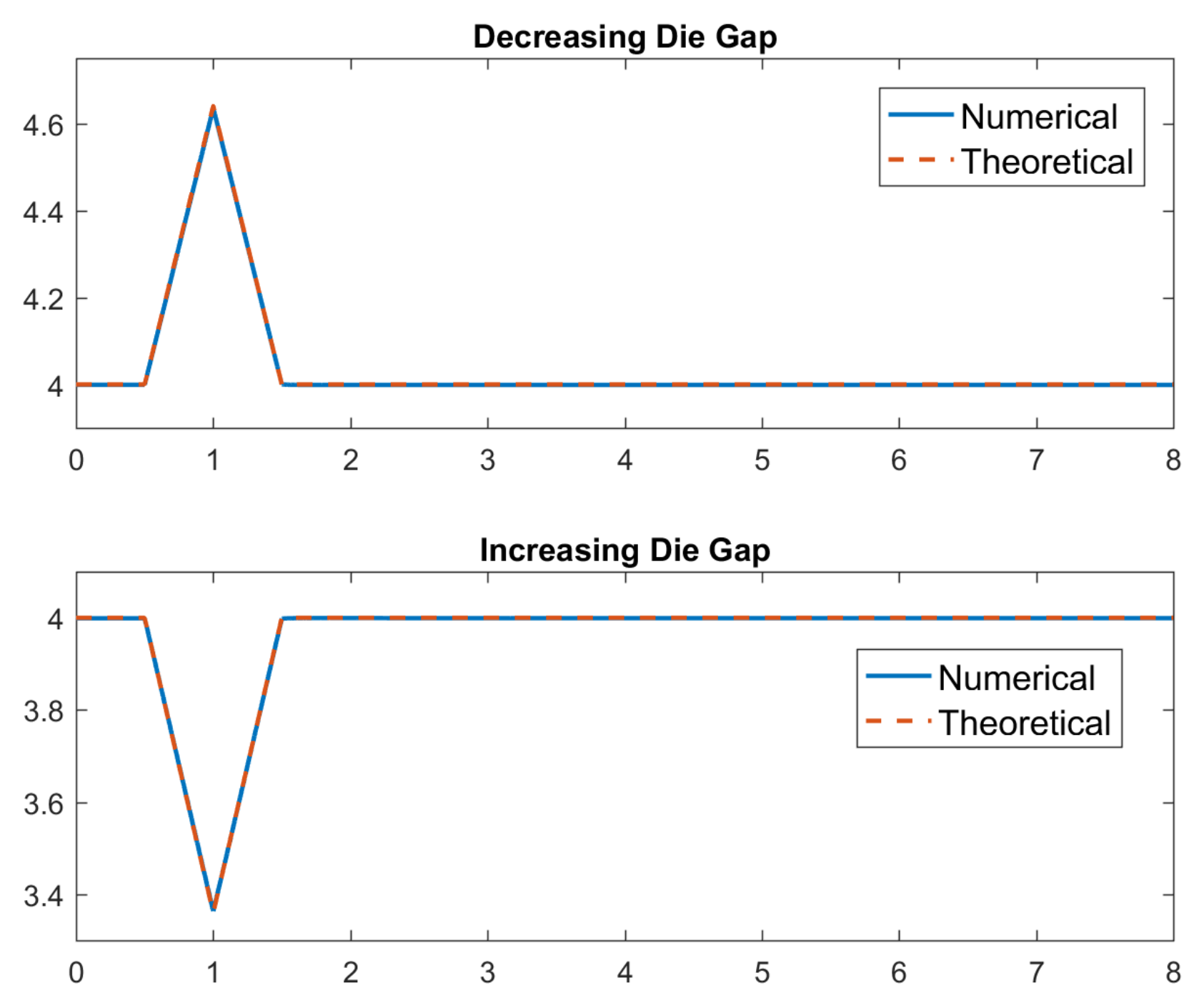}
	\centering
	\caption{Volumetric flow rate.}
	\label{fig:flow}
\end{figure}

In this section, the factors contributing to the bulging and necking are studied. These phenomena are mainly due to the change in the volumetric flow rate at the die $Q_2(t)$ during the die gap change. Theoretically, $Q_2(t) = Q_1 - \frac{\partial V(t)}{\partial t}$ where $Q_1$ is the constant volumetric flow rate at $x=0$ and is equal to $4$. $V(t)$ is the volume inside the barrel and the die. It can be shown that the derivative of the volume is equal to $2  L_D  u_D(t)$ where $L_D$ is equal to length of the die $L_5+L_6$. Replacing the derivative of the volume in the expression of $Q_2(t)$ results in 
\begin{equation}
   \label{eq:Q21}
   Q_2(t) = Q_1 - 2  L_D  u_D(t).
\end{equation}
In order to verify (\ref{eq:Q21}), $Q_2$ can also be calculated numerically by integration. 
\begin{equation}
   \label{eq:Q22}
   Q_2(t)=2 \int_{0}^{y_D(t)} u_{xD}(t,y)dy,
\end{equation}
where $u_{xD}(t,y)$ is the fluid velocity at the die exit $(x=4)$ in the $x-$direction. In Figure \ref{fig:flow}, the results of simulations 22 and 23 are used to show the flow rate during the decrease and the increase of the die gap. The integral in (\ref{eq:Q22}) is calculated numerically using the Simpson's integration rule at every time step. It can be observed in Figure \ref{fig:flow} that the numerical results of (\ref{eq:Q22}) are in close agreement with the theoretical results of (\ref{eq:Q21}).

Due to the excess material pushed out of the die while decreasing the die gap, a bulge on the extrudate is observed. On the other hand, when the die gap increases, the volume within the die increases causing a decrease in flow rate. The decrease in flow rate causes thinning of the extrudate. The extra volume of material that causes a bulge or a neck is calculated by integrating with respect to time the second term in the right hand side of (\ref{eq:Q21}). $- 2 L_D \int_{t_1}^{t_2} u_D(\tau)d\tau = -2 L_D  \Delta y_D$ where $t_1$ is the time when the die gap started changing and $t_2$ is the time when the die gap stopped changing. Therefore, the volume of extra material due to the change in volumetric flow rate is dependent on the amount of die gap change $2 \Delta y_D$ and die length $L_D$. In addition to the extra volume of material, the time in which this extra material is extruded $(\Delta T = t_2-t_1)$ affects the shape of the bulge or neck.

\begin{figure}
    \centering
    \begin{subfigure}[b]{0.95\linewidth}
        \includegraphics[width=1\linewidth]{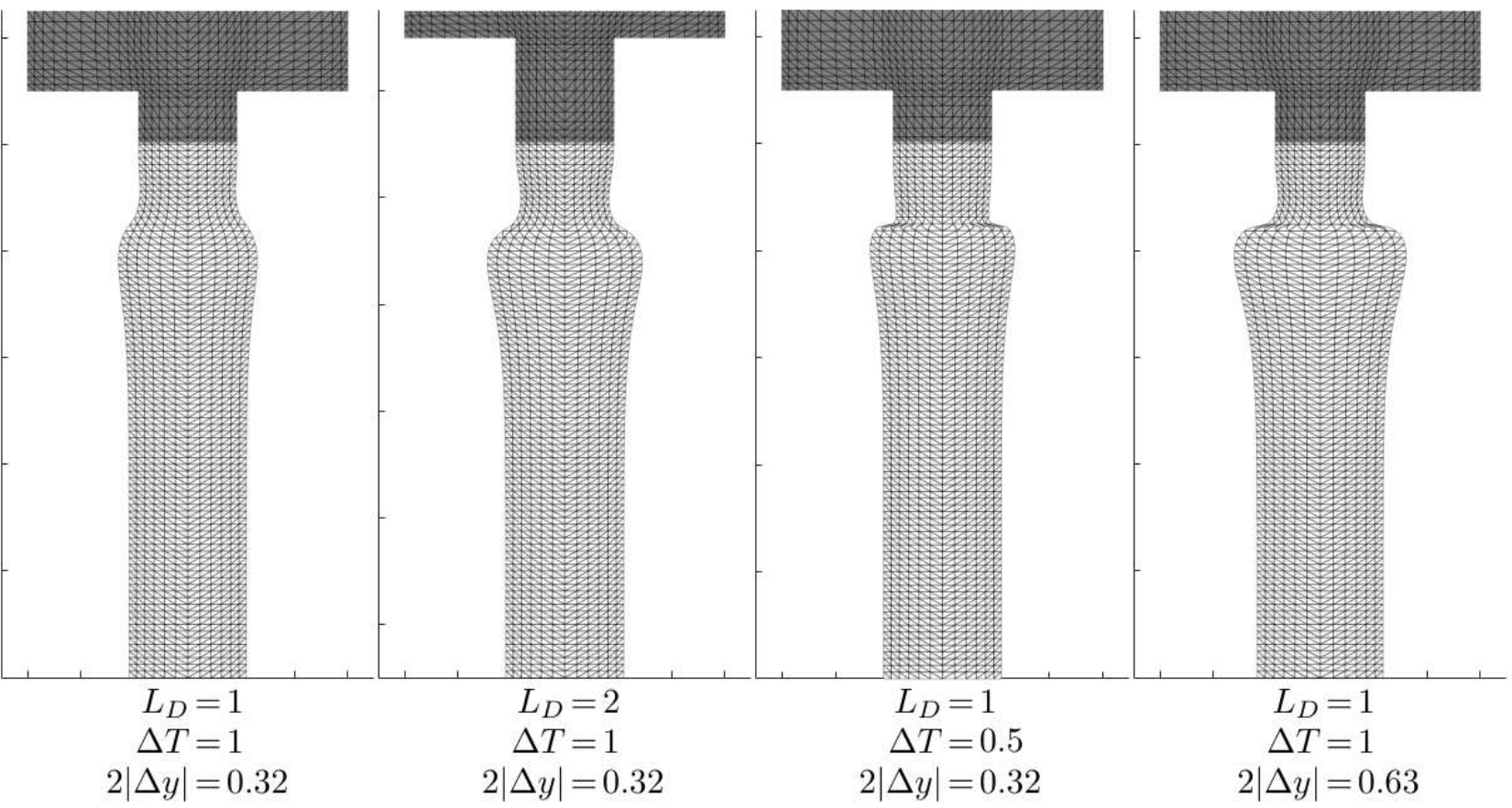}
	    \caption{Extrudate shape in case of sudden decrease in die gap.}
	    \label{fig:ExtC}
    \end{subfigure}
    
    \begin{subfigure}[b]{0.95\linewidth}
        \includegraphics[width=1\linewidth]{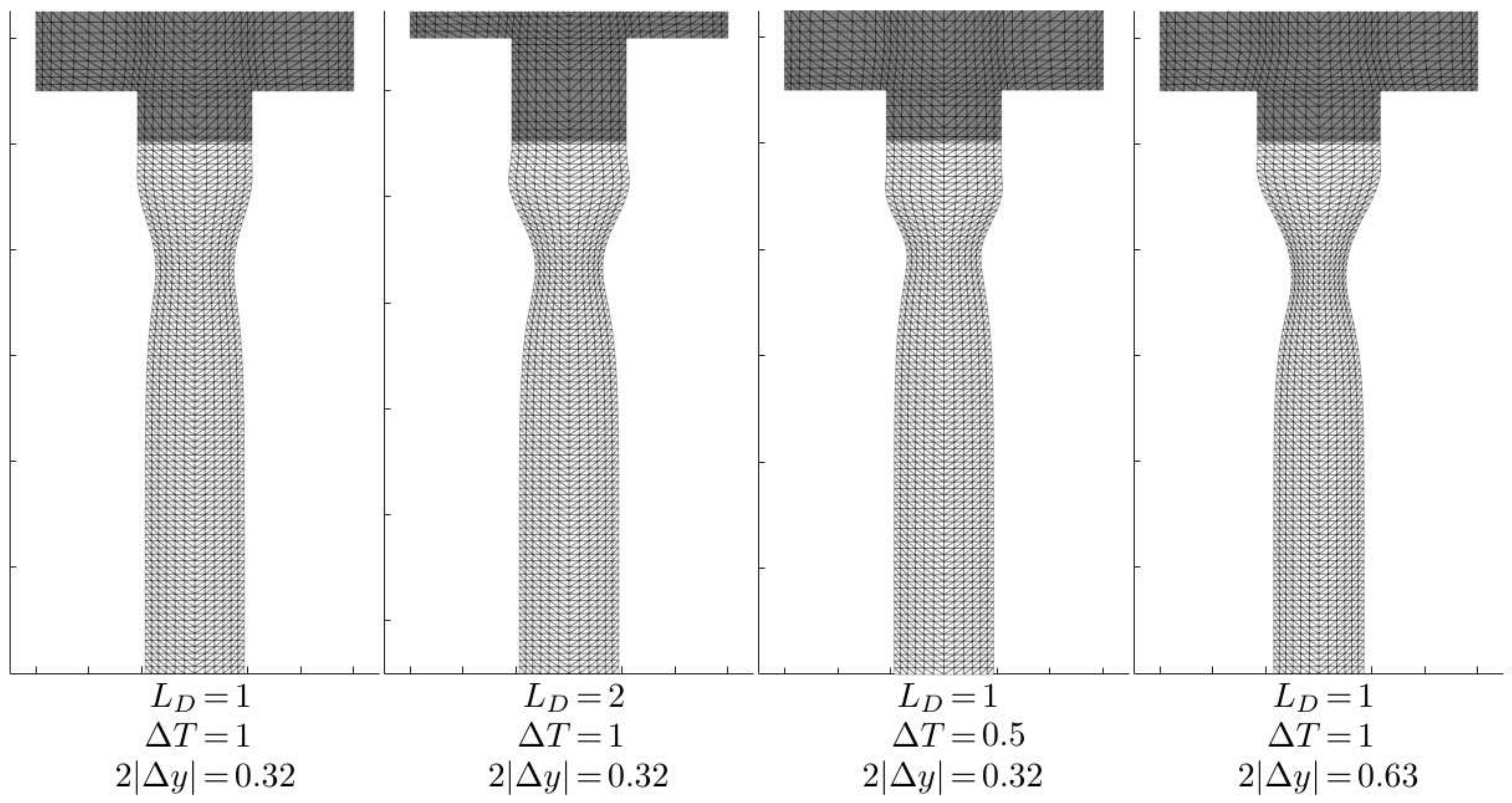}
	    \caption{Extrudate shape in case of sudden increase in die gap.}
	    \label{fig:ExtO}
    \end{subfigure}
    \caption{Extrudate shape with bulging and necking effect.}
    \label{fig:ExtCO}
\end{figure}

The results of simulations 13 to 20 were used in the construction of Figures \ref{fig:ExtC} and \ref{fig:ExtO} to show the extrudate shape under different configurations at $t=2$ for decreasing  and increasing the die gap, respectively. The extrudates are shown in light gray while the fluid inside the barrel and the die are shown in darker gray. The figures show only part of the reservoir. The first extrudates from the left in the two figures show the default extrusion configuration used in this paper. It is used as a reference to compare the rest of the extrudates in Figure \ref{fig:ExtCO}. For the second extrudate, the die length is doubled. For the third, the value of $\Delta T$ is reduced by half. For the fourth, the die gap change is doubled. It can be observed that increasing $L_D$, decreasing $\Delta T$ or increasing $|\Delta y|$ will make the bulging and necking effect more severe. 

\section{Diminishingly Small $Re$ Values}
\label{sec:SmallRe}
\begin{figure}
    \centering
    \begin{subfigure}[b]{.85\linewidth}
        \includegraphics[width=1\linewidth]{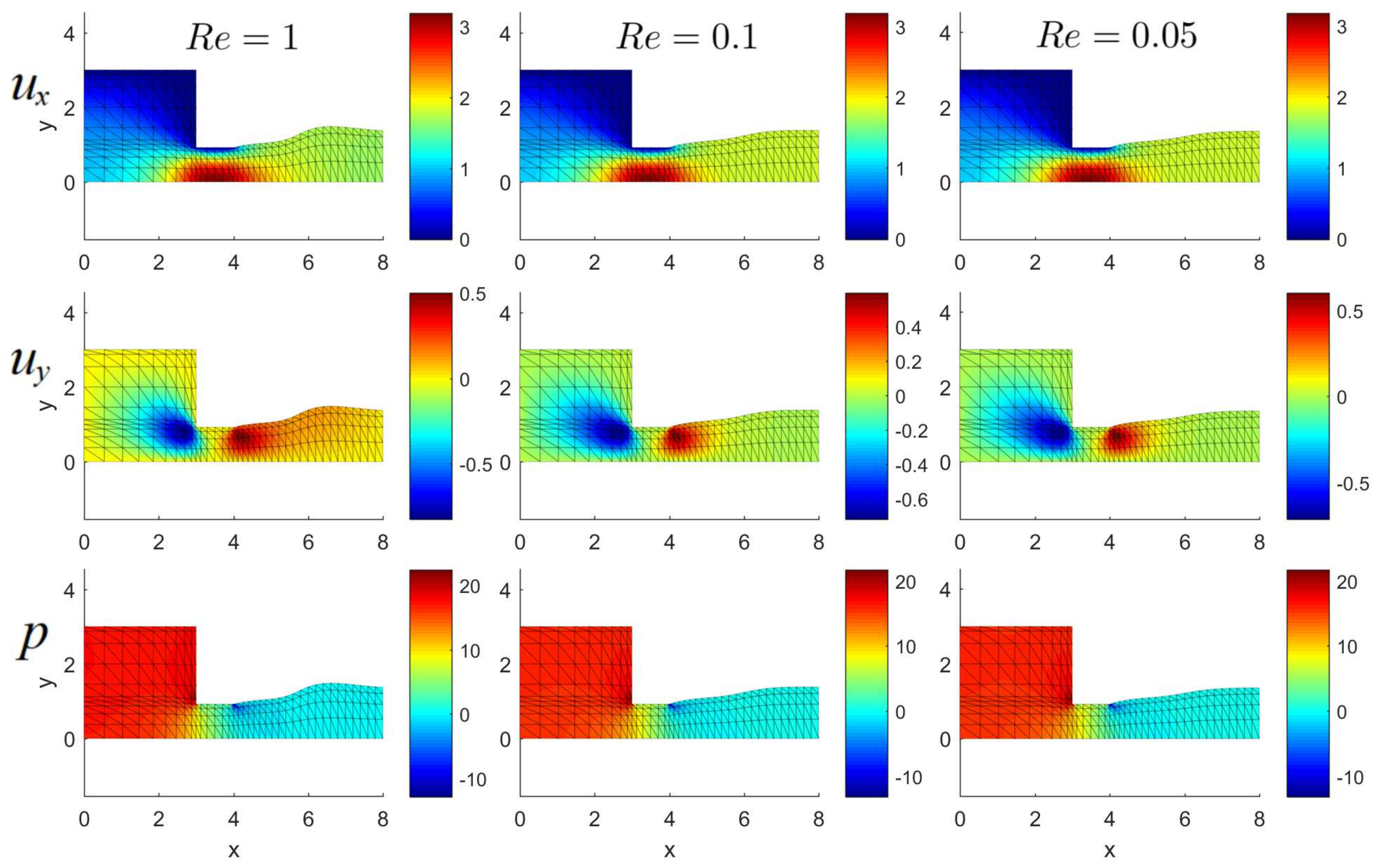}
	    \caption{Low $Re$ die gap decrease.}
	    \label{fig:LReC}
    \end{subfigure}
    
    \begin{subfigure}[b]{.85\linewidth}
        \includegraphics[width=1\linewidth]{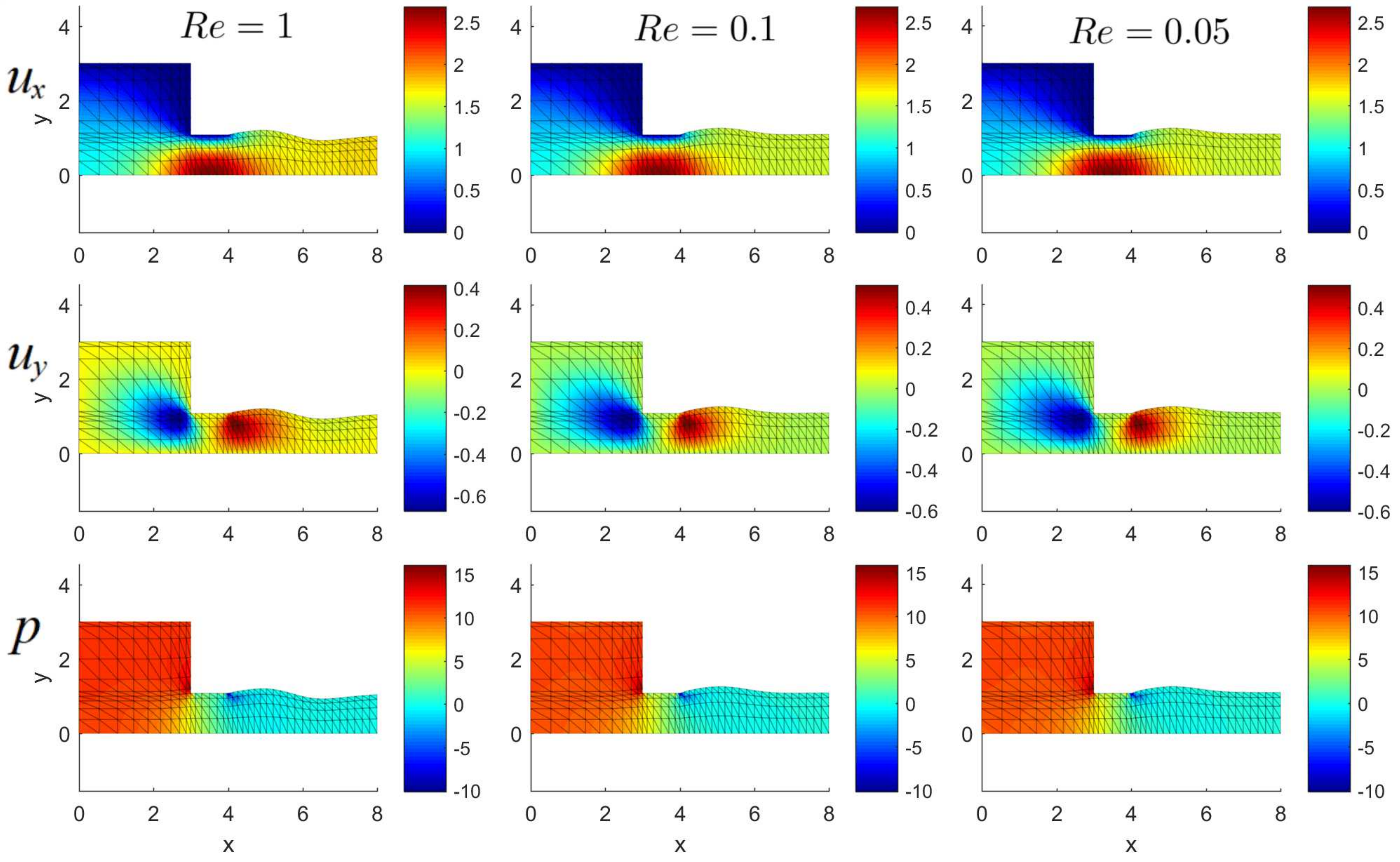}
	    \caption{Low $Re$ die gap increase.}
	    \label{fig:LReO}
    \end{subfigure}
    \caption{$u_x$, $u_y$ and $p$ for $Re=1$, $Re=0.1$ and $Re=0.05$.}
    \label{fig:LRe}
\end{figure}

In practical applications, polymer melts are very viscous fluids and should be modeled using small $Re$ values \cite{VinodPoF,Vinod2}. As the value of $Re$ decreases, the largest possible $\Delta t$ decreases making the simulation computationally expensive. In Figure \ref{fig:LRe}, $Re = 1$, $Re = 0.1$ and $Re = 0.05$ are considered. To be able to simulate such $Re$ values, the simulation time is reduced to $3$, extrudate length is reduced to $4$, and Mesh2.5 is used. The results of Simulations 29, 30, 31, 32, 33 and 34 at $t=2$ are used in Figure \ref{fig:LRe}. $\Delta t$ as small as $5\times 10^{-6}$ is used for $Re=0.05$ as shown in Table \ref{table:3}.  

It can be observed that the results of $Re=0.1$ and $Re=0.05$ are similar to the results of $Re=1$. As values of $Re$ decrease, the most significant difference observed is the thickness variation of the extrudate. The free surface variation gets smoother as $Re$ decreases. Moreover, the gradient of $u_y$ becomes smaller for lower values of $Re$. These observations are also true when comparing $Re=1$ and $Re=10$ in Appendix \ref{sec:Re1} and Appendix \ref{sec:Re10}.

\section{Conclusion} 

FEM simulation results for varying die gap extrusion are presented. To validate the simulator, mesh convergence analysis and literature comparison is performed for the steady simulator. Moreover, the new free surface interpolation scheme is shown to eliminate free surface oscillations. The oscillations occur when higher values of $Re$ are considered and a coarse mesh is used. Eliminating oscillations makes cheap coarse mesh simulations possible. In addition, the effects of varying the die gap on pressure, velocity and extrudate shape is presented and described. Varying the die gap leads to the formation of bulge and neck shapes on the extrudate. Factors contributing to bulging and necking are studied and reported. Finally, simulations for diminishingly small values of $Re$, which model viscous molten polymer, are presented in the last section. 

\chapter{Parameter Identification Based Modeling}

\section{Introduction}
The FEM model discussed in Chapters 2 and 3 is not suitable for simulating practical setups and controller design. Spatially discretizing the governing equations results in very high order nonlinear ODEs, which are complex and computationally very expensive. In addition, several simplifications are performed such as assuming isothermal conditions and considering Newtonian fluids. To address these issues, a parameter identification based approach for modeling time dependent extrudate thickness is proposed in this chapter. The proposed model has a transport PDE/nonlinear ODE cascade structure. The structure is designed to replicate observed characteristics of the extrudate thickness of the FEM simulator's step response. 

\begin{figure}
\centering
	\includegraphics[width=0.35\linewidth]{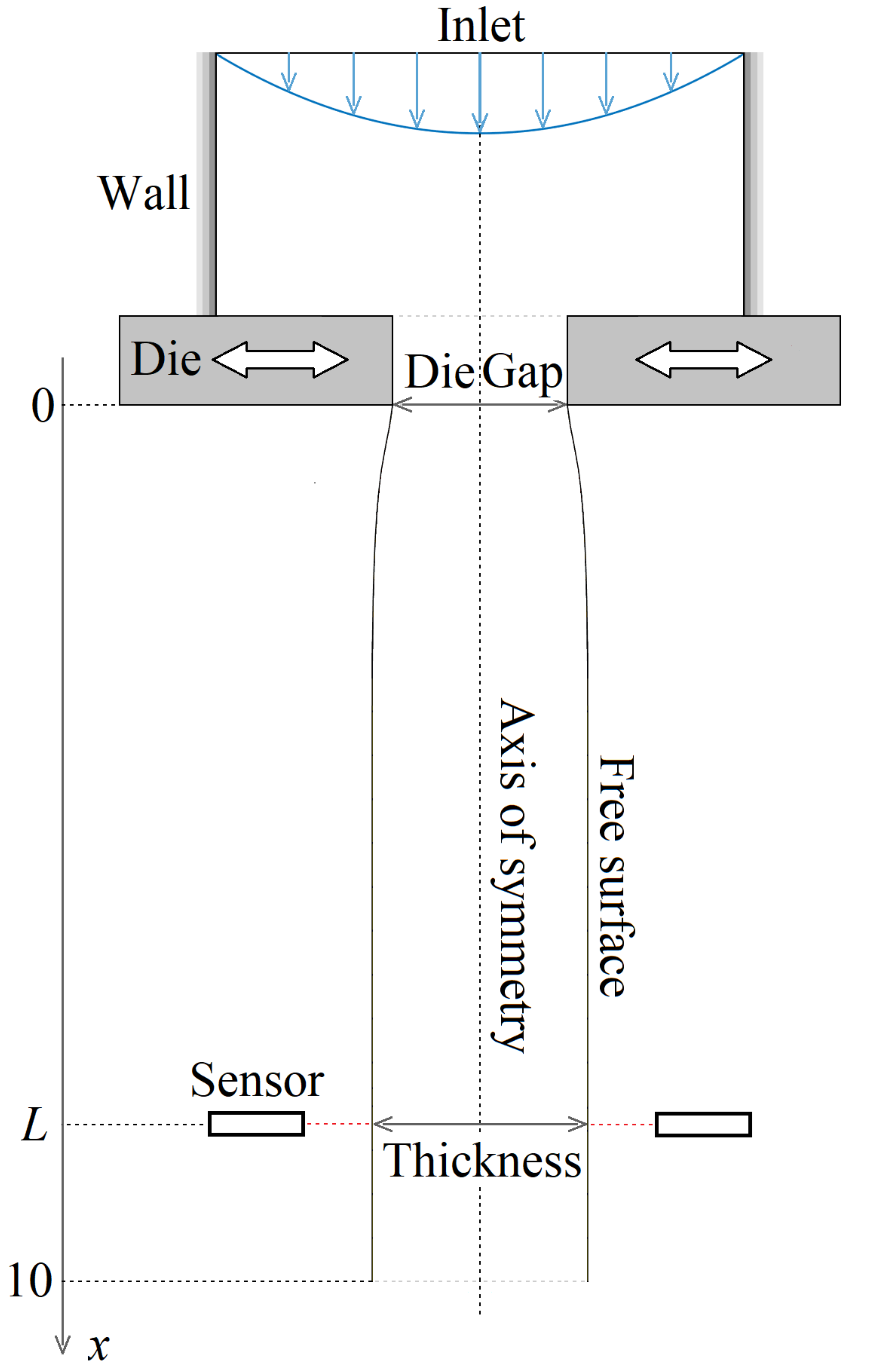}
	\caption{Extruder diagram.}
	\label{fig:Ext3}
\end{figure}

The FEM simulator presented in Chapters 2 and 3 is used as a reference model to obtain thickness data used to identify the parameters of the proposed model. The same configuration used in previous chapters is used throughout the chapter. The origin $(x=0)$ of the coordinate system is at the die exit and $x$ is the distance away from the die as shown in Figure \ref{fig:Ext3}. In addition, Reynolds number of 2.5 and Mesh2 is used. Note that lower values of $Re$ mimic polymer melts flow closer, however, $Re$ values and mesh used in the simulation are chosen by taking into consideration the excessive computational requirements of the FEM simulator.

Note, real setups can be used as a reference model where measured thickness can be used to identify the model. Different types of plastic thickness measurement sensors are available. Proximity sensors from both sides could be used to measure the thickness given that the distance between the sensors is known. 

\section{Model Description and Identification}
\label{sec:MDisc}

\begin{figure}
\centering
	\includegraphics[width=0.6\linewidth]{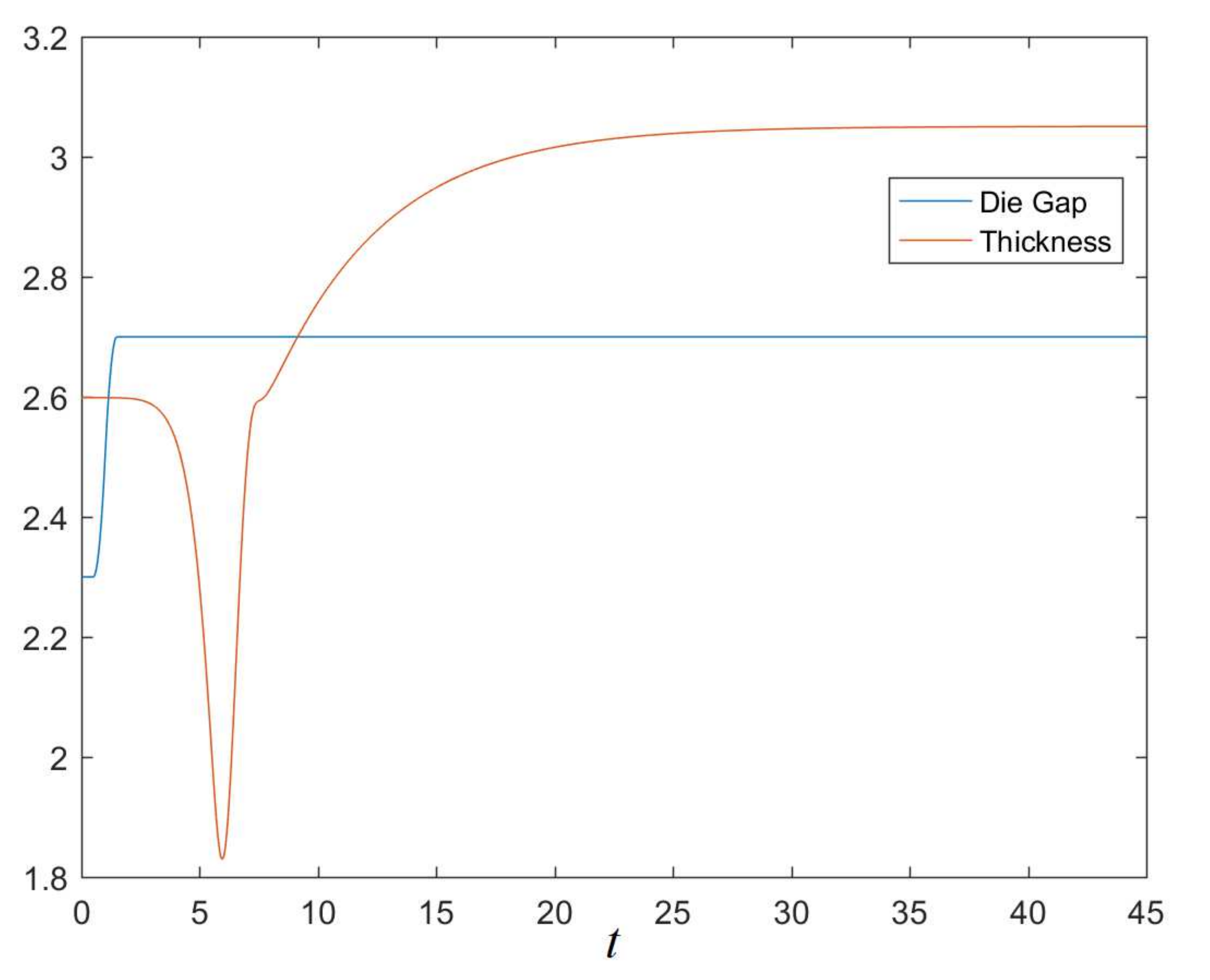}
	\caption{Extrudate thickness versus time.}
	\label{fig:Step1}
\end{figure}

The structure of the model is designed to replicate the observed characteristics of the step response. Figure \ref{fig:Step1} shows the extrudate thickness $y(t)$ versus time  at $x=8$ due to a step change in the die gap $g$. It is assumed that $g$ holds its initial value for a long time before $t=0$. The die gap changes from 2.3 to 2.7 during a period $\Delta T = 1$. The die gap increases with a constant second derivative of value $a$ for the first half of $\Delta T$ and then increases with a constant second derivative of value $-a$ during the second half. It can be observed that the thickness has a steady value at the beginning ($t = 0$) and at the end ($t = 45$). The steady value is larger than the die gap value due to the swell of the fluid as it is extruded out of the die. It can  also be observed that the thickness does not change instantly when the die gap changes. The thickness decreases before it starts increasing during the transient change due to necking and bulging described in previous chapters.

A Hammerstein model with delay will be used to replicate the extrudate thickness at a certain distance away from the die. The delay models the lag due to the distance between the actuator (die gap) and output (thickness). The Hammerstein model is composed of two cascaded elements. The first element is a static non-linear function that gives the steady thickness value for a given die gap. The second element is a non-minimum-phase dynamic linear transfer function with a unity gain that replicates the transient change of the thickness from one steady value to the next. Therefore, the three main elements that constitute the model are a delay block, a static nonlinear function and a linear dynamic function. 

\subsection{Nonlinear Static Function}
Consider extrusion with a constant die gap $g$ for a long period of time. The extrudate will attain a steady shape which has different thicknesses for different values of $g$. The value of the nonlinear function $f_s(x,g)$ is the thickness of the extrudate at a point $x$ and for a certain value of $g$. $f_s(x,g)$ is an interpolation function created by using the steady thickness at the combination of $g = 1.8, 2.2, 2.3, 2.7, 2.8, 3.2$ ($6$ points) and $x = 0, 0.0625, 0.125, \cdots, 10$ ($161$ points). Let each location point be denoted as $x_0=0$, $x_1=0.0625$, $\cdots$, and $x_{160}=10$. The derivation of the function could be done using curve fitting as done in \cite{Vinod4}. 
\begin{figure}
\centering
	\includegraphics[width=0.75\linewidth]{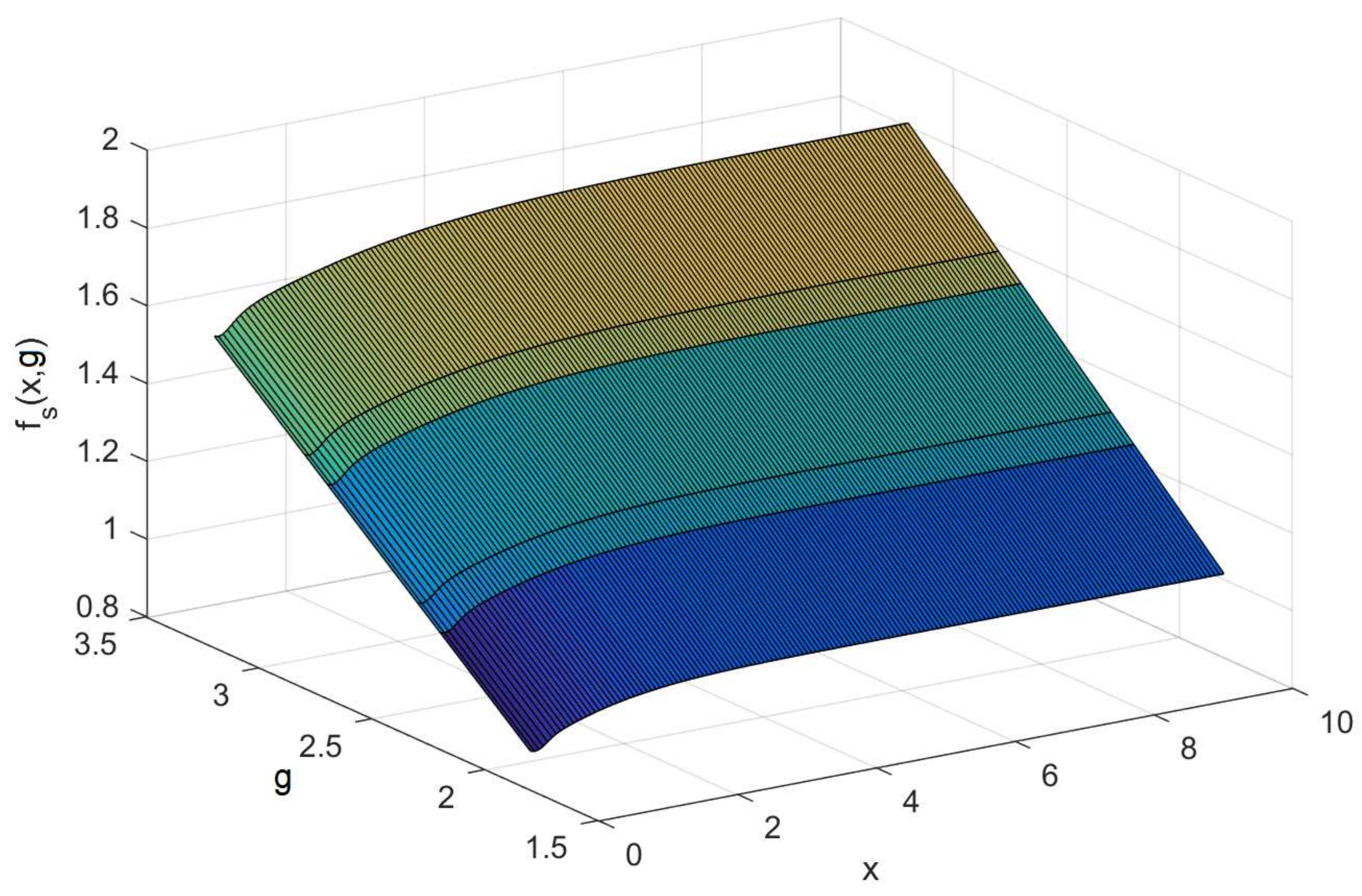}
	\caption{Steady thickness with respect to $x$ and $g$.}
	\label{fig:fs}
\end{figure}

Figure \ref{fig:fs} shows $f_s(x,g)$ which is created by using $6\times 161$ thickness values. The number of points needed depends on the shape of $f_s(x,g)$. It can be observed that $161$ points on the $x-$axis are more than necessary to obtain $f_s(x,g)$. A larger number of points is chosen to observe the parameter variation of the dynamic function with respect to $x$ in later parts of the chapter.

\subsection{Linear Dynamic Function}
\label{sec:ldf}
Low order linear transfer functions with delay are widely used to model a wide range of processes due to their simplicity and practicality \cite{Benoit2,Lopd1,Lopd2}. These models have proven to be effective in model based control algorithms \cite{LopdC1,LopdC2,LopdC3}. Step response tests are widely used to identify low order systems with delay \cite{Step1,Step2,Step3}. In this section, a fourth order transfer function with delay is used to reproduce the major features of the step response of the FEM simulator.  

\subsubsection{Transfer Function}
\begin{figure}
\centering
	\includegraphics[width=0.7\linewidth]{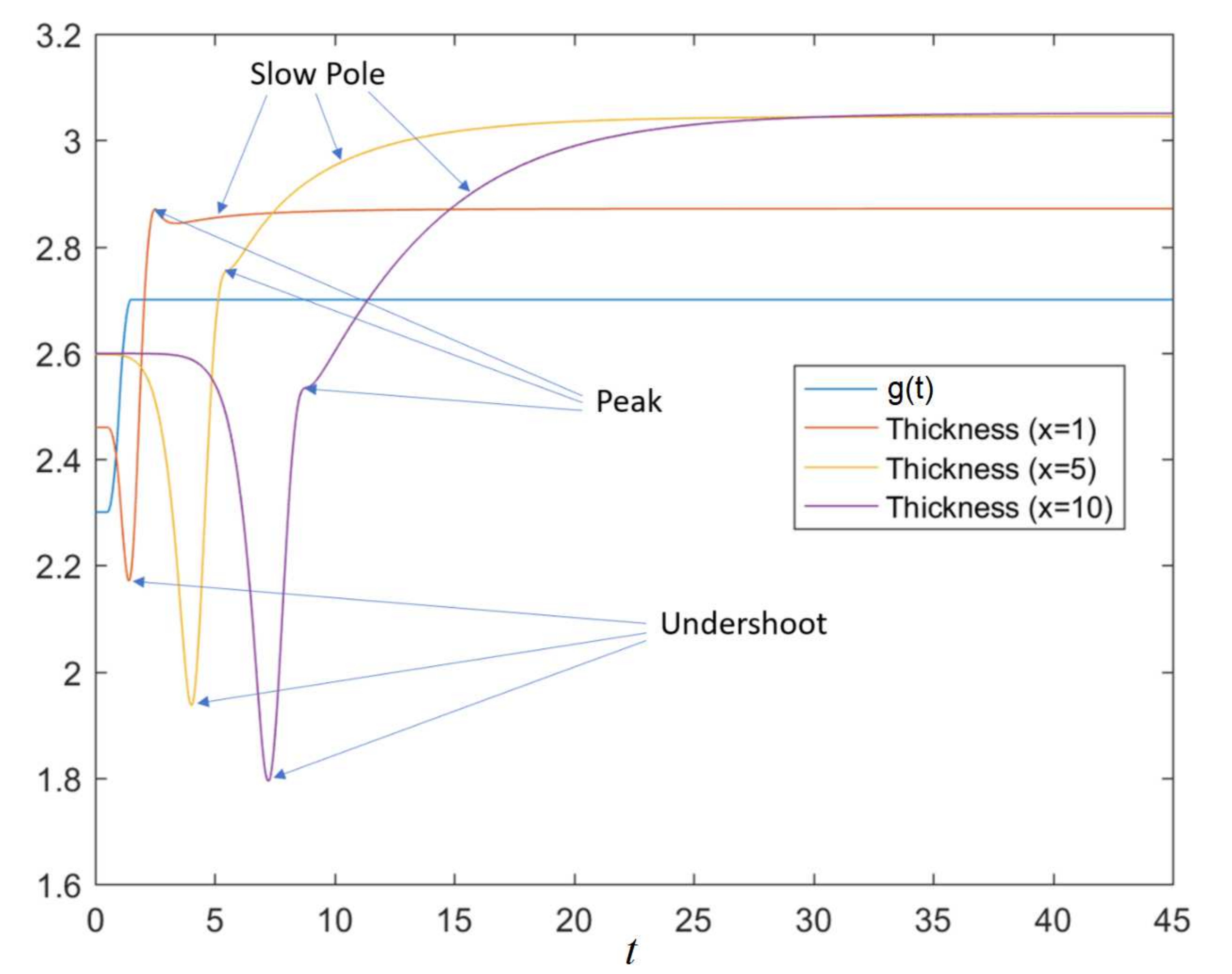}
	\caption{Extrudate thickness versus time.}
	\label{fig:Step2}
\end{figure}

Figure \ref{fig:Step2} depicts the transient thickness change at $x=1$, $x=5$, and $x=10$. The signals at different values of $x$ share common characteristics. The signals undergo an undershoot then a small peak before settling into the final values. The signals also show delay with respect to $g(t)$ especially for larger values of $x$. Therefore, the structure of the linear dynamic function is chosen to replicate these signals for different values of $x$.

\begin{equation}
   \label{eq:Hs}
   H(s,x) = \left(\frac{-G_1(x)p_1(x)}{s+p_1(x)}+\frac{G_2(x)\omega_n^2(x)}{s^2+2\zeta(x)\omega_n(x)+\omega_n^2(x)}+\frac{G_3(x)p_3(x)}{s+p_3(x)}\right)e^{-s\tau(x)}
\end{equation}

$H(s,x)$ is the delayed sum of three transfers $H_1(s,x)$, $H_2(s,x)$, and $H_3(s,x)$ from left to right in (\ref{eq:Hs}). Since the steady-state values are defined by $f_s(x,g)$, the gain of $H(s,x)$ should be equal to 1. Therefore, the gains of the summed transfer functions should satisfy $-G_1(x)+G_2(x)+G_3(x)=1$. $H_1(s,x)$ is a first order system with a negative gain and a fast pole $p_1(x)$. The negative gain creates the undershoot in the signal. $H_2(s,x)$ is a second order transfer function with a natural frequency $\omega_n(x)$ and a damping ratio $\zeta(x)$. The second order transfer function causes the peak in the final signal. $H_3(s,x)$ is a first order transfer function with a slow pole $p_3(x)$ which causes the slow change after the peak. Finally, $\tau(x)$ is the overall delay of the system. 

\begin{figure}
\centering
	\includegraphics[width=0.6\linewidth]{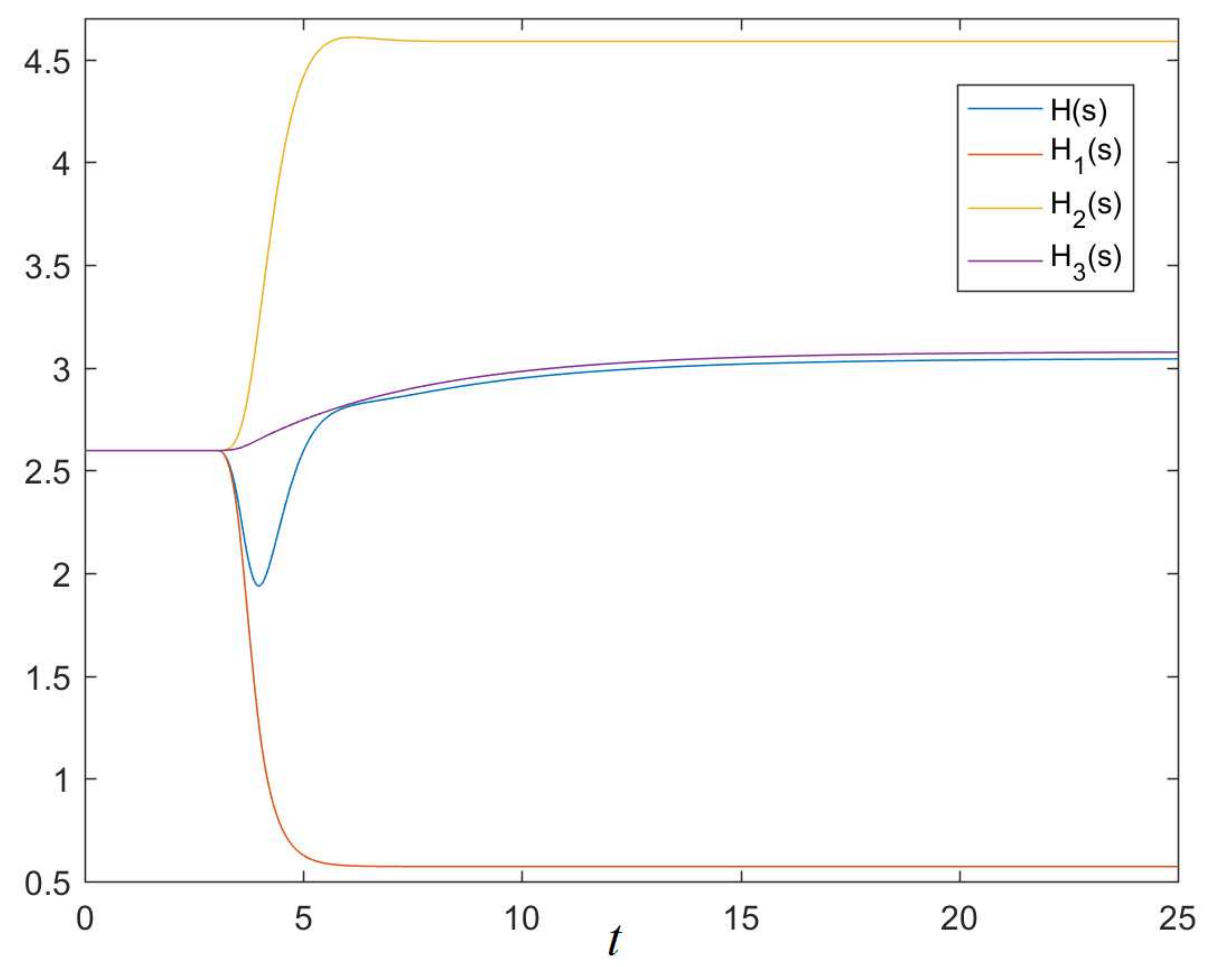}
	\caption{Step response of transfer functions.}
	\label{fig:Step3}
\end{figure}

The overall transfer function is a fourth order transfer function with 3 zeros. The transfer function has 8 parameters that are a function of $x$. Figure \ref{fig:Step3} shows the delayed step response of a sample set of $H_1(s)$, $H_2(s)$, and $H_3(s)$ and the step response of their sum $H(s)$. 

\subsubsection{Transfer Function Parameter Identification}
\label{sec:ID}
The parameter identification algorithm is based on minimizing the integral of the response error between the two systems. Let
\begin{equation}
\label{eq:e1}
\begin{aligned}
   e(G_1,G_3,P_1,P_3,\omega_n,\zeta,\tau,g(t)) &:=\int_0^T |y_{ID}(t)-y_r(t)|dt\\
   &+\int_0^T |\frac{\partial y_{ID}(t)}{\partial t}-\frac{\partial y_{r}(t)}{\partial t}| dt.
\end{aligned}
\end{equation}
 Although the terms in (\ref{eq:e1}) are dependent on $x$, it is not indicated for conciseness. $y_{ID}(t)$ is the response of $H(s)$ for a given set of parameters $G_1$, $G_3$, $P_1$, $P_3$, $\omega_n$, $\zeta$, and $\tau$ due to the input signal $f_s(x,g(t))$. $y_{r}(t)$ is the reference signal that needs to be replicated. In this case $y_{r}(t)$ is the response of the FEM simulator for input signal $g(t)$. $y_{r}(t)$ could be a signal obtained using sensors from experimental \cite{Vinod4} or industrial setups. The derivative of the signal is included in the right hand side of (\ref{eq:e1}) to preserve the main characteristics of $y_{r}(t)$ such as the moment when peaks and minima occur. From $0$ to $T$ is the time interval in which the step response is considered. $T$ should be large enough for the response to reach a steady value. $G_2$ is not considered in (\ref{eq:e1}) since $G_2$ is unique for a given set of $G_1$ and $G_3$. The parameter identification of the transfer function becomes a minimization problem as stated in (\ref{eq:min1}).
\begin{equation}
\label{eq:min1}
\min_{G_1,G_3,P_1,P_3,\omega_n,\zeta,\tau}e(G_1,G_3,P_1,P_3,\omega_n,\zeta,\tau,g(t))
\end{equation}

The minimization problem is solved using the function fminsearch in Matlab\textsuperscript{\textregistered}, which will be used to solve all the minimizations throughout the chapter. The minimization is an iterative process that includes 7 arguments. It is important to have a good initial guess for iterative techniques having a large number of arguments. 

Several assumptions are made to obtain an initial guess for \ref{eq:min1}. The step signals used for parameter identification are not ideal. It takes $\Delta T$ for the step to change from one value to the other. In order to approximate some of the parameters of $H(s)$, the step is assumed to be ideal. This assumption becomes more valid as $\Delta T$ becomes smaller with respect to the settling time of the step response. Therefore, the initial guess of the parameters of $H(s)$ with largest settling time ($x = 10$) are estimated first. 

In a step response of an underdamped second order system the peak occurs at $t_p=\pi/w_n\sqrt{1-\zeta^2}$ assuming the step occured at $t=0$ \cite{Ogata}. The instant at which the peak occurs in the step response of $H(s)$ is approximated to be equal to the instant at which it occurs for $H_2(s)$. Therefore, the initial guess for the natural frequency of $H_2(s)$ is approximated by 
\begin{equation}
\label{eq:omega}
\hat{\omega}_n=\frac{\pi}{t_p\sqrt{1-\zeta^2}}.
\end{equation}
Assuming that the second order system is underdamped, the initial guess of the damping ratio is set to $\hat{\zeta}=0.65$. The initial guess of $\tau$ is estimated by 
\begin{equation}
\label{eq:tau}
\hat{\tau} = t_m - 1.1\times(t_p-t_m)
\end{equation}
where $t_m$ is the instant when the minimum value is reached. In (\ref{eq:tau}), it is assumed that the time interval between $t_m$ and the instant when the signal starts to change $\tau$ is slightly greater than the time interval between $t_p$ and $t_m$. 

\begin{figure}
\centering
	\includegraphics[width=0.7\linewidth]{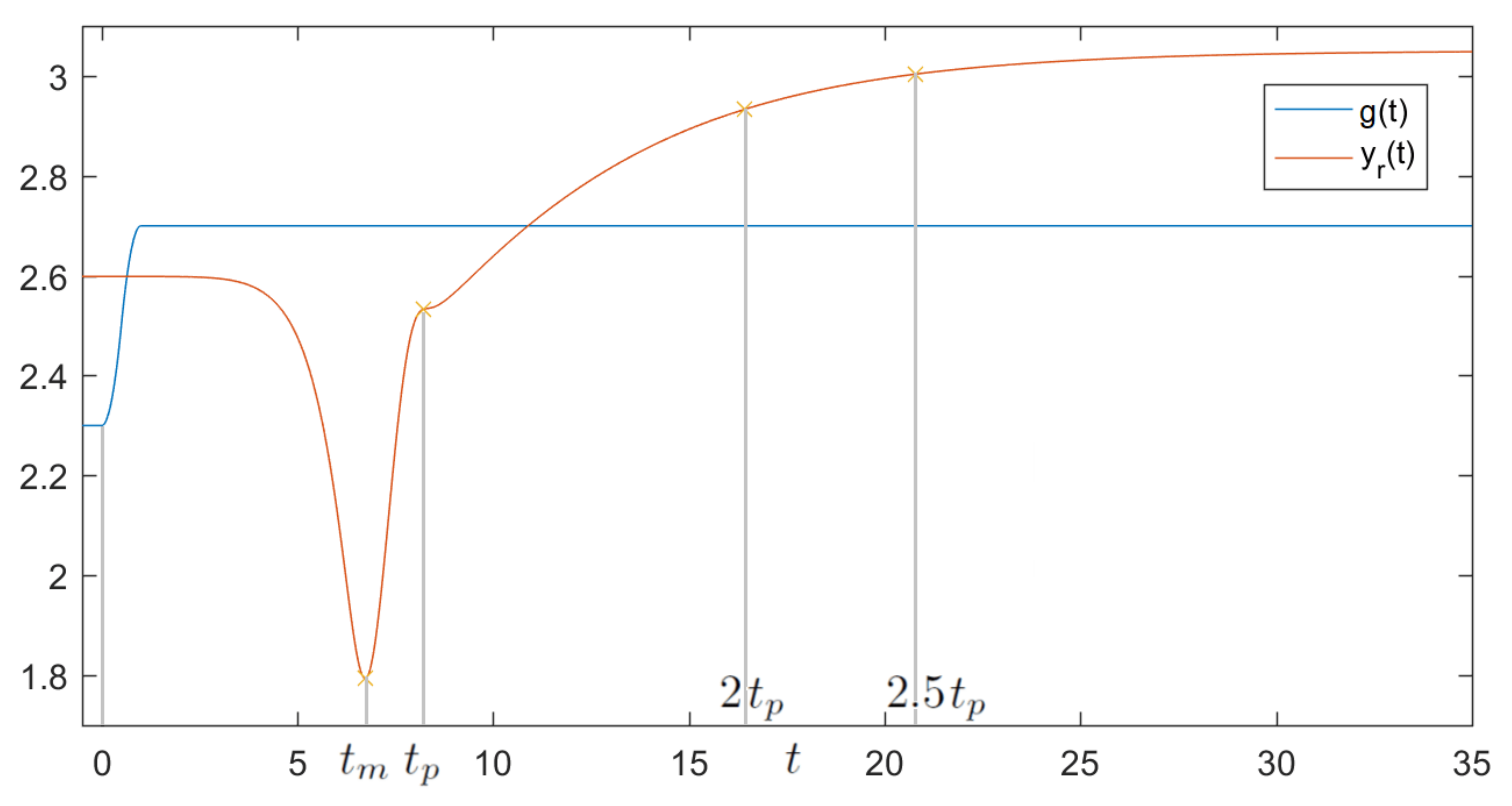}
	\caption{Time instances used to estimate parameters.}
	\label{fig:Step15}
\end{figure}

The fact that $H_3(s)$ has the longest settling time is exploited to approximate the parameters of $H_3(s)$. It is assumed that $H_1(s)$ and $H_2(s)$ settle before $H_3(s)$. Therefore, the derivative of the responses of $H_1(s)$ and $H_2(s)$ are assumed to be zero at $t_1=2t_p$ and $t_2=2.5t_p$ and $\partial y_{3}(t)/\partial t$ is approximated by $\partial y_{r}(t)/\partial t$. Figure \ref{fig:Step15} shows the instances $t_m$, $t_p$, $2t_p$ and $2.5t_p$ used to evaluate some of the initial guesses of the parameters. Note, $t=0$ is set to the instant the input starts to change. The derivative of the response of $H_3(s)$ for an ideal step is
\begin{equation}
\label{eq:dy3}
 \frac{\partial y_{3}(t)}{\partial t} = AG_3P_3e^{-p_3t}
\end{equation}
where $A$  is the step amplitude. $\partial y_{r}(t)/\partial t$ is obtained by numerically differentiating $y_{r}(t)$. By replacing $t_1$, $t_2$, $\partial y_{3}(t_1)/\partial t$, and $\partial y_{3}(t_2)/\partial t$ in (\ref{eq:dy3}), the initial guess of $P_3$ can be written as 
\begin{equation}
\label{eq:P3}
 \hat{P_3} = \frac{-\ln(\frac{\partial y_{3}(t_1)/\partial t}{\partial y_{3}(t_2)/\partial t})}{t_1-t_2}
\end{equation}
The initial guess of $G_3$ is set to
\begin{equation}
\label{eq:G3}
 \hat{G}_3 = \frac{1}{2}\left(\frac{\partial y_{3}(t_1)/\partial t}{AP_3e^{-P_3t_1}}+\frac{\partial y_{3}(t_2)/\partial t}{AP_3e^{-P_3t_2}}\right).
\end{equation}
$\hat{G}_3$ could be set to the left or right term within the parenthesis in (\ref{eq:G3}). The average of the terms is used in this paper as presented in \ref{eq:G3}. 

The initial guesses of the gain and pole of $H_1(s)$ are denoted by $\hat{G}_1$ and $\hat{P}_1$ respectively. $\hat{G}_1$ and $\hat{P}_1$ are computed by solving the minimization problem
\begin{equation}
\label{eq:min2}
\min_{G_1,P_1}e(G_1,\hat{G}_3,P_1,\hat{P}_3,\hat{\omega}_n,\hat{\zeta},\hat{\tau},g(t)).
\end{equation}
The initial guesses of the minimization are denoted by $\bar{G}_1$ and $\bar{P}_1$. $\bar{P}_1$ is estimated by $\hat{\omega}_n$ and $\bar{G}_1$ is estimated to be equal to $-G_2$. Given that $H(s)$ has a unity gain, $\bar{G}_1$ is equal to $|1-\hat{G}_3|/2$. The minimization in (\ref{eq:min2}) is performed because the initial guess $\bar{G_1}$ and $\bar{P_1}$ are very rough estimates. $\bar{G}_1$ and $\bar{P}_1$ are improved to $\hat{G}_1$ and $\hat{P}_1$ as a better initial guess for (\ref{eq:min1}).

\begin{figure}
\centering
	\includegraphics[width=0.85\linewidth]{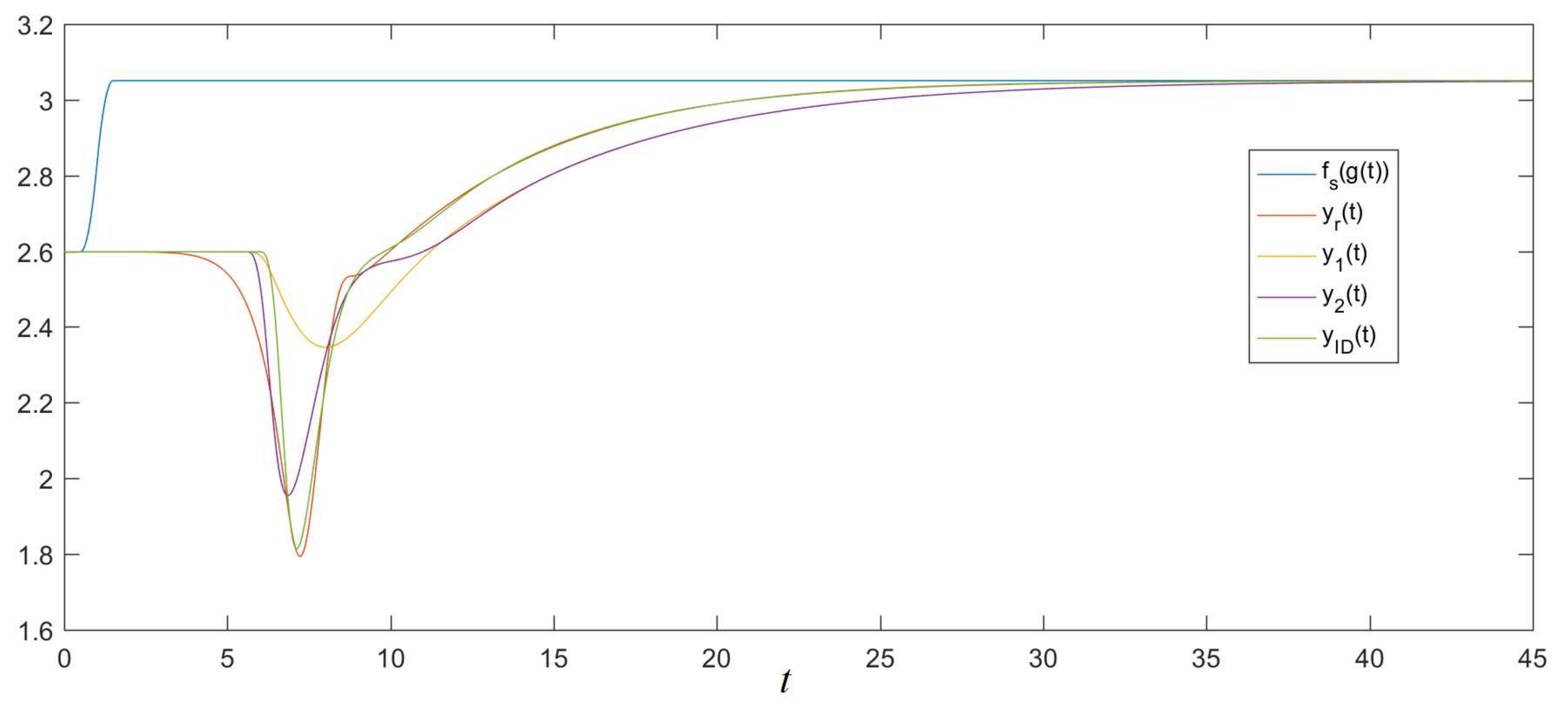}
	\caption{Step response of identified transfer functions.}
	\label{fig:Step4}
\end{figure}

In Figure \ref{fig:Step4}, the step responses of the identified transfer functions at $x = 10$ are plotted. $y_1(t)$ is the step response of transfer function with the set of parameters $\bar{G}_1$, $\hat{G}_3$, $\bar{P}_1$, $\hat{P}_3$, $\hat{\omega}_n$, $\hat{\zeta}$, and $\hat{\tau}$. $y_2(t)$ is the step response of transfer function with the set of parameters $\hat{G}_1$, $\hat{G}_3$, $\hat{P}_1$, $\hat{P}_3$, $\hat{\omega}_n$, $\hat{\zeta}$, and $\hat{\tau}$. $y_{ID}(t)$ is the response of transfer function having the final set of parameters which are computed by solving (\ref{eq:min1}). For transfer functions having $x<10$, its parameters are identified in a decreasing order of $x$. The identified transfer function parameters for a certain value of $x$ are used as a first guess for the next transfer function parameter identification. 

\begin{figure}
\centering
	\includegraphics[width=0.7\linewidth]{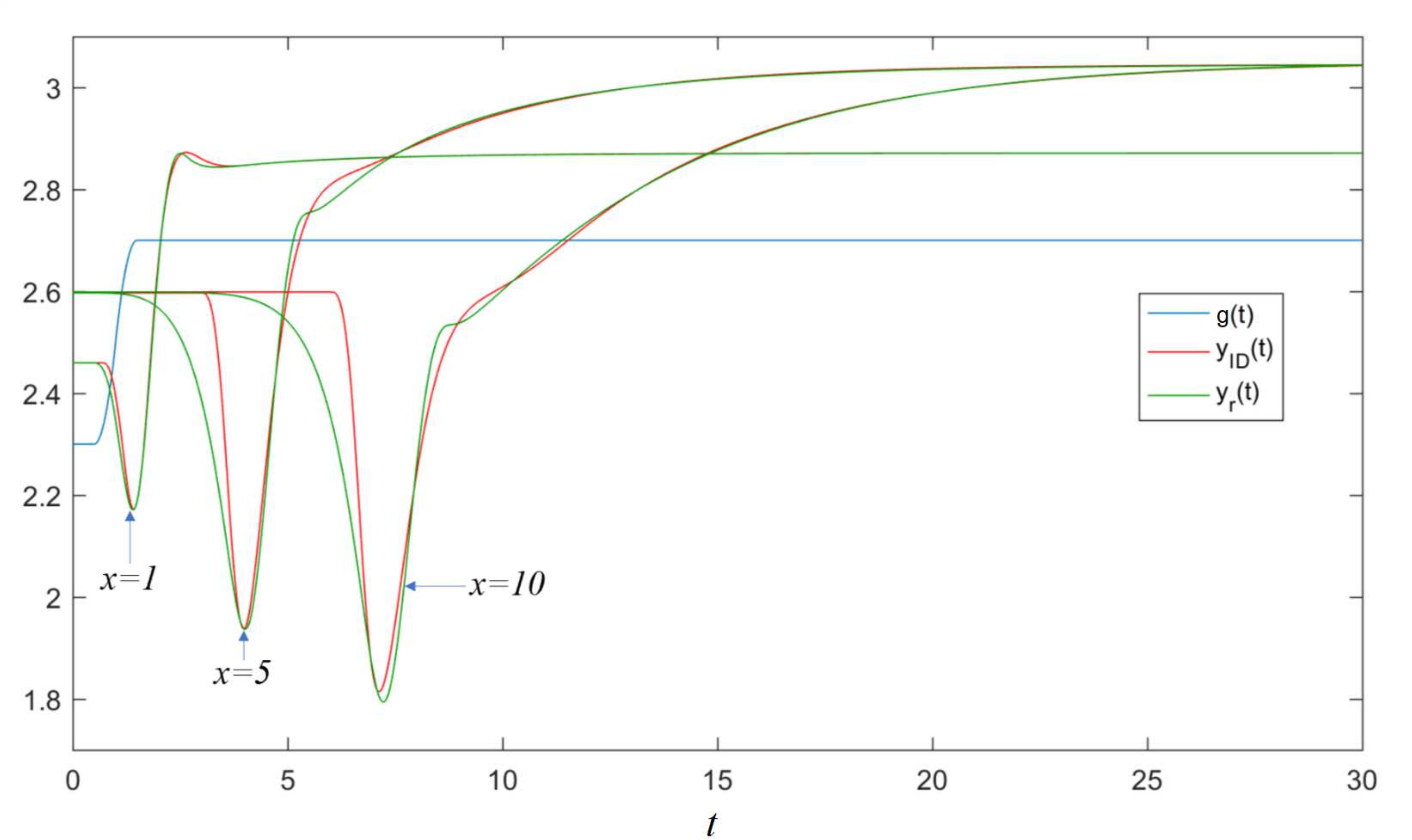}
	\caption{Step response of identified transfer functions.}
	\label{fig:Step5}
\end{figure}

In Figure \ref{fig:Step5}, the responses from the identified transfer function $y_{ID}(t)$ and the FEM simulator $y_{r}(t)$ are plotted for $x=1$, $x=5$, and $x=10$. It can be observed that the response of transfer function having $x=10$ has the biggest error due to the large delay and the slow decrease at the beginning of the FEM response. The error could be decreased by using higher order transfer functions that can replicate the slow decrease.

Note, extrusion of different fluids or different extrusion configurations may result in different responses. Depending on the response, different transfer function identification techniques and/or different transfer function orders can be used. The identification technique presented in this paper preserves the step response's main characteristics of the FEM simulator having $Re=2.5$.

\subsubsection{Transfer Function Parameters Dependency on the Input}
The Navier Stokes equations solved in the FEM simulator are nonlinear. Different parameters for the linear transfer function can be obtained if different inputs are used. Therefore, the proposed model simulates best inputs close to the inputs used to identify the model. The input operation range for the proposed model is chosen in the range $2<g<3$. In this subsection, different step inputs within the operation range are used to identify the parameters of the model. The dependency of the parameters on the input signal used is studied and discussed. 

\begin{figure}
\centering
	\includegraphics[width=1\linewidth]{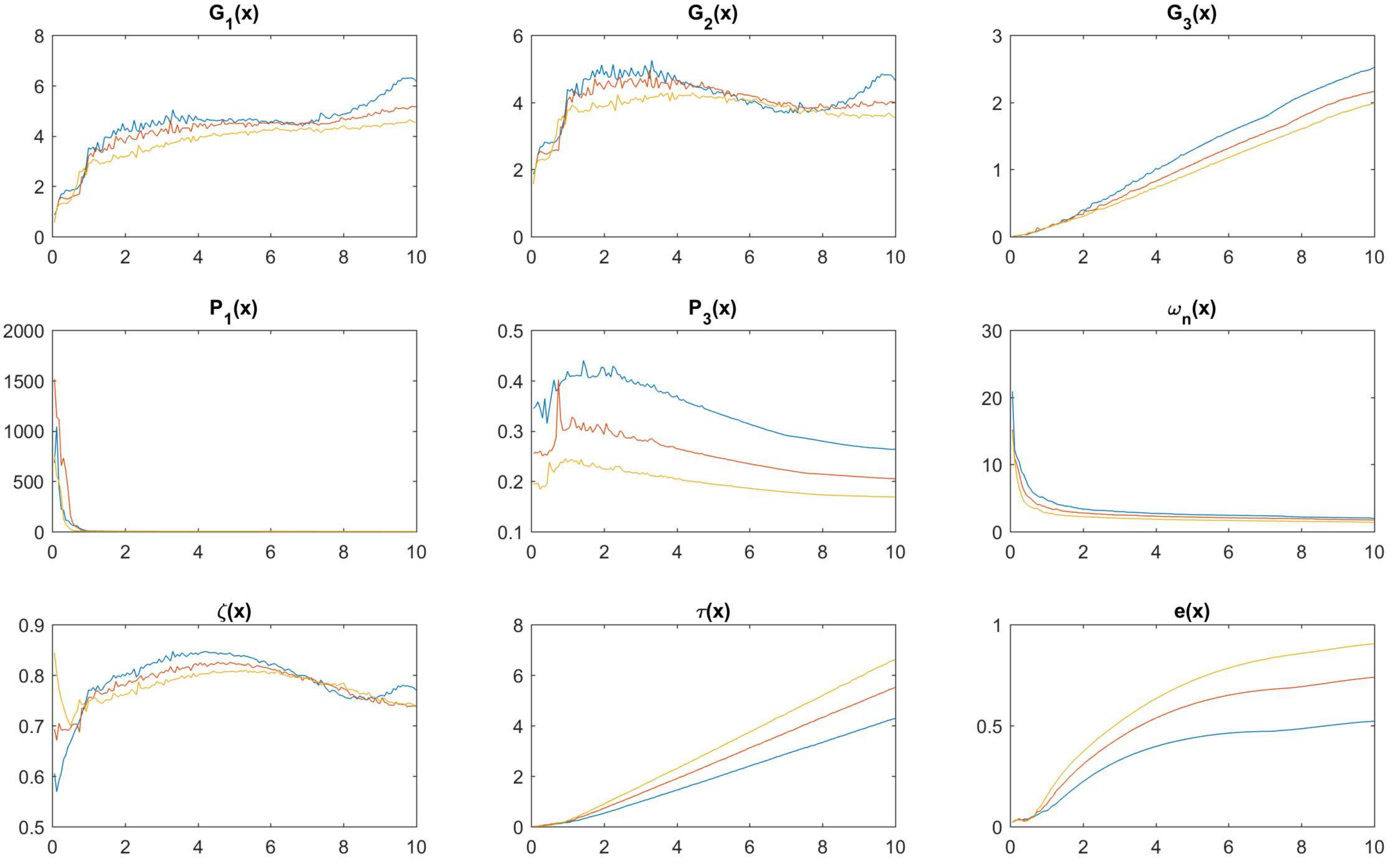}
	\caption{Transfer function parameters and $e$ versus $x$.}
	\label{fig:Para1}
\end{figure}

Figure \ref{fig:Para1} shows the identified parameters and $e$ versus $x$ obtained by solving \ref{eq:min1} for different input signals. The blue, red, and yellow plots show parameters and $e$ obtained using input signals $g_1(t)$, $g_2(t)$, and $g_3(t)$ respectively. $g_2(t)$ is equivalent to input signal in Figure \ref{fig:Step1}. $g_1(t)$ and $g_3(t)$ are similar but have different initial and final values. The initial and final values of $g_1(t)$ are 1.8 and 2.2 respectively, while the initial and final value of $g_3(t)$ are 2.8 and 3.2 respectively. It can be observed that parameters identified using $g_1(t)$, $g_2(t)$, and $g_3(t)$ have similar trends.

\begin{figure}
\centering
	\includegraphics[width=0.9\linewidth]{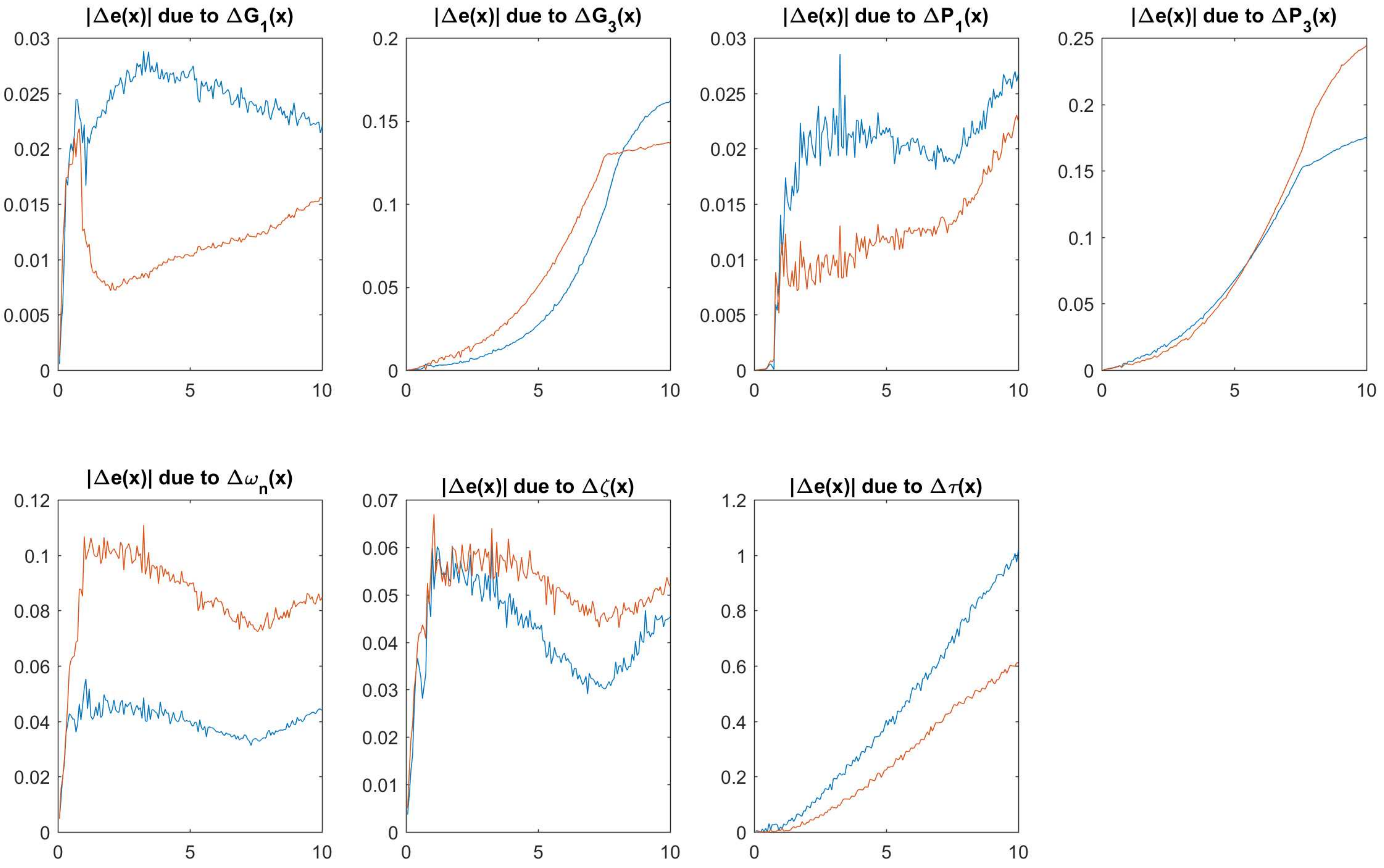}
	\caption{$|\Delta e(x)|$ due to parameter change.}
	\label{fig:ParaE}
\end{figure}

In Figure \ref{fig:ParaE}, the sensitivity of $e(x)$ is studied with respect to each parameter. Each parameter is increased and decreased by $10\%$. The blue plot is $|\Delta e(x)|$ when the parameter is increase while the red plot is when the parameter is decrease. $\Delta e(x)$ is the absolute change in $e(x)$ when the parameter is varied. It can be observed that the error is most sensitive to $\tau$. Moreover, in controller design, the model delay is a crucial parameter for system stability \cite[p.~411]{BouletBook}. An overpredicted model delay may lead to slow closed loop response and an underpredicted delay model may lead to an unstable closed loop system. Based on the discussion and analysis in this subsection, the transfer function parameters of the proposed model are set to the parameters identified using $g_2(t)$ except for the delay $\tau$. In addition, the parameters are chosen to be independent of the input except for $\tau$. 

\subsection{Transport Delay}
\label{sec:TD}
During the extrusion process, the observed delay is caused due to different factors. The major factor is the time needed for the material to transport from the die to the location of measurement. Another factor could be that the die gap change could create a wave that travels over the material. In this work, it is assumed that the delay is entirely due to material transport. Two material transport models will be proposed to estimate an input dependent delay. The first model assumes that the material being extruded is a solid. The particle on different locations of the extrudate have the same velocity and will be referred to as solid extrusion model. In the second model, improvements on the solid extrusion model are applied to better model the identified delay.  The velocity of the particles on the extrudate in the second model are not uniform and will be referred to as fluid extrusion model. 

\subsubsection{Solid Extrusion Model}
\label{sec:sem}

It is assumed that when the material exits the die, the extrudate formed does not change shape. Therefore, the velocity of the particles on the extrudate is uniform $v_s(t)=Q(t)/g(t)$, where $Q(t)$ is the volumetric flow rate at the die exit ($x=0$). It is shown that $Q(t) = Q_0-L_D\times \partial g(t)/ \partial t$ for this die configuration in Chapter 3, where $Q_0$ is the volumteric flow rate at the inlet and $L_D$ is the length of the die. Combining the two equations results in
\begin{equation}
\label{eq:v1}
v_s(t)= \frac{Q_0-L_D\times \partial g(t)/ \partial t}{g(t)}.
\end{equation}
$Q_0$ and $L_D$ are equal to $4$ and $1$ respectively. The delay is the time needed for a particle to travel with velocity $v_s(t)$ from the die exit to the location of the output considered. Therefore, the delay is dependent on $g(t)$ for $t_1 \leq t \leq t_2$. $t_1$ is the time when the particle was at the die exit and $t_2$ is the instant when the particle reaches the point of measurement at $x$. Thus the delay is equal to $t_2-t_1$ and the delayed input $g_d(t_2,x)$ is equal to $g(t_1)$. 

The delay at $x$ could be represented by two different functions denoted by input and output delay functions. Input delay function is defined with respect to the time reference of the input $g(t)$. The input delay function $\tau_{in}(t_1,x)$ is how much $g(\cdot)$ will be delayed at $t_1$ where $\tau_{in}(t_1,x)=t_2-t_1$. Output delay function is defined with respect to the time reference of the delayed input $g_d(t,x)$. The output delay function $\tau_o(t_2,x)$ is how much $g_d(\cdot,x)$ has been delayed at time $t_2$ where $\tau_o(t_2,x)=t_2-t_1$. 

\begin{figure}
\centering
	\includegraphics[width=0.7\linewidth]{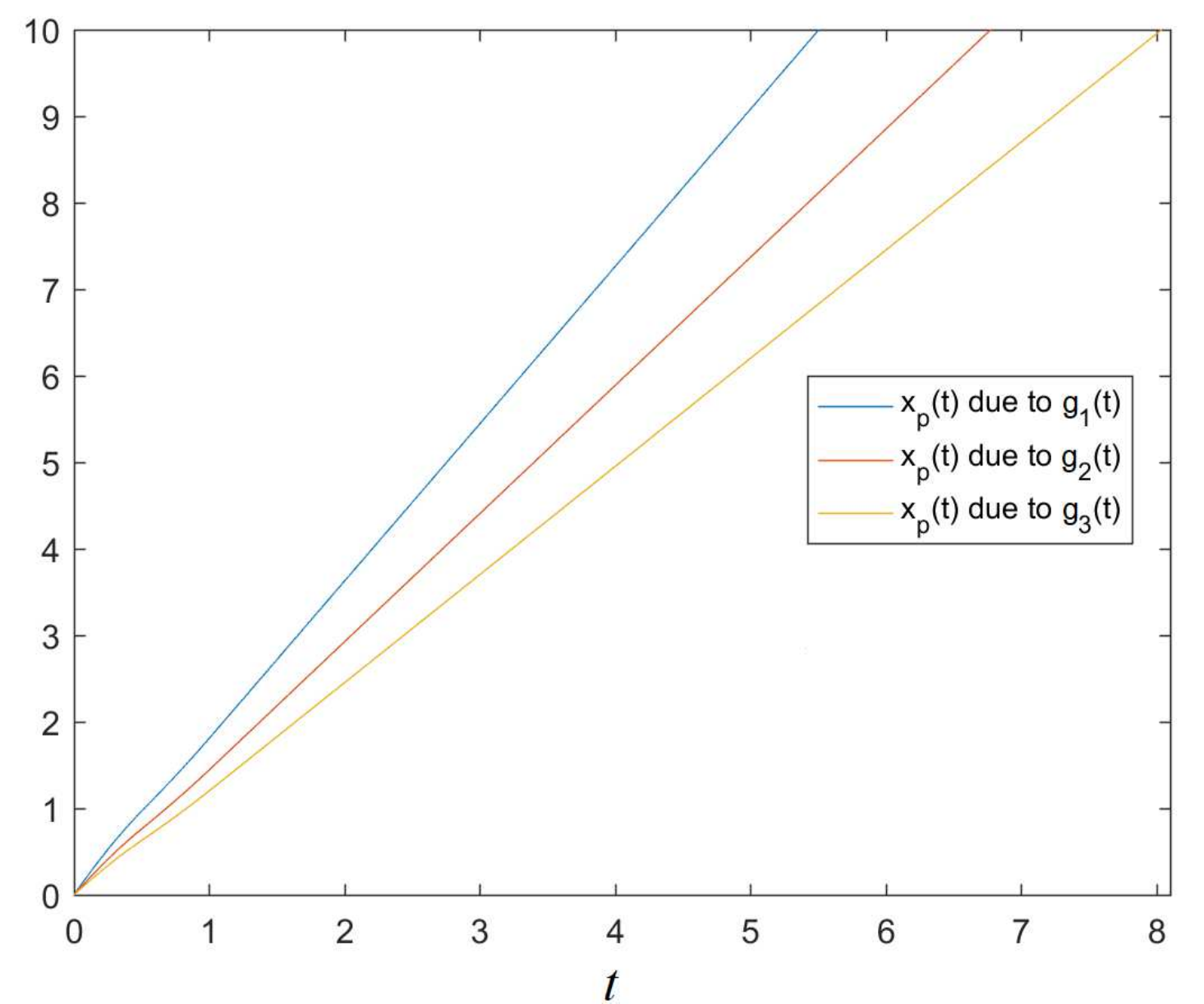}
	\caption{Particle travel versus time.}
	\label{fig:xp}
\end{figure}

In subsection \ref{sec:ID}, the identified delay $\tau_{ID}(x)$ is the input delay at the instant when the input starts to change denoted by $t_0$. $\tau_{ID}(x)$ is modeled by $\tau_{in}(t_0,x)$ using the transport model. $\tau_{in}(t_0,x)$ can be computed by tracking the position of a particle $x_p(t)$ located at the die exit ($x=0$) at $t_0$ as time progresses. $x_p(t)$ can be computed by integrating the velocity $v(t)$ as follows.
\begin{equation}
\label{eq:xp}
x_p(t) = \int_{t_0}^{t}v(\tau)d\tau + x_p(t_0).
\end{equation}
Equation (\ref{eq:xp}) is solved numerically by discretization with respect to time. The velocity is numerically integrated until $x_p(t)$ reaches a value greater than the considered extrudate length ($x=10$). Figure \ref{fig:xp} shows the particle travel due to $g_1(t)$, $g_2(t)$, and $g_3(t)$ for $t_0=0$. It can be observed that the particle travels faster for smaller values of $g$. The position of the particle increase nearly linearly for the given model.

Given that $x_p(t)$ is the location of the particle at a certain time and the delay is the time needed for the material to travel from the die to the location of the output, $\tau_{in}(t_0,x)$ is equal to the inverse of the computed function $x_p(t)$. The inverse of the function $x_p^{-1}(x)$ exists if $v_s(t)>0$ for a given time interval. If $v_s(t)>0$, $x_p(t)$ is monotonically increasing. Thus, the inverse of $x_p(t)$ has a unique solution for any value in the given time interval. Note that, inputs that cause negative values of $v_s(t)$ are considered to be out of this model's operational range. Negative flow rate values at the die exit (equivalently $v_s(t)<0$) may be caused by a sudden increase in the die gap. Negative values of flow rate at the die exit cause highly nonlinear behavior on the extrudate surface and are usually avoided in industry.

\begin{figure}
\centering
	\includegraphics[width=0.65\linewidth]{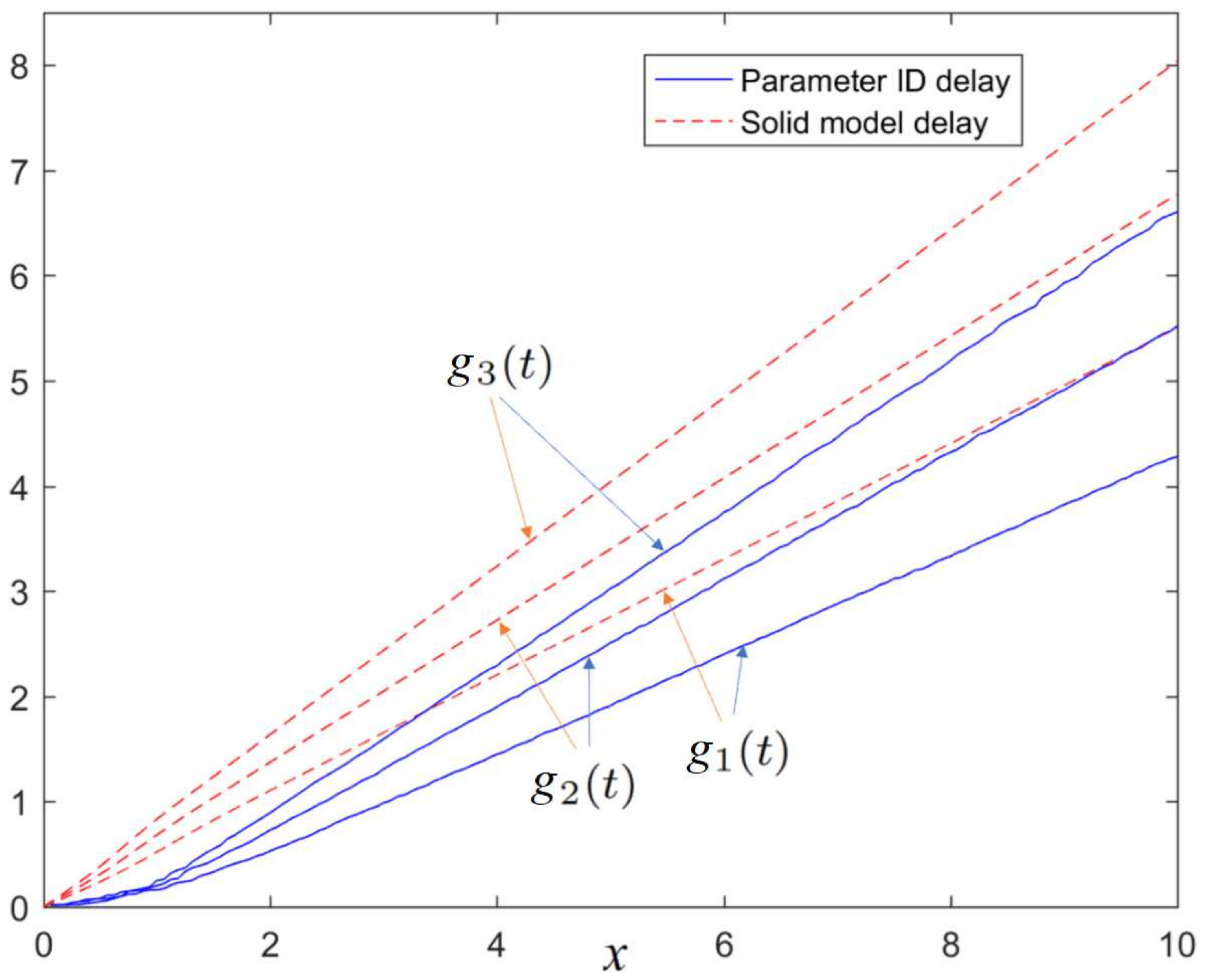}
	\caption{Solid model delay due to $g_1(t)$, $g_2(t)$ and $g_3(t)$.}
	\label{fig:taus}
\end{figure}

$x_p^{-1}(x)$ is obtained by interpolating $t$ versus $x_p$ at every discrete value of $x_p$ obtained during numerical integration. Figure \ref{fig:taus} shows the identified delay and solid model delay due to $g_1(t)$, $g_2(t)$ and $g_3(t)$ with respect to $x$. Given that $g_1(t)$, $g_2(t)$ and $g_3(t)$ have equivalent derivatives, the delay due to $g_1(t)$ is the smallest while the delay due to $g_3(t)$ is the largest. It can be observed that the solid model overpredicts the delay and increases nearly linearly as $x$ increases. On the other hand, the derivative of the identified delay with respect to $x$ increases gradually until it reaches a constant value after a certain value of $x$. 

\subsubsection{Fluid Extrusion Model}
\begin{figure}
\centering
	\includegraphics[width=0.7\linewidth]{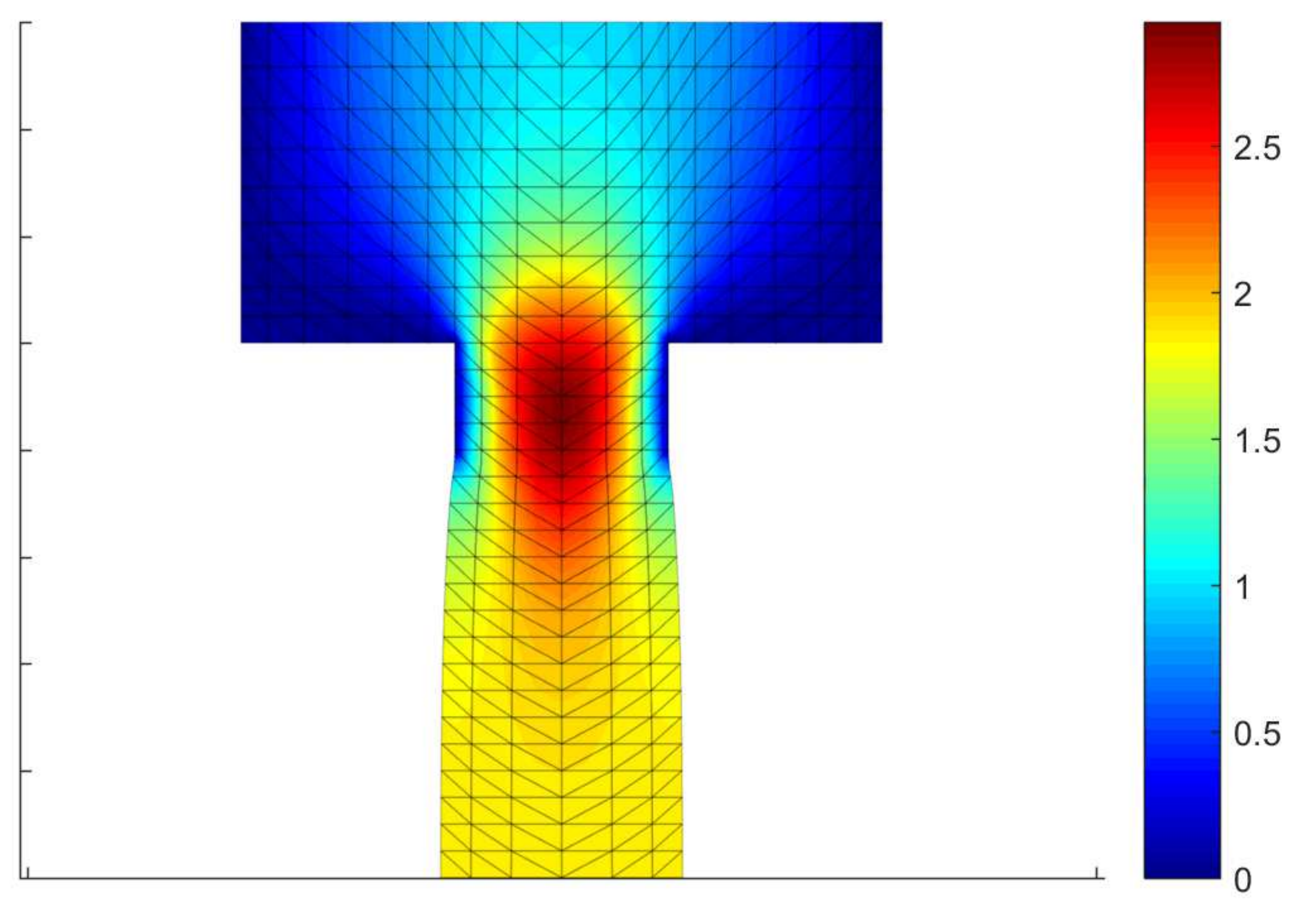}
	\caption{Fluid velocity in the downward direction.}
	\label{fig:v}
\end{figure}

During fluid extrusion, particles on the extrudate do not have uniform velocity. Therefore, the extrudate changes shape as it is being extruded. Figure \ref{fig:v} shows the downward component of the velocity vector of the fluid  around the die exit obtained using the FEM simulator. It can be observed that as soon as the fluid leaves the die, the extrudate will have a thickness greater than the die gap due to extrusion swell. It can also be observed that the velocity in the downward direction is not uniform at the die $(x=0)$. The fluid velocity reaches its maximum value in the center and decreases until it reaches its minimum value zero at the wall due to the no slip boundary condition assumed. After the fluid flows away from the die, the velocity becomes uniform (plug velocity profile). Particles at the center of the die, travel faster and need less time to reach a certain point away from the die compared to other particles at the die. In the case of a step die gap change, the change at the output will start as soon as the fastest particle reaches the output (sensor) location. This explains part of the reasons why the solid extrusion model overpredicts the delay, especially for small values of $x$. 

\begin{figure}
\centering
	\includegraphics[width=0.7\linewidth]{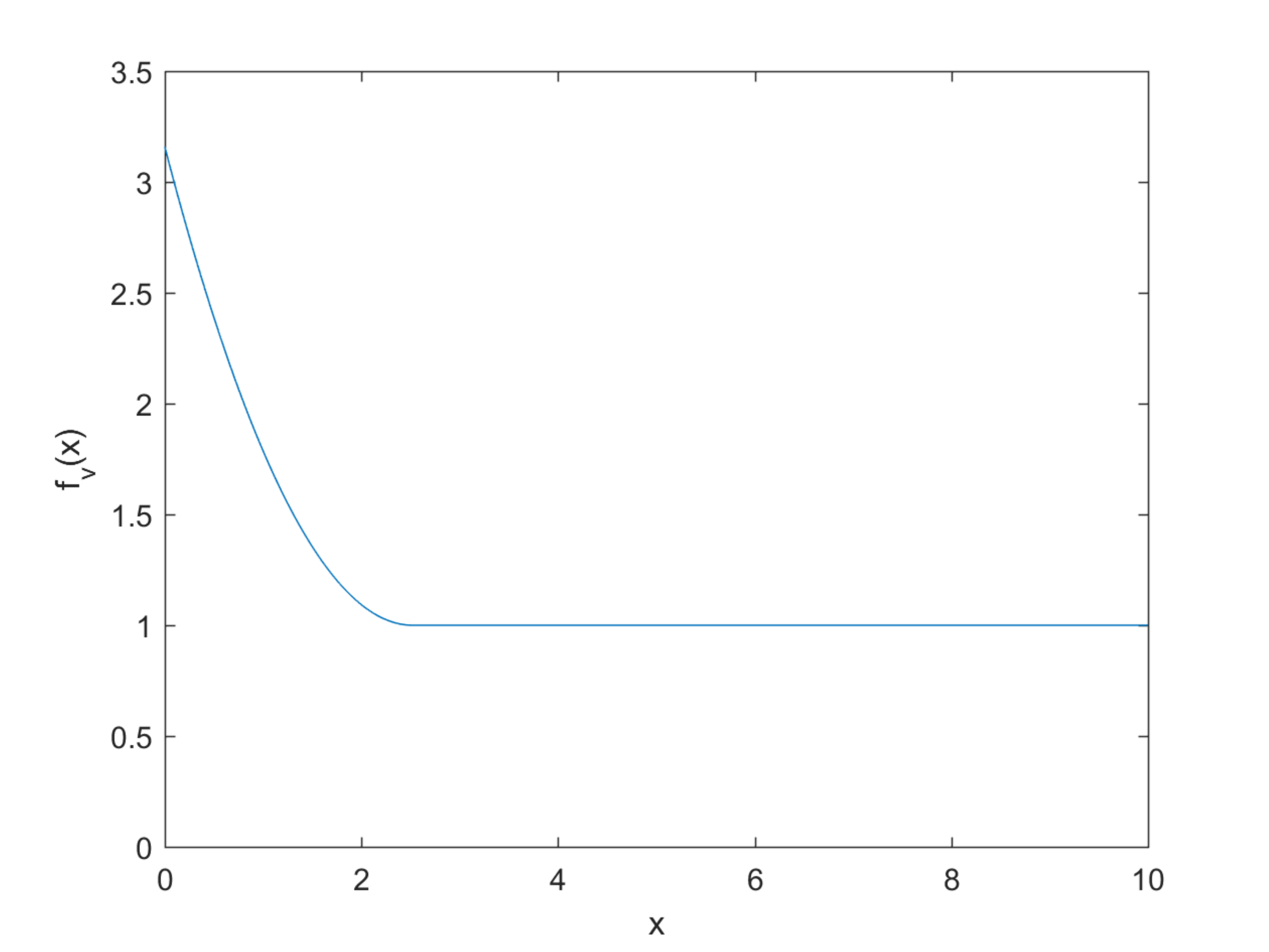}
	\caption{$f_{v}(x).$}
	\label{fig:fv}
\end{figure}

In this subsection, a new particle velocity $v_f(t,x)$ is proposed by applying some changes on $v_s(t)$ to better fit the identified delay. $v_s(t)$ is multiplied by an $x$ dependent function $f_v(x)$. At $x=0$, $f_{v}(0)=f_{v0}$ where $f_{v0}>1$. $f_{v}(x)$ decreases until it reaches $x_0$ where $f_{v}(x_0)=1$ and $\partial f_v(x_0)/\partial x$ is set to 0. $f_v(x)$ is chosen to be a second order polynomial. Note that the described polynomial's coefficients are unique for a given set of $x_0$ and $f_{v0}$. Figure \ref{fig:fv} depicts $f_v(x)$ for a given $x_0$ and $f_{v0}$.

\begin{figure}
\centering
	\includegraphics[width=0.65\linewidth]{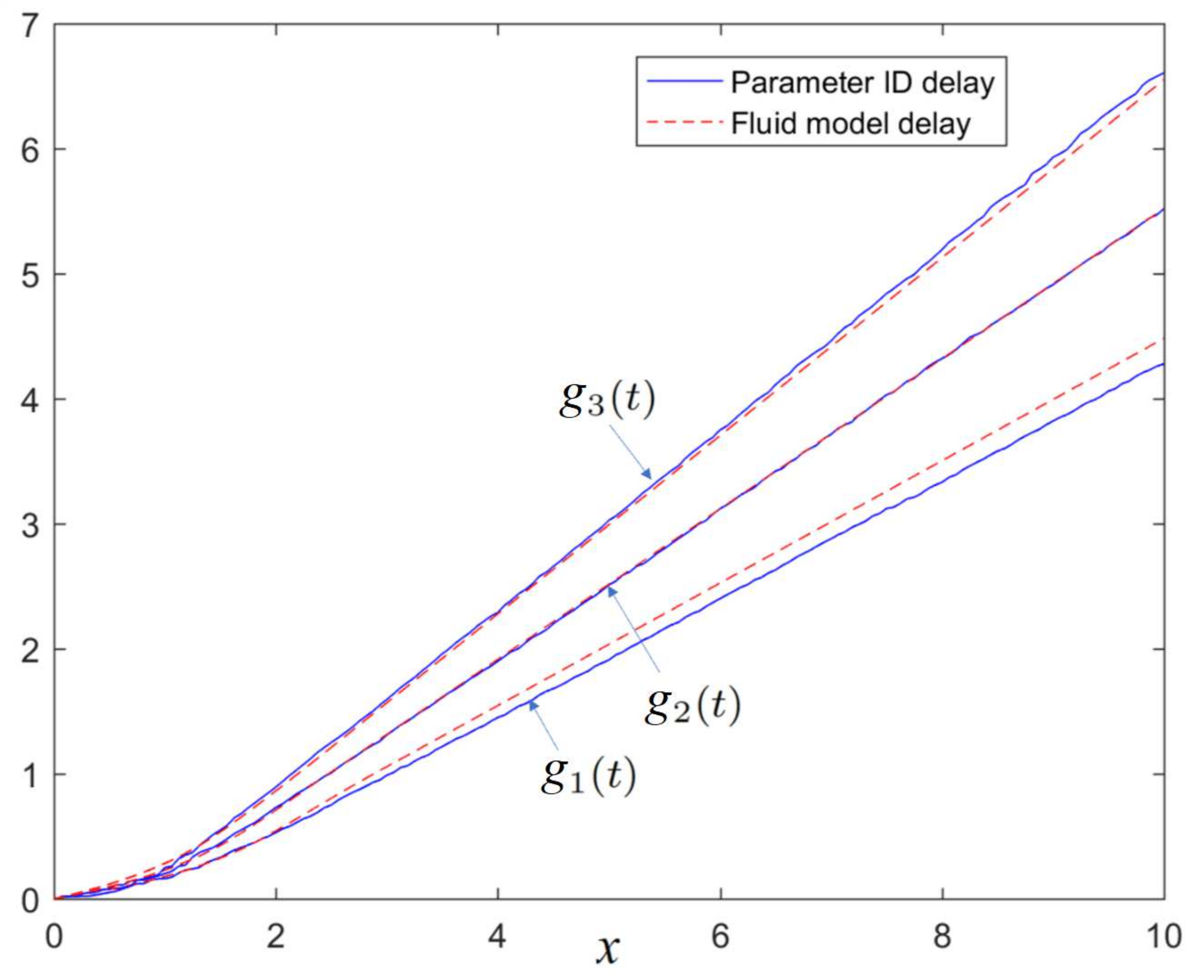}
	\caption{Fluid model delay due to $g_1(t)$, $g_2(t)$ and $g_3(t)$.}
	\label{fig:tauf}
\end{figure}

In addition, the velocity is further multiplied by a coefficient $\beta$ to better fit the slope of the delay with respect to $x$ as $x$ increases beyond $x_0$. Thus, the proposed fluid model velocity is 
\begin{equation}
\label{eq:v2}
v_f(t,x)= \beta \times f_v(x) \times \frac{Q_0-L_D\times \partial g(t)/ \partial t}{g(t)}.
\end{equation}
$\beta$, $x_0$ and $f_{v0}$ are identified by minimizing the delay error between the model and identified delay. Let
\begin{equation}
\label{eq:delaye}
e_\tau(\beta,x_0,f_{v0},g(t))=\sum_{n=1}^{160}|\tau_{ID}(x_i)-\tau_{in}(t_0,x_i)|
\end{equation}
where $\tau_{ID}$ is the identified delay for input $g(t)$. $\tau_{in}(t_0,x_i)$ is the delay at $x_i$ obtained using $v_f(t,x)$ for input $g(t)$. $\tau_{in}(t_0,x_i)$ is calculated as described in subsection \ref{sec:sem} but by substituting the new velocity expression in (\ref{eq:xp}). $\beta$, $x_0$ and $f_{v0}$ are obtained by minimizing the delay error for inputs $g_1(t)$, $g_2(t)$ and $g_3(t)$.
\begin{equation}
\label{eq:min3}
\min_{\beta,x_0,f_{v0}}\left[ e_\tau(\beta,x_0,f_{v0},g_1(t))+e_\tau(\beta,x_0,f_{v0},g_2(t))+e_\tau(\beta,x_0,f_{v0},g_3(t)) \right]
\end{equation}
Figure \ref{fig:tauf} shows the fluid model delay compared to the identified delay for values of $\beta$, $x_0$, and $f_{v0}$ that minimize the error in (\ref{eq:min3}). The minimum error is obtained for $\beta=0.8897$, $x_0=2.5219$, and $f_{v0}=3.158$. The delay is slightly overpredicted for $g_1(t)$ and slightly underpredicted for $g_3(t)$.

\subsubsection{Transport Equation}
So far the input delay is computed only for the moment when the input begins to change ($t_0$) by tracking the position of a particle. To simulate the model, the delayed input $g_d(t,x)$ should be computed for different values of $x$ and $t$. $g_d(t,x)$ could be computed by solving the transport equation in (\ref{eq:PDE1}).
\begin{equation}
\label{eq:PDE1}
\frac{\partial g_d(t,x)}{\partial t}=v(t)\frac{\partial g_d(t,x)}{ \partial x}
\end{equation}
The boundary condition at $x=0$ is $g_d(t,0)=g(t)$. It is assumed that the initial value of $g$ has remained  constant for a long time before $t=0$. Therefore, the initial condition is $g_d(0,x)=g(0)$. $v(t)$ could be replaced by $v_s$ or $v_f$ in (\ref{eq:PDE1}). To solve  (\ref{eq:PDE1}), the equation is discretized with respect to space and time. The solution is found at spatially equidistant points 0.003125 away from each other. The time step used is equal to 0.0025. Crank-Nicolson Finite Difference Method (FDM) is used to solve the equation. The delay at a certain point $x$ with velocity independent of $x$ could be also solved using the numerical method described in \cite{Matlab}.

Note that the fluid model is established based on two fundamental assumptions. First, the effects of the die gap change on the extrudate travel with the material down the extrudate with a velocity dependent on the die gap. The first assumption is due to the variation of the material velocity at the die exit with the die gap value given a constant volumetric flow rate at the inlet. Second, the effects due to the die gap change travel faster close to the die mainly due to non-uniform downward velocity profile at the die as shown in Figure \ref{fig:v}. These assumptions are common to most fluid extrusion setups.

\subsection{Parameter Identification Second Round}
\begin{figure}
\centering
	\includegraphics[width=0.75\linewidth]{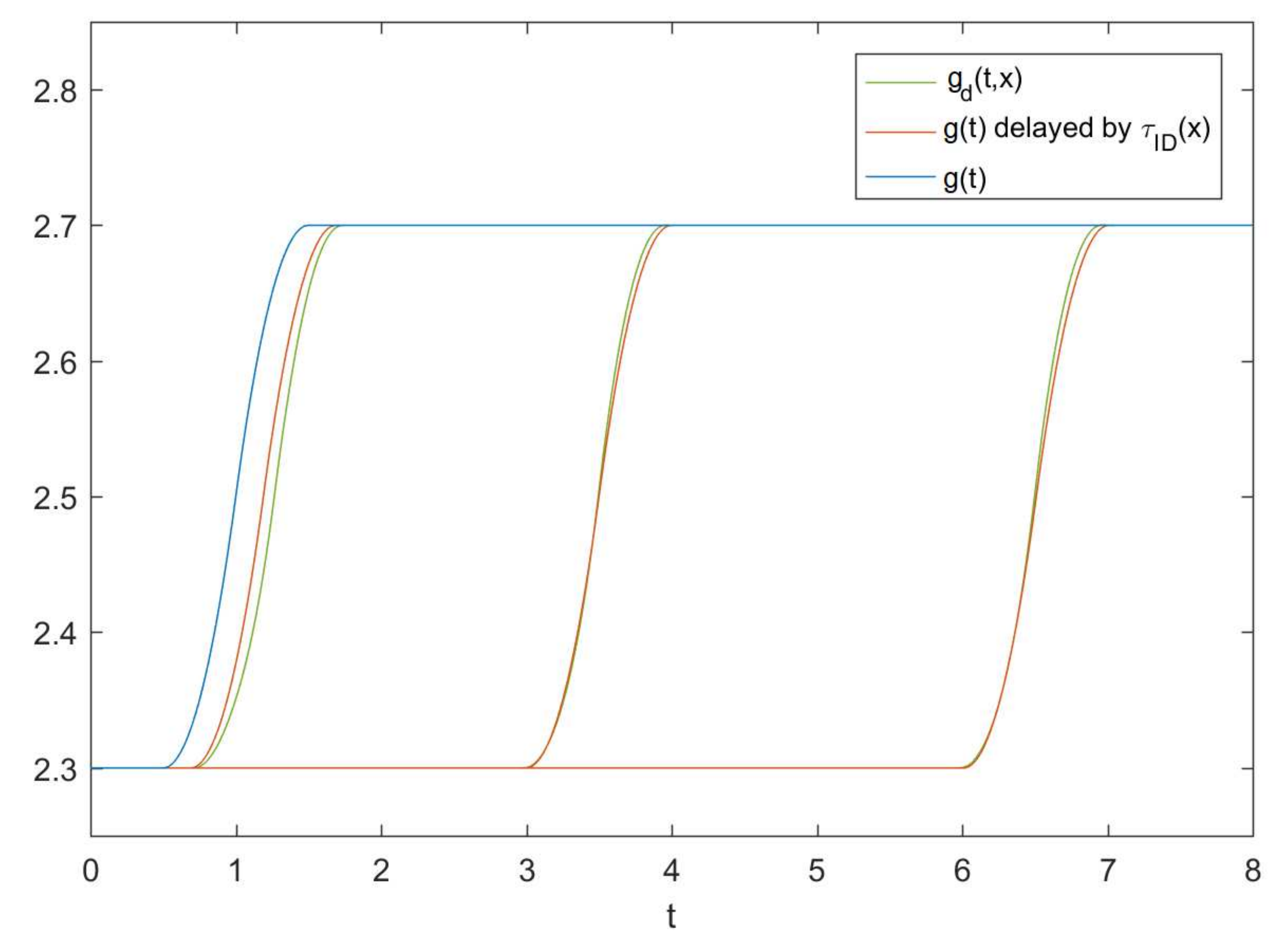}
	\caption{Constant delayed input and $g_d(t,x)$.}
	\label{fig:Delay}
\end{figure}
\begin{figure}
\centering
	\includegraphics[width=1\linewidth]{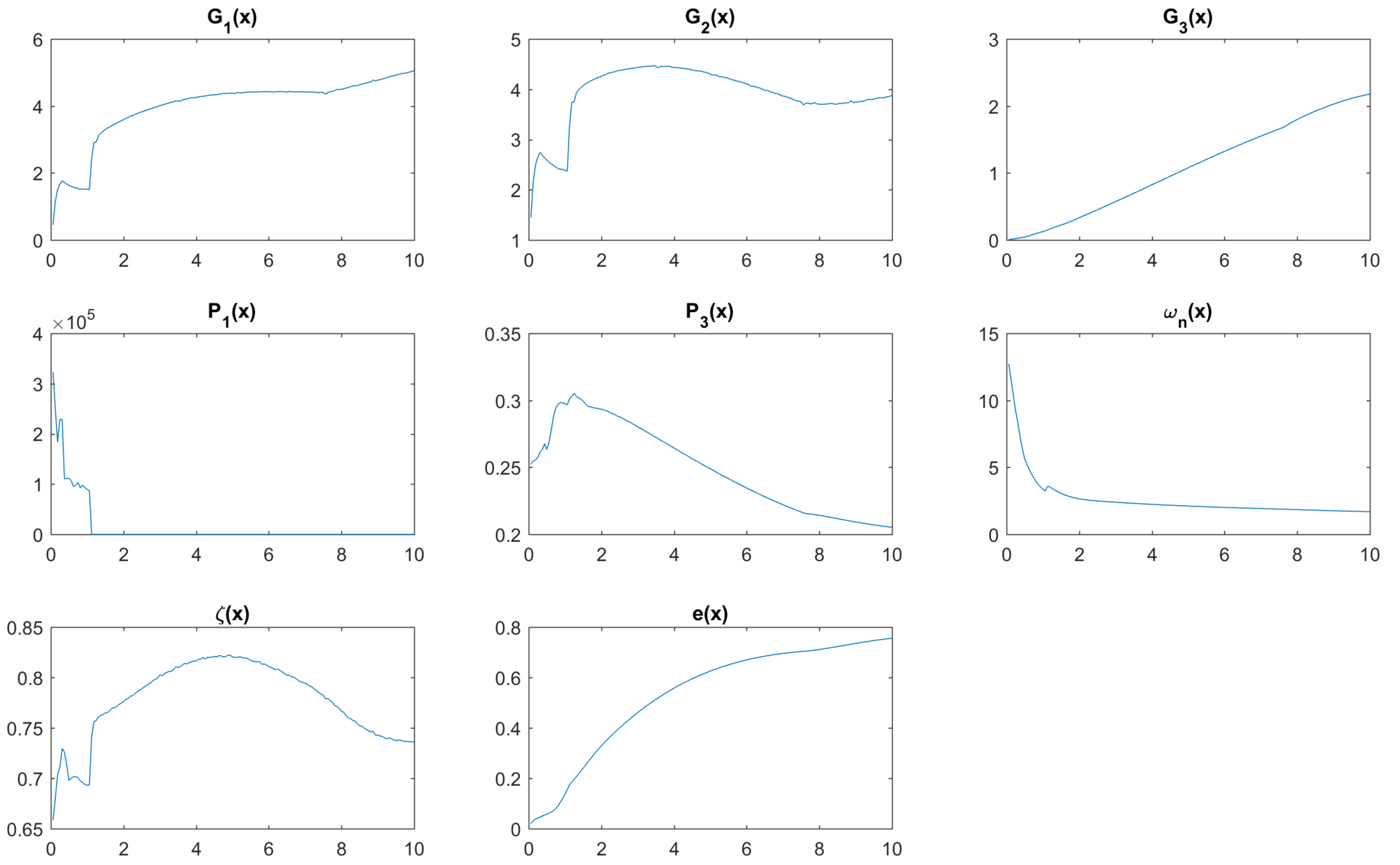}
	\caption{Transfer function parameters and $e$ versus $x$.}
	\label{fig:Para2}
\end{figure}

In subsection \ref{sec:ldf}, the input is delayed by a constant value $\tau_{ID}(x)$ as time changes and the transfer function parameters are identified accordingly. In subsection \ref{sec:TD}, the delay is time varying. Figure \ref{fig:Delay} shows $g_d(t,x)$ obtained by solving the fluid model compared to the input delayed by a constant value $\tau_{ID}(x)$. The delayed inputs are plotted for $x=1$, $x=5$, and $x=10$. The delayed input signals with the largest delays are for $x=10$ while the signals with the smallest delays are for $x=1$. $\beta$, $x_0$ and $f_{v0}$ are identified by minimizing the delay difference at the point where the input begins to change $(t_0)$. Therefore, the delay difference is very small when the delayed signal starts to change. The delay difference slightly increases at the moment when the signal stops changing altering the step signal's increase. The parameters of $H(s,x)$ are identified by using inputs delayed with a constant value. Therefore, the parameters of $H(s,x)$ are identified again after identifying $\beta$, $x_0$, and $f_{v0}$. Let
\begin{equation}
\label{eq:e2}
\begin{aligned}
   e'(G_1,G_3,P_1,P_3,\omega_n,\zeta,\beta,x_0,f_{v0},g(t)) &:=\int_0^T |y'_{ID}(t)-y_r(t)|dt\\
   &+\int_0^T |\frac{\partial y'_{ID}(t)}{\partial t}-\frac{\partial y_{r}(t)}{\partial t}| dt.
\end{aligned}
\end{equation}
The identified system output $y'_{ID}(t)$ is obtained by delaying $g(t)$ using the fluid model having parameters $\beta$, $x_0$, and $f_{v0}$ then inputting the delayed input to $f_s(x,\cdot)$ cascaded with $H(s,x)$ having parameters $G_1$ ,$G_3$ ,$P_1$, $P_3$, $\omega_n$, $\zeta$, and $\tau=0$. The new parameters of $H(s,x)$ are identified by solving the minimization problem below.
\begin{equation}
\label{eq:min4}
\min_{G_1,G_3,P_1,P_3,\omega_n,\zeta,}e'(G_1,G_3,P_1,P_3,\omega_n,\zeta,\beta,x_0,f_{v0},g(t)).
\end{equation}
The initial guess of the parameters are estimated as described in subsection \ref{sec:ID}. The transfer function parameters identified by solving (\ref{eq:min4}) and error $e'$ are shown in Figure \ref{fig:Para2}. 

\subsection{Model Structure}
\begin{figure}
\centering
	\includegraphics[width=\linewidth]{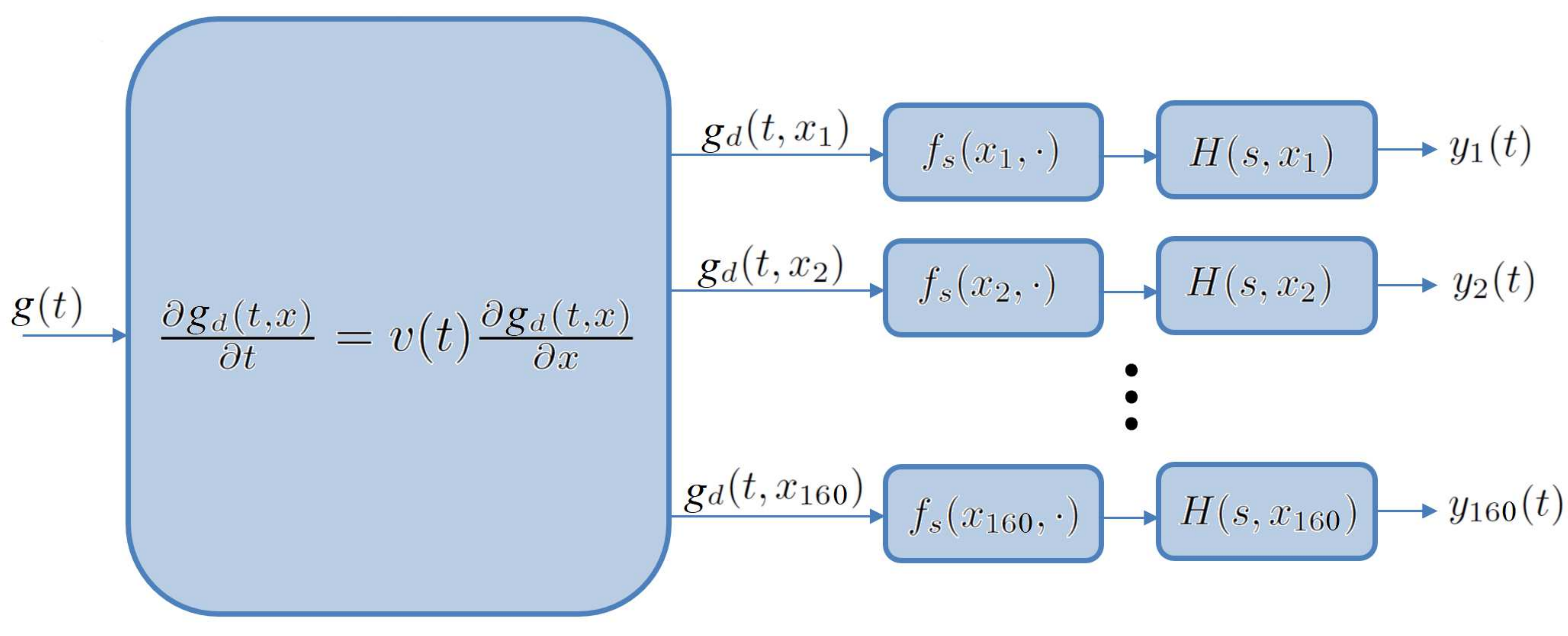}
	\caption{Overall block diagram.}
	\label{fig:struc}
\end{figure}

Figure \ref{fig:struc} shows the block diagram of the overall model. The transport equation outputs the delayed input for different values of $x$. The delayed input is the input for a Hammerstein model composed of $f_s(x_i,\cdot)$ and $H(s,x_i)$. $f_s(x_i,\cdot)$ determines the value that the output tends to if the input is constant. $H(s,x_i)$ determines the transient change of the output. The output thickness at $x_1$, $x_2$, $\cdots$ and $x_{160}$ are denoted by $y_1(t)$, $y_2(t)$, $\cdots$ and $y_{160}(t)$.

\section{Results and Discussion}

The proposed model is derived by minimizing the response error due to step inputs. The transfer function parameters are derived by minimizing the response error due to input $g_2(t)$. The velocity parameters are derived by minimizing the delay error for inputs $g_1(t)$, $g_2(t)$, and $g_3(t)$. Therefore, the operational region of the model is centered around $g=2.5$. In this section, the responses of the derived model and FEM simulator are compared for different inputs in the input operational region. 

\subsection{Step Inputs}
\begin{figure}
\centering
	\includegraphics[width=\linewidth]{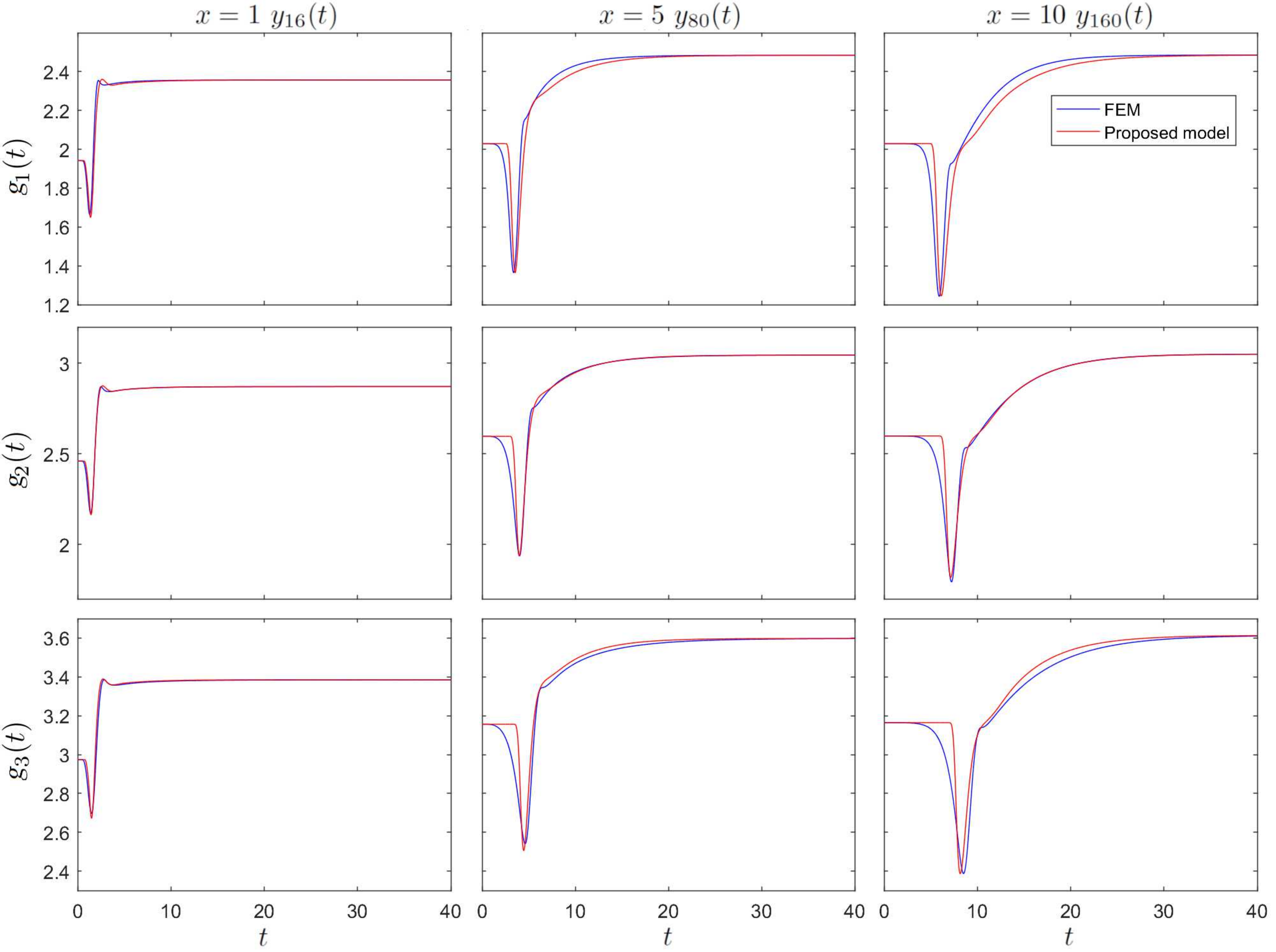}
	\caption{Thickness change with respect to time.}
	\label{fig:StepS}
\end{figure}

Figure \ref{fig:StepS} shows the step response of the two models at $x=1$, $x=5$, and $x=10$ due to $g_1(t)$, $g_2(t)$ and $g_3(t)$. The responses due to $g_2(t)$ match best since $g_2(t)$ is used to identify the transfer function parameters. The main features of the step response are captured by the proposed model. It can be observed that the delay is slightly overestimated for $g_1(t)$ and slightly underestimated for $g_3(t)$. This observation is  consistent with the plot in Figure \ref{fig:tauf}. 

\begin{figure}
\centering
	\includegraphics[width=0.9\linewidth]{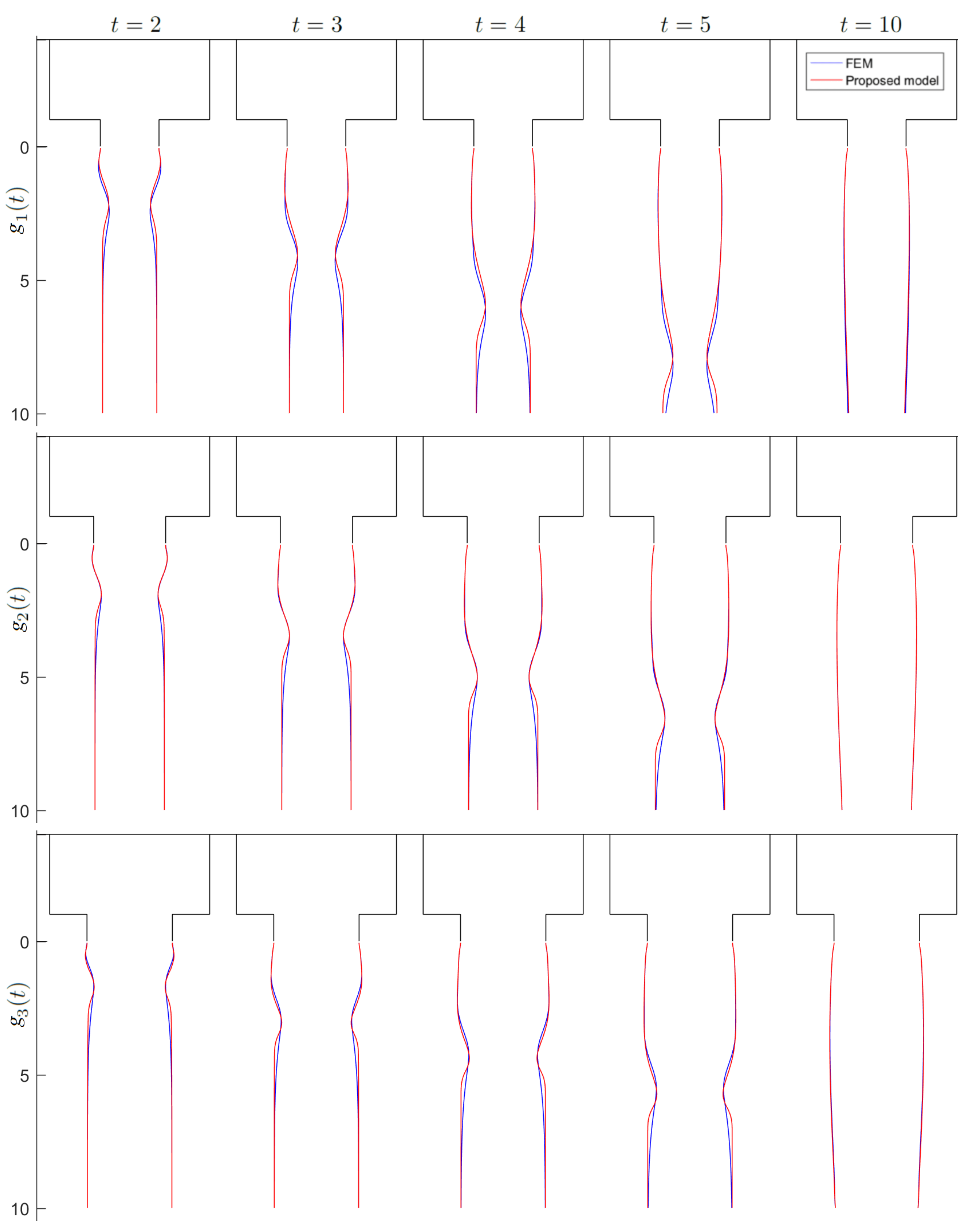}
	\caption{Extrudate shape change with respect to time.}
	\label{fig:StepE}
\end{figure}

Figure \ref{fig:StepE} shows the extrudate shape at $t=2$, $t=3$, $t=4$, $t=5$ and $t=10$ for $g_1(t)$, $g_2(t)$ and $g_3(t)$. The extrudates develop a neck like shape due to the die gap increase. The neck shape travels downstream fastest for $g_1(t)$ and slowest for $g_3(t)$ since the material velocity is higher for smaller die gaps. The extrudate shape is in good agreement for the two models.

\subsection{Die Gap Programming}

\begin{figure}
\centering
	\includegraphics[width=0.75\linewidth]{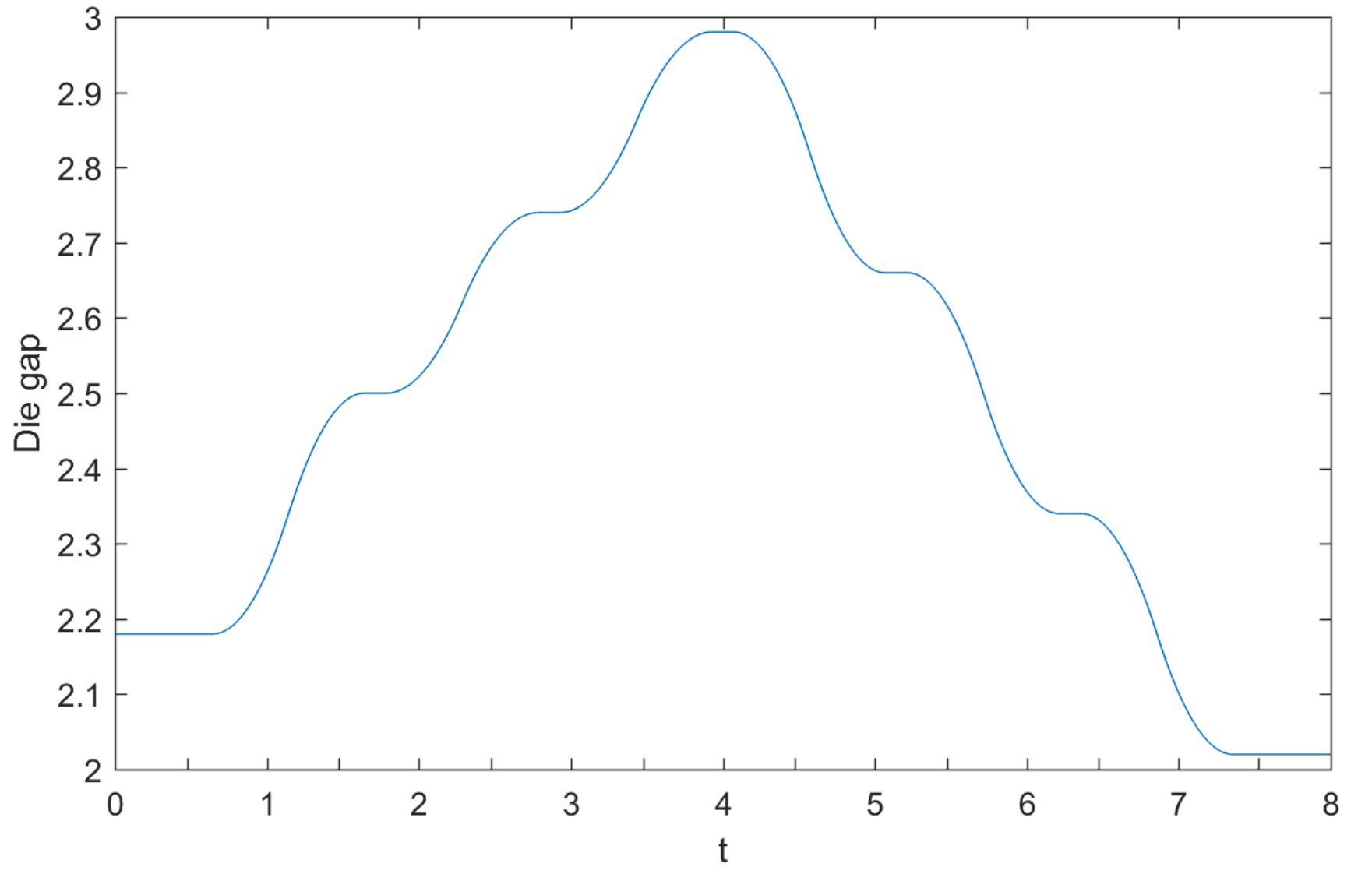}
	\caption{Die gap versus time.}
	\label{fig:ut}
\end{figure}

Die gap programming is used in the EBM industry to control the thickness of the extrudate and the final product. Extrusion time is divided into intervals in which the die gap is set to a constant setpoint. In this subsection, a hypothetical toy example is considered to show the effectiveness of the proposed model. In this example, the extrusion time is divided into 8 equal intervals. The desired extrudate is chosen to be thick in the middle where usually most of the stretching occurs during molding. On the other hand, it is chosen to be thin at the top and bottom where the molds clamp on each other. To obtain the desired shape at the end of the extrusion cycle $(t=8)$, die gap programming is done by trial and error. The proposed model is computationally much less expensive, therefore, it is used for die gap programming instead of the FEM model. Figure \ref{fig:ut} shows the die gap change as time progresses for the chosen setpoints. The die gap changes from one value to the other by steps that are similar to the steps used to identify the proposed model. After choosing the setpoints, the input is run on the FEM simulator.

\begin{figure}
\centering
	\includegraphics[width=\linewidth]{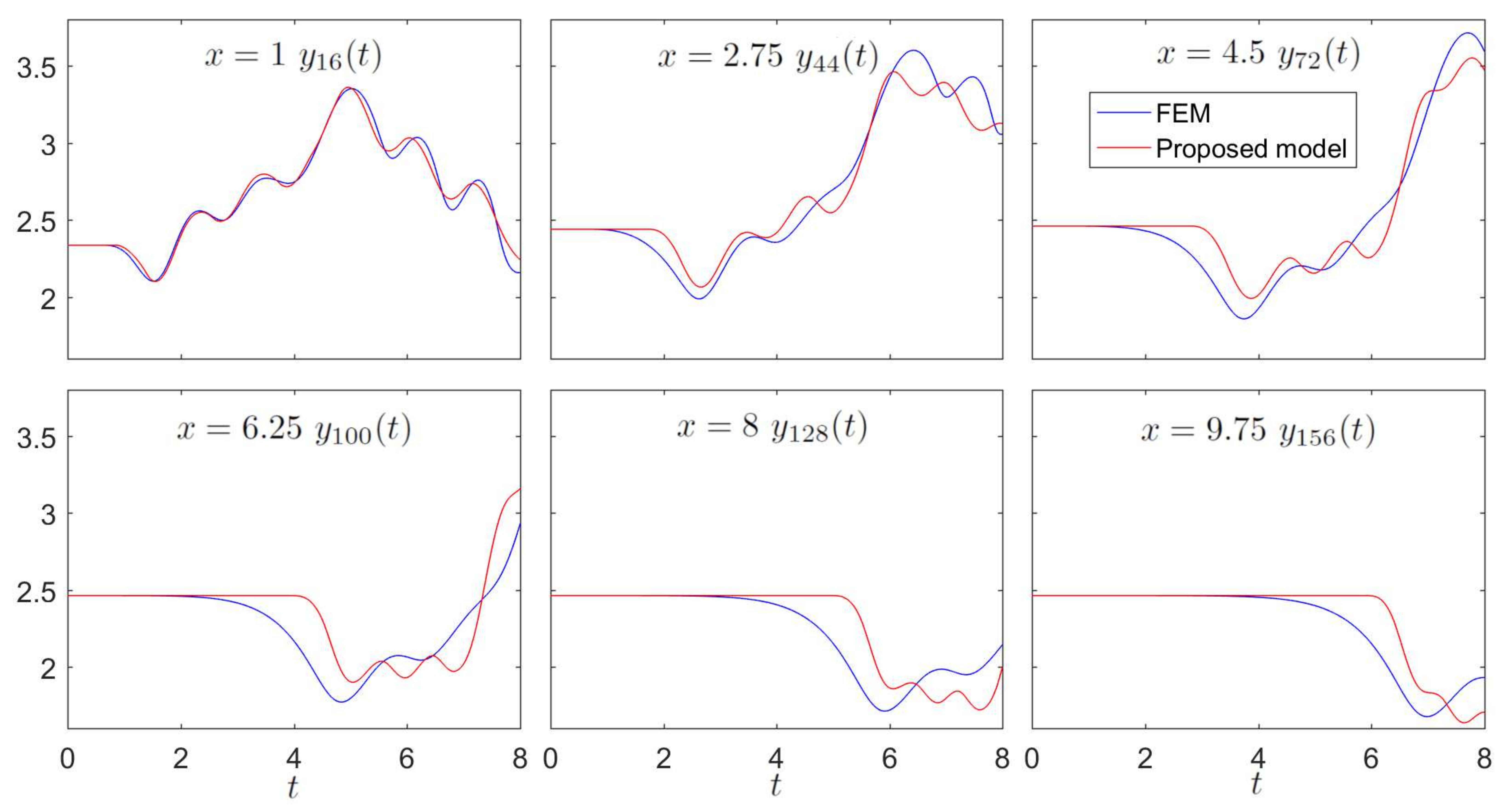}
	\caption{Thickness change with respect to time.}
	\label{fig:Prac1}
\end{figure}

Figure \ref{fig:Prac1} shows the thickness change with respect to time at $x=1$, $x=2.75$, $x=4.5$, $x=6.25$, $x=8$ and $x=9.75$. First the two models start from the same point. The FEM simulator begins to slowly decrease before the proposed model. The slower decrease for the FEM simulator when the output starts to change is similar to the slow change observed in step responses in Figures \ref{fig:Step4} and \ref{fig:Step5}. Once the proposed model catches up with the FEM after the initial decrease, small peaks and valleys do not match perfectly for larger values of $x$. Nevertheless, the overall low frequency signal trend overlay as observed for $x=2.75$, $x=4.5$ and $x=6.25$.

\begin{figure}
\centering
	\includegraphics[width=0.95\linewidth]{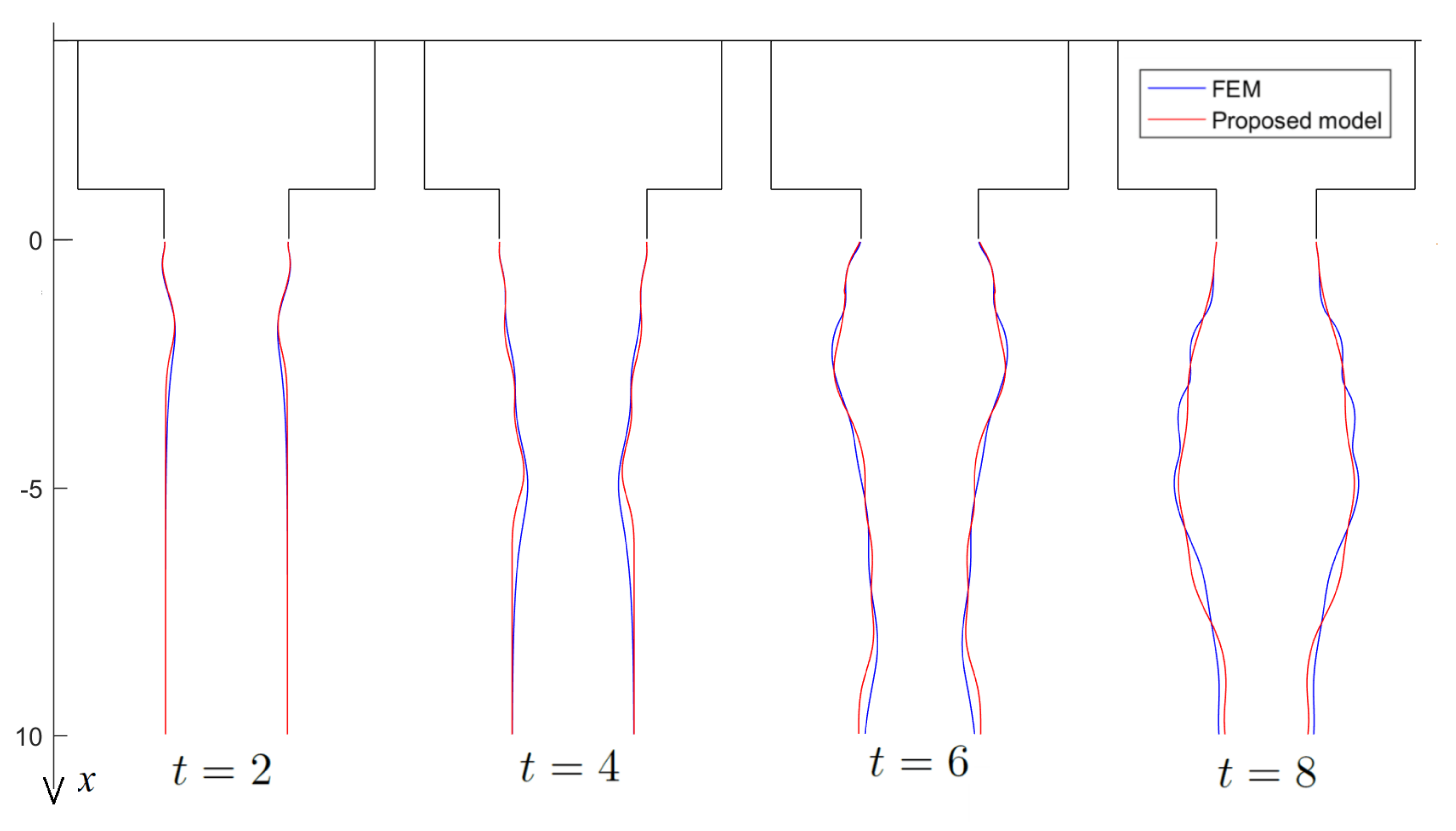}
	\caption{Extrudate shape change with respect to time.}
	\label{fig:Prac2}
\end{figure}

Figure \ref{fig:Prac2} shows the extrudate shapes obtained by the two models. The small peaks and bumps on the extrudate do not match perfectly at $t=6$ and $t=8$. However, the intended shape  was obtained using the proposed model (thick in the middle and thin at the top and bottom). In addition, the time needed for die gap programming is reduced significantly by using the proposed model instead of the FEM simulator during trial and error.

\subsection{Sinusoidal Inputs}
So far, the input signals used are composed of single or several step changes. To check the validity of the model for different types of inputs, sinusoidal inputs with different frequencies are considered. The signals are centered at $g=2.5$ and have an amplitude of $0.5$ so that the signal varies from 2 to 3. The frequencies used are $f=1$, $f=2$, $f=3$, and $f=4$. 

\begin{figure}
\centering
	\includegraphics[width=0.95\linewidth]{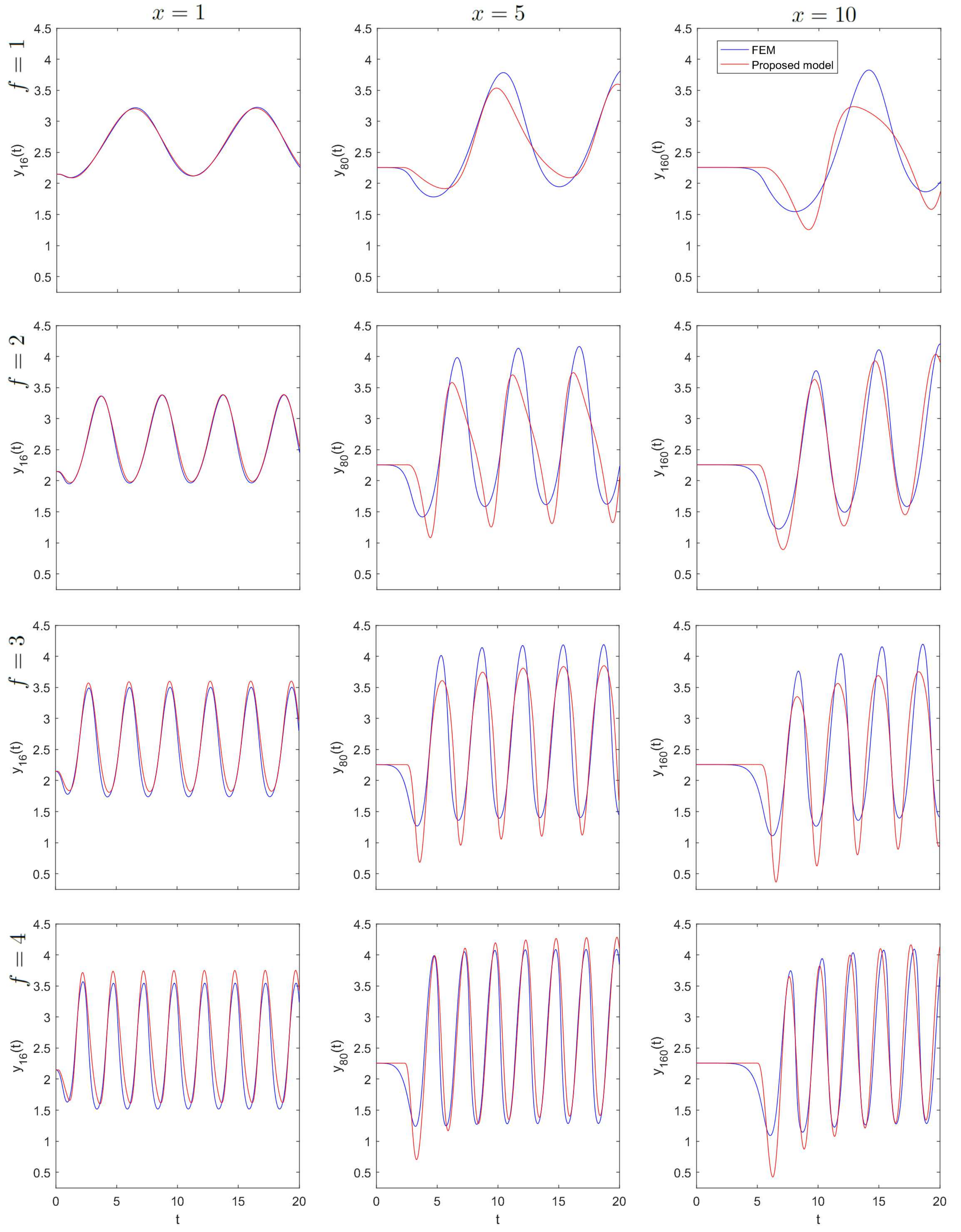}
	\caption{Thickness change with respect to time.}
	\label{fig:Signal}
\end{figure}

\begin{figure}
\centering
	\includegraphics[width=0.7\linewidth]{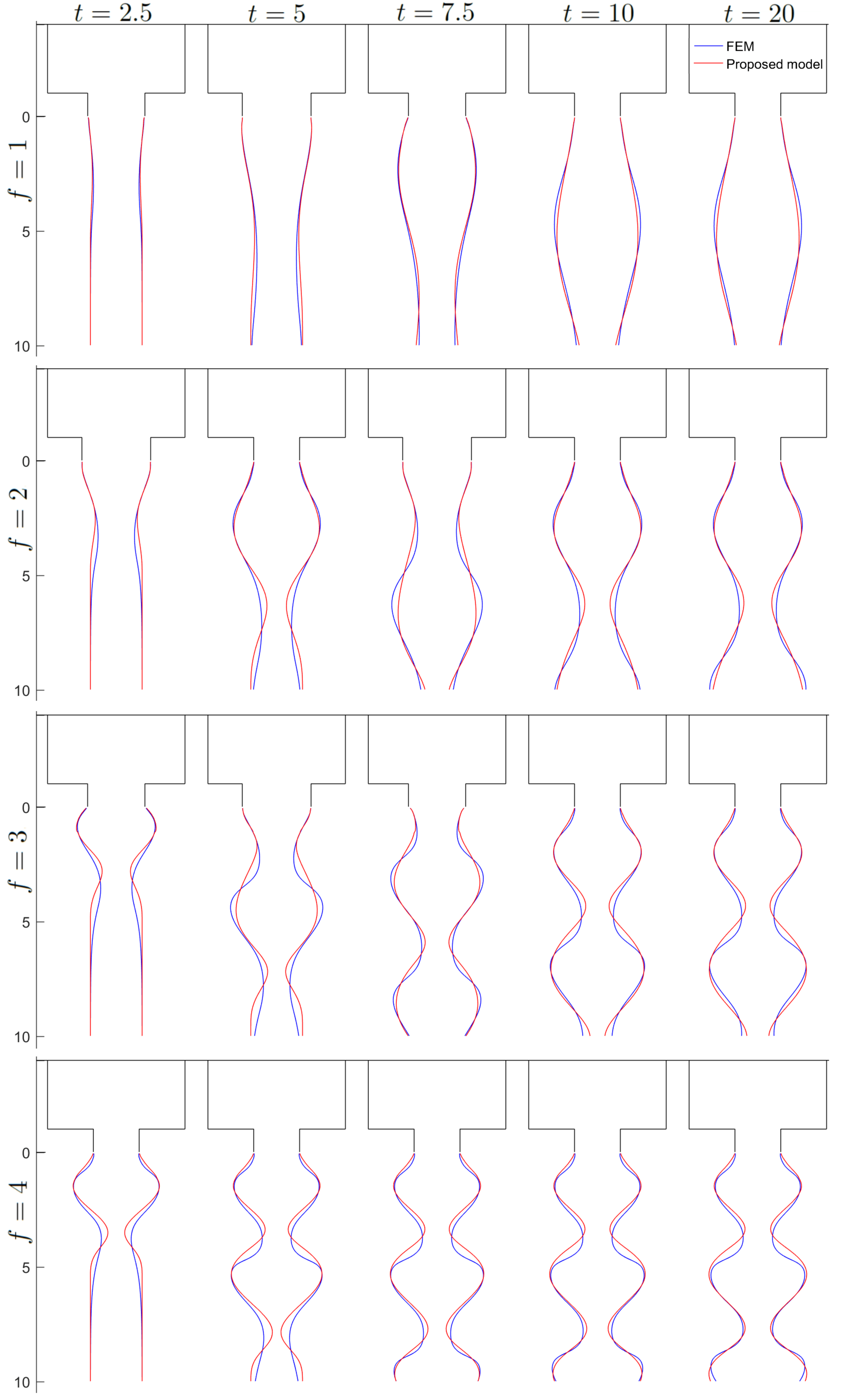}
	\caption{Extrudate shape for sinusoidal inputs.}
	\label{fig:Extrudate}
\end{figure}

Figure \ref{fig:Signal} shows the extrudate thickness varying with time at $x=1$, $x=5$, and $x=10$. The results of the FEM simulator and the proposed model are in close agreement for $x=1$. The output of the FEM simulator slowly starts to change before the proposed model as observed for step inputs. The sinusoidal output signals are generally in phase which shows that the delay is well predicted by the proposed model. Some errors are present at the peaks and valleys of the signals specifically for $f=1$.
Figure \ref{fig:Extrudate} shows the shape of the extrudate at $t=2.5$, $t=5$, $t=7.5$, $t=10$ and $t=20$. The shape of the extrudates are similar for the two models. The proposed model simulates the main features and shape of the extrudate simulated by the FEM simulator.

\subsection{Discussion}

The reference FEM simulator used in this work is computationally expensive. The position of the mesh, velocity and pressure is calculated at every time step in a two dimensional domain. Every time step involves a Lagrangian sub-step and an Arbitrary Eulerian Lagrangian sub-step. Furthermore, at each sub-step the variables are computed using an iterative method. On the other hand, the proposed method includes the transport equation, 160 transfer functions and 160 algebraic functions. The transport equation used is a one dimensional PDE that has single variable $g_d(t,x)$ evaluated for different values of $t$ and $x$. As mentioned before, the number of transfer functions and algebraic functions used can be reduced. 160 transfer functions were identified to observe the parameter variation of the transfer functions. In this section, the proposed model was able to replicate the results of a much more complicated reference model.

Replicating the output of a real extrusion setup by parameter identification could be more effective than a physics based approach. Understanding and modeling the physics behind complex behavior that is dependent on many factors may be difficult and produce unsatisfactory results. Although the reference system in this chapter is a physics based FEM simulator, real setups could be used in future work. The parameter identification could be done using sensor measured signals from a real setup.

\section{Conclusion}

A novel model is proposed to simulate fluid extrusion with varying die gap. The physics based FEM simulator is used as a reference for the proposed model to replicate its time dependent extrudate thickness profile. The proposed model has predefined transport PDE/ nonlinear ODE structure. First, a uniform material velocity is presented for the transport PDE. The presented transport PDE with uniform material velocity models solid extrusion. Then, a non-uniform material velocity that is dependent on the location on the extrudate is proposed. The non-uniform material velocity is proposed to better simulate fluid extrusion. The structure of the model is suitable for controller design and provides a computationally cheap alternative for physics based models. The parameters of the PDE and the non-linear ODE are identified by minimizing the response error between the proposed and reference model due to several step inputs. The proposed model is able to replicate the response of the reference model for various inputs within the operational range of the proposed model. 
\chapter{Conclusion}

Simulations, experiments, and mathematical models for fluids extruded out of a die are widely discussed in the literature. However, the effects of varying the die gap during extrusion on the extrudate are rarely addressed. Two models using different approaches are presented to simulate the effects of varying the die gap on the extrudate. The first model is a physics based FEM simulator. The model is used to provide an insight into the effects of varying the die gap on the material velocity, extrudate shape and pressure. The time dependent extrudate thickness results are used as a reference to derive the second model. The parameters of the second model are derived using a parameter identification scheme that minimizes the response error of the two models due to step inputs. The parameter identification based model is computationally cheap and suitable for feedback controller design in contrary to the FEM model. In this chapter, a summary of the thesis is presented in Section \ref{sec:sum} and recommendations for future work are suggested in Section \ref{sec:FW}.

\section{Summary}
\label{sec:sum}
The work in this thesis is part of a larger project that aims to improve the EBM process. The introduction starts with a background introducing the overall project. The background continues by describing the different steps involved in EBM plastic forming and the different challenges faced during manufacturing. Afterwards, the different governing equation and numerical methods used to model fluid extrusion in the literature are presented. The introduction ends with presenting the configuration considered and the different models proposed in the thesis. 

In Chapter 2, the Navier Stokes equations are used to model molten polymers. The Navier Stokes equations are spatially and temporally discretized using standard GFEM and trapezoidal rule respectively. The discretized equations are solved in two sub-steps using the Predictor Multi-Corrector method. In the first sub-step the equations are solved in the Lagrangian form. In the Lagrangian sub-step the mesh moves with the material and results in the shape of the free surface. The $y-$position of free surface mesh node is estimated using an interpolation scheme at the original $x$-position of the mesh node to prevent the mesh from flowing downstream with the material. The ALE velocity of the mesh is obtained according to the estimated free surface mesh position. In the second sub-step, the equations are solved in the ALE form using the obtained mesh velocity. Extrusion with varying die gap results in large free surface variations. Therefore, a novel free surface interpolation method with extra nodes is proposed to define the free surface with higher resolution. Finally, the mesh generator used and inner node movement is described in the last two sections.

Simulation results of the FEM simulator are presented in Chapter 3. Die swell ratio is used to show mesh convergence and to compare the obtained results with results from the literature. Then, the results of the proposed interpolation scheme presented in Chapter 2 are compared to the results of an interpolation scheme from the literature. The proposed scheme eliminates spatial free surface oscillations that develop when a higher $Re$ values are simulated using coarse meshes. In section \ref{sec:Results}, material velocity, pressure and extrudate shape are reported and discussed. The extrudate shape shows bulging and necking effects during sudden die gap decrease and increase, respectively. The different factors affecting bulging and necking are presented and analyzed. Finally, fluids with diminishingly small $Re$ values that mimic molten polymers are presented in the last section.

Chapter 4 starts with presenting the characteristics of the step response of the physics based model. The different blocks forming the parameter identification based model are chosen to replicate the step response main characteristics. The three main blocks of the proposed model are the nonlinear static function, the linear dynamic function and the delay. After presenting the identification procedure of the nonlinear static function and the linear dynamic function, two delay models are proposed. The first delay model is denoted by solid delay model. The solid delay model has a uniform material velocity along the extrudate. The second delay model is proposed after applying several improvements on the solid delay model. Afterwards, the transport equation used to calculate the delay at different values of $x$ and $t$ is presented. Finally, comparisons between the responses of the FEM simulator and the proposed model for different inputs are presented in the last section.

\section{Future Work}
\label{sec:FW}

Several recommendations for future work are discussed in this section. The recommendations include model improvements and applications for the the proposed models. 

The FEM simulator could be upgraded to model additional factors such as gravity and temperature. Moreover, non-Newtonian constitutive equations could be used to model the viscoelastic fluid behavior observed in molten polymers. The FEM simulator includes a Lagrangian sub-step that facilitates solving history dependent constitutive equations. The FEM simulator upgrade choice depends mainly on the simulated setup and is limited by the computational cost. 

Although factors such as fluid type and temperature are not taken into account while developing the parameter identification based model, the proposed model may replicate the response of a wide range of setups. Nevertheless, the model could be improved by taking into consideration additional factors such as gravity. The weight of the extrudate below a certain point could be taken into consideration while modeling the thickness at that point. Moreover, assumptions such as conservation of mass or volume can be used to better estimate material velocity. 

Finally, the parameter identification based model is suitable for controller design. EBM In-Cycle and Cycle-to-Cycle feedback controllers discussed in the introduction may be designed using the proposed model. The In-Cycle controller could limit machine drift while the Cycle-to-Cycle controller could be used to automate die gap programming.

\addcontentsline{toc}{chapter}{Bibliography}
\bibliographystyle{plain}

\bibliography{mybibfile}

\begin{appendices}
\chapter{Simulations Table}
\label{AppA}
\myappendices{Appendix \ref{AppA} \byname{AppA}}

\begin{table}
\centering
\resizebox{!}{.35\paperheight}{%
  \begin{tabular}{ | c | c | c | c | c | c | c | }
  
    \hline
          \# & Mesh & $Re$ & Extra Nodes & Die Gap Change & $\Delta T $ & $\Delta t \times 10^{-3}$ \\ \hline    \hline
            1& Mesh1 & 1 & 0 & -0.32 & 1 & 0.5 \\ \hline
            2& Mesh1 & 5 & 0 & -0.32 & 1 & 1 \\ \hline
            3& Mesh1 & 10 & 0 & -0.32 & 1 & 4 \\ \hline
            4& Mesh1 & 10 & 1 & -0.32 & 1 & 4\\ \hline
            5& Mesh1 & 10 & 3 & -0.32 & 1 & 4\\ \hline
            6& Mesh2 & 1 & 0 & -0.32 & 1 & 0.25\\ \hline
            7& Mesh2 & 5 & 0 & -0.32 & 1 & 1\\ \hline
            8& Mesh2 & 10 & 0 & -0.32 & 1 & 2\\ \hline
            9& Mesh3 & 1 & 0 & -0.32 & 1 & 0.1\\ \hline
            10& Mesh3 & 1 & 9 & -0.32 & 1 & 0.1\\ \hline
            11& Mesh3 & 1 & 9 & 0.32 & 1 & 0.05\\ \hline
            12& Mesh3 & 5 & 0 & -0.32 & 1 & 0.5\\ \hline
            13& Mesh3 & 5 & 9 & -0.32 & 1 & 0.5\\ \hline
            14& Mesh3 & 5 & 9 & 0.32 & 1 & 0.25\\ \hline
            15& Mesh3L & 5 & 9 & -0.32 & 1 & 0.25\\ \hline
            16& Mesh3L & 5 & 9 & 0.32 & 1 & 0.25\\ \hline
            17& Mesh3 & 5 & 9 & -0.32 & 0.5 & 0.1\\ \hline
            18& Mesh3 & 5 & 9 & 0.32 & 0.5 & 0.05\\ \hline
            19& Mesh3 & 5 & 9 & -0.64 & 1 & 0.1\\ \hline
            20& Mesh3 & 5 & 9 & 0.64 & 1 & 0.05\\ \hline
            21& Mesh3 & 10 & 0 & -0.32 & 1 & 0.5\\ \hline
            22& Mesh4 & 10 & 9 & -0.32 & 1 & 0.2\\ \hline
            23& Mesh4 & 10 & 9 & 0.32 & 1 & 0.125\\ \hline
            24& Mesh1 & 1  & 9 & -0.32 & 1 & 0.5\\ \hline
            25& Mesh1 & 10 & 9 & -0.32 & 1 & 4\\ \hline                        
            26& Mesh2 & 1 & 9 & -0.32 & 1 & 0.25\\ \hline
            27& Mesh2 & 10 & 9 & -0.32 & 1 & 2\\ \hline
            28& Mesh3 & 10 & 9 & -0.32 & 1 & 0.5\\ \hline
            29& Mesh2.5 & 1 & 9 & -0.32 & 1 & 0.05\\ \hline
            30& Mesh2.5 & 1 & 9 & 0.32 & 1 & 0.05\\ \hline
            31& Mesh2.5 & 0.1 & 9 & -0.32 & 1 & 0.01\\ \hline
            32& Mesh2.5 & 0.1 & 9 & 0.32 & 1 & 0.01\\ \hline
            33& Mesh2.5 & 0.05 & 9 & -0.32 & 1 & 0.005\\ \hline
            34& Mesh2.5 & 0.05 & 9 & 0.32 & 1 & 0.005\\ \hline
                                                 
  \end{tabular}
}

\caption{Simulations' Parameters.}
\label{table:3}
\end{table} 
\chapter{$Re = 1$}
\label{sec:Re1}
\begin{figure}
	\includegraphics[width=.9\linewidth]{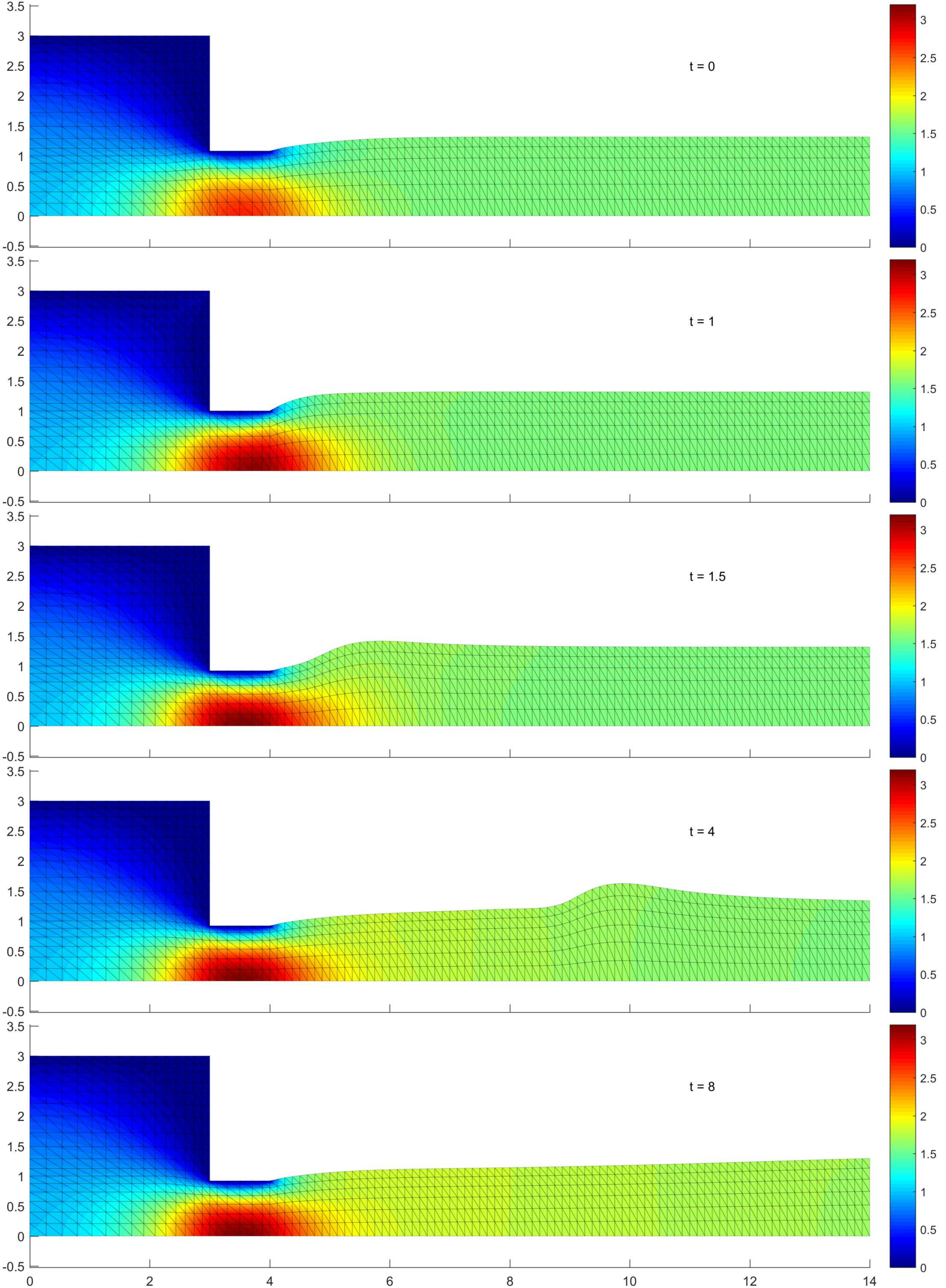}
	\centering
	\caption{$u_x$ of closing die for $Re=1$.}
	\label{fig:cRey1u}
\end{figure}

\begin{figure}
	\includegraphics[width=.9\linewidth]{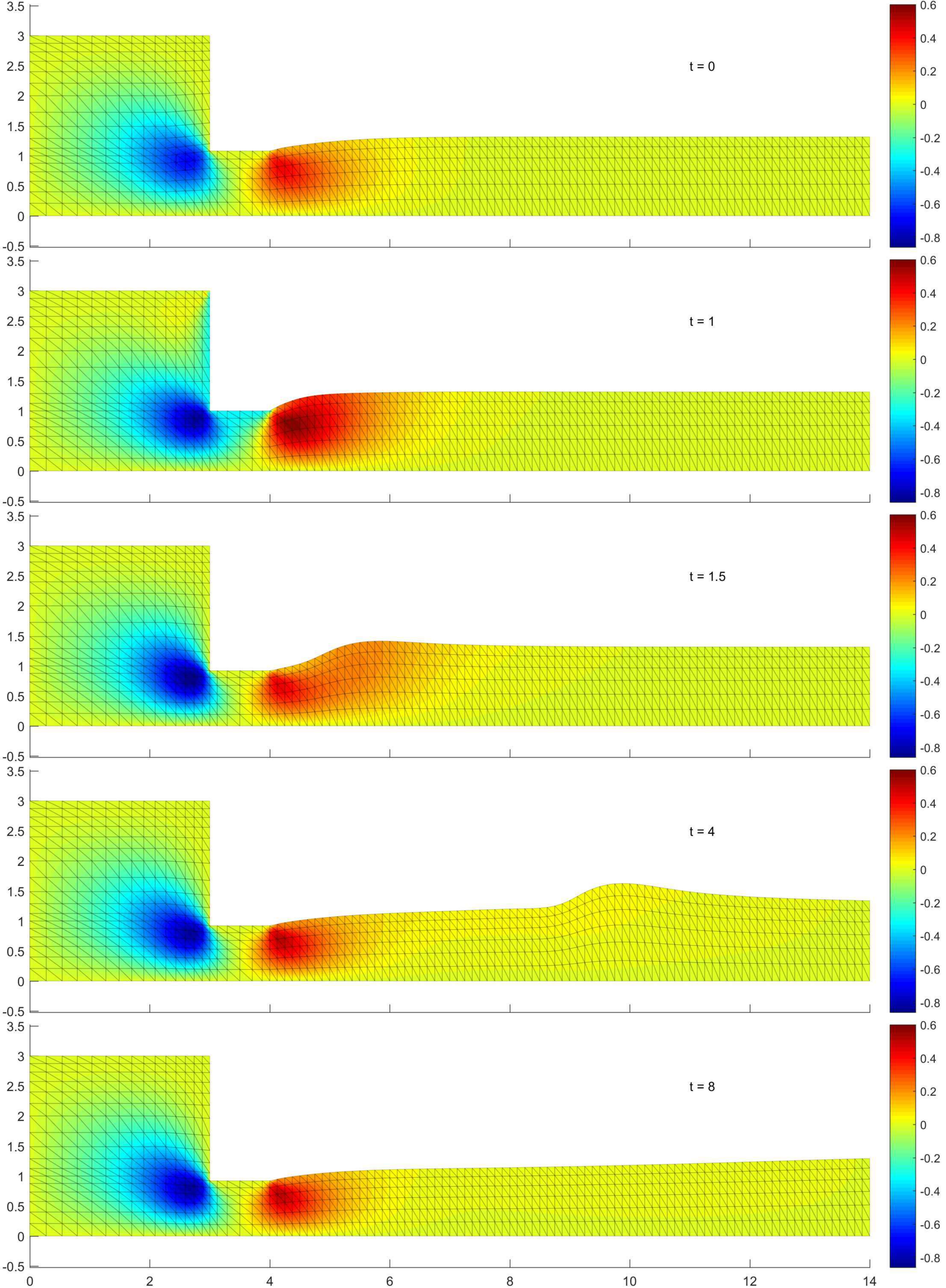}
	\centering
	\caption{$u_y$ of closing die for $Re=1$.}
	\label{fig:cRey1v}
\end{figure}

\begin{figure}
	\includegraphics[width=.9\linewidth]{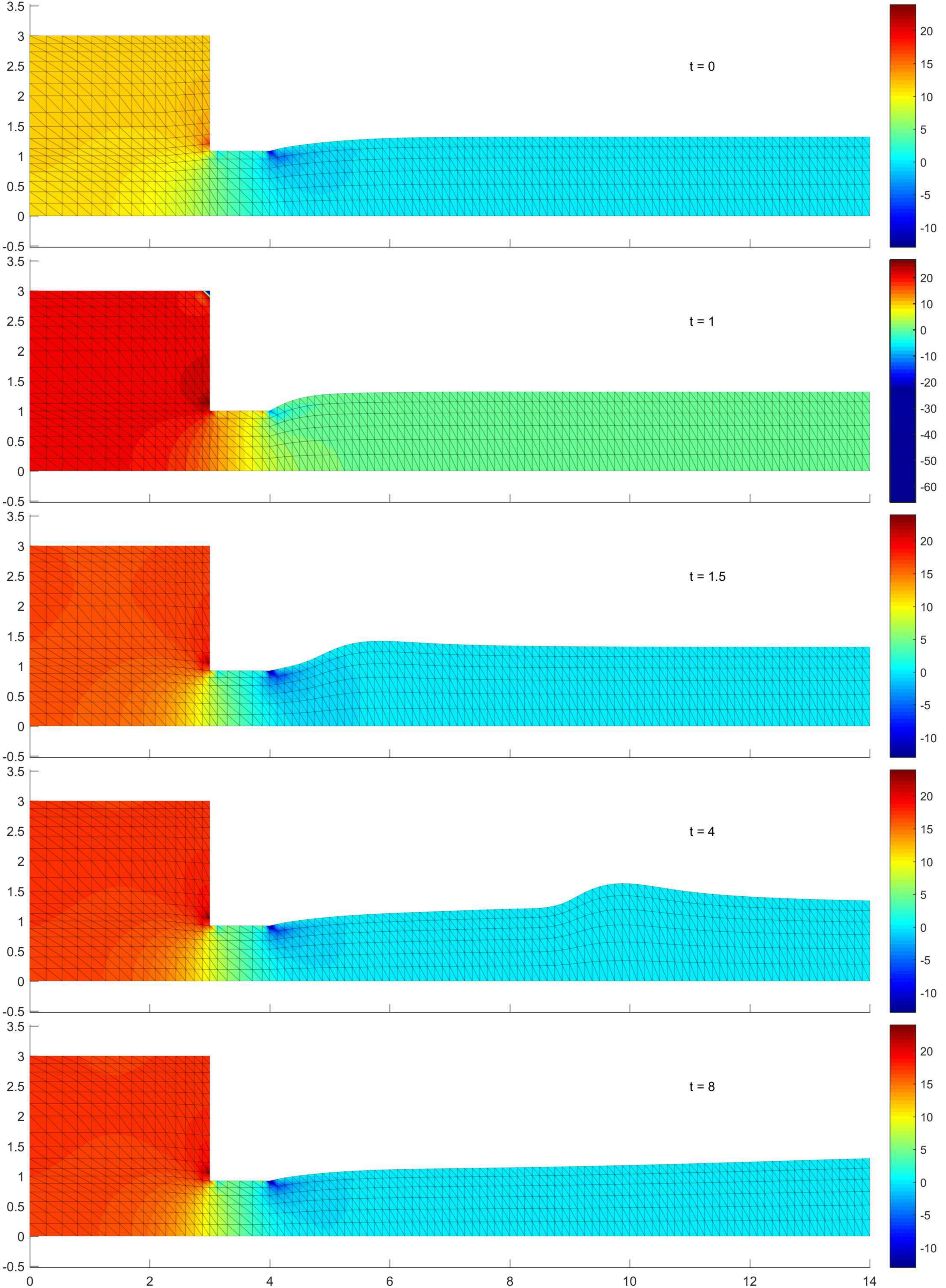}
	\centering
	\caption{$p$ of closing die for $Re=1$.}
	\label{fig:cRey1p}
\end{figure}

\begin{figure}
	\includegraphics[width=.9\linewidth]{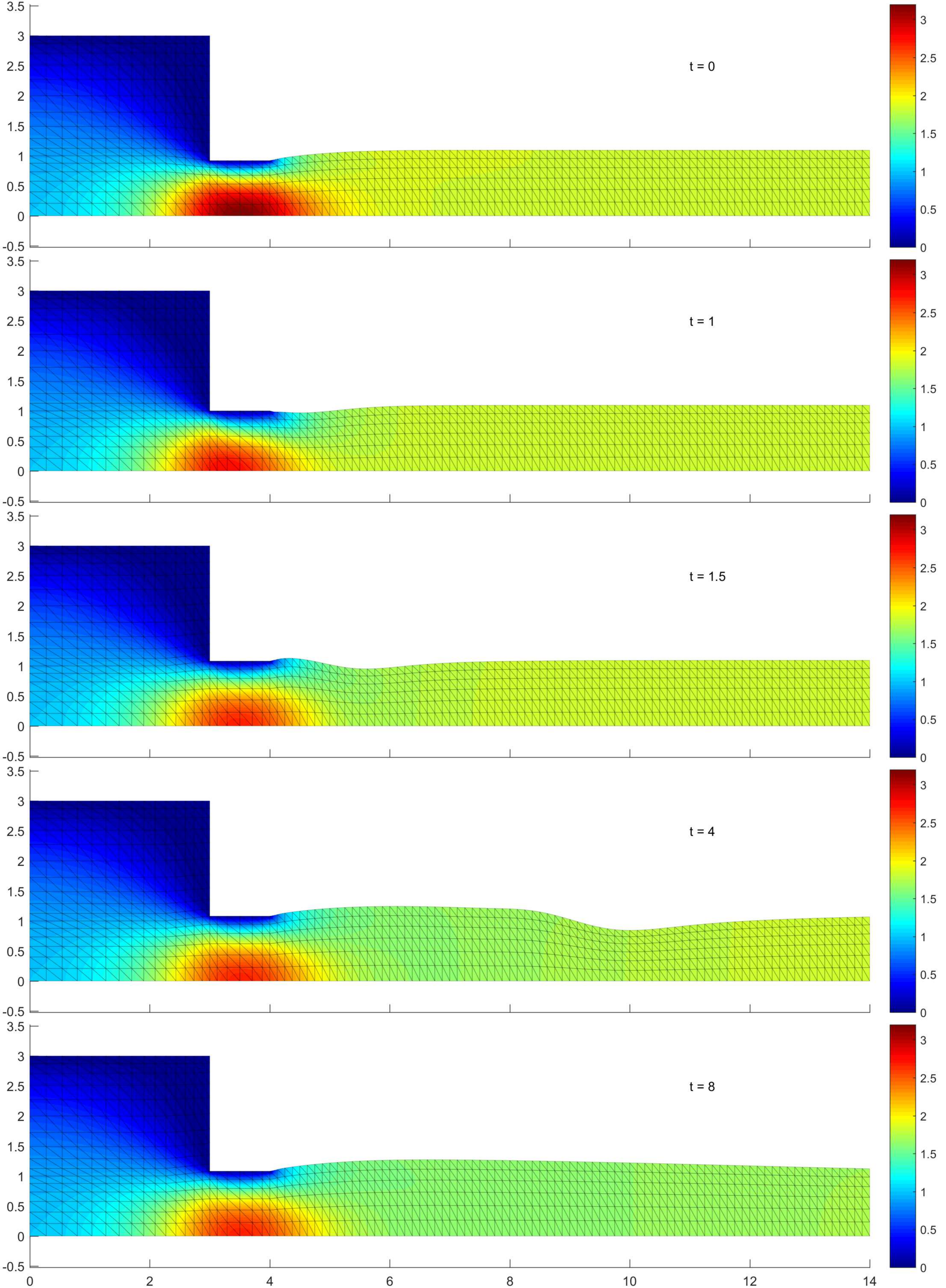}
	\centering
	\caption{$u_x$ of opening die for $Re=1$.}
	\label{fig:oRey1u}
\end{figure}

\begin{figure}
	\includegraphics[width=.9\linewidth]{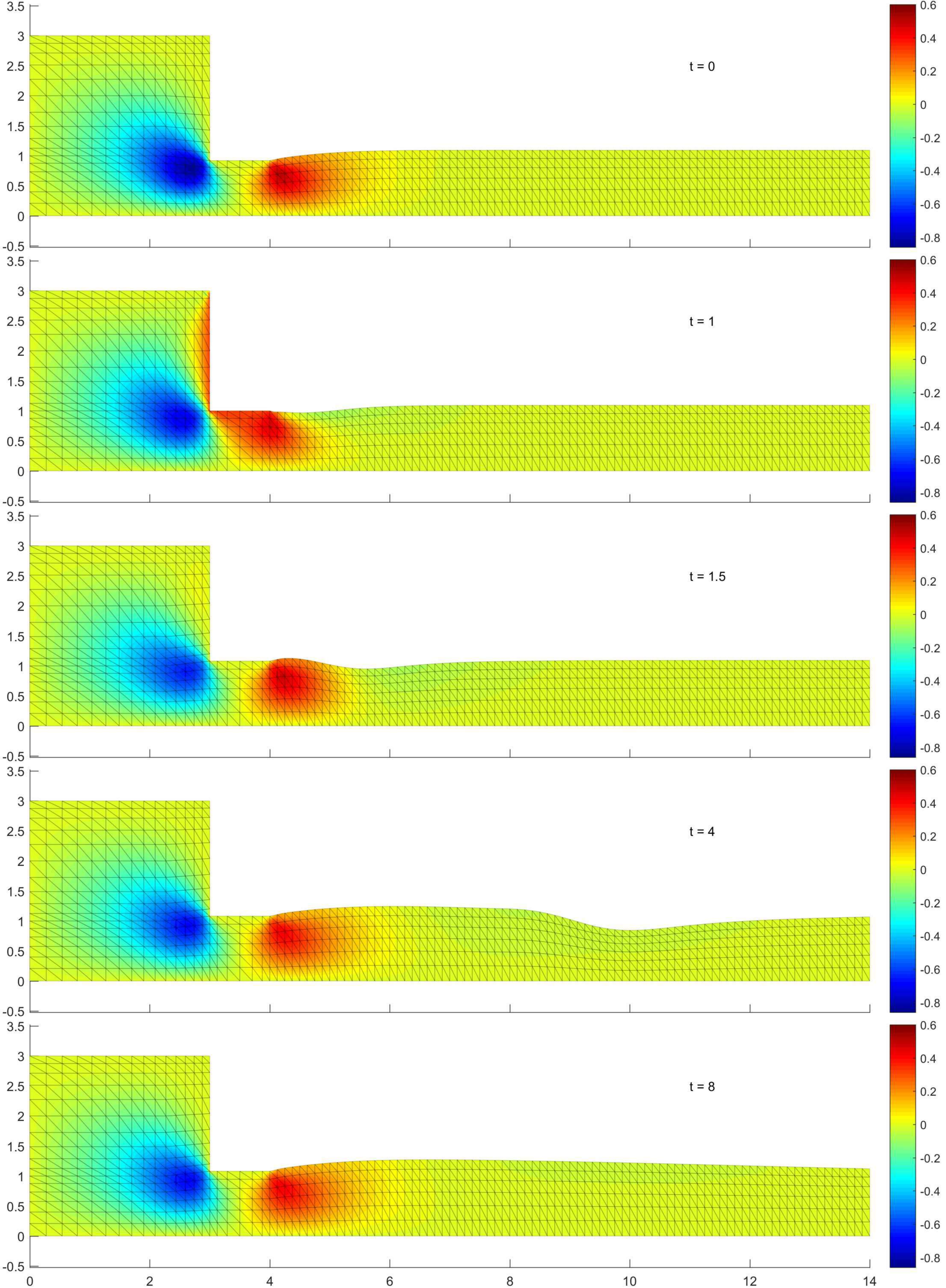}
	\centering
	\caption{$u_y$ of opening die for $Re=1$.}
	\label{fig:oRey1v}
\end{figure}

\begin{figure}
	\includegraphics[width=.9\linewidth]{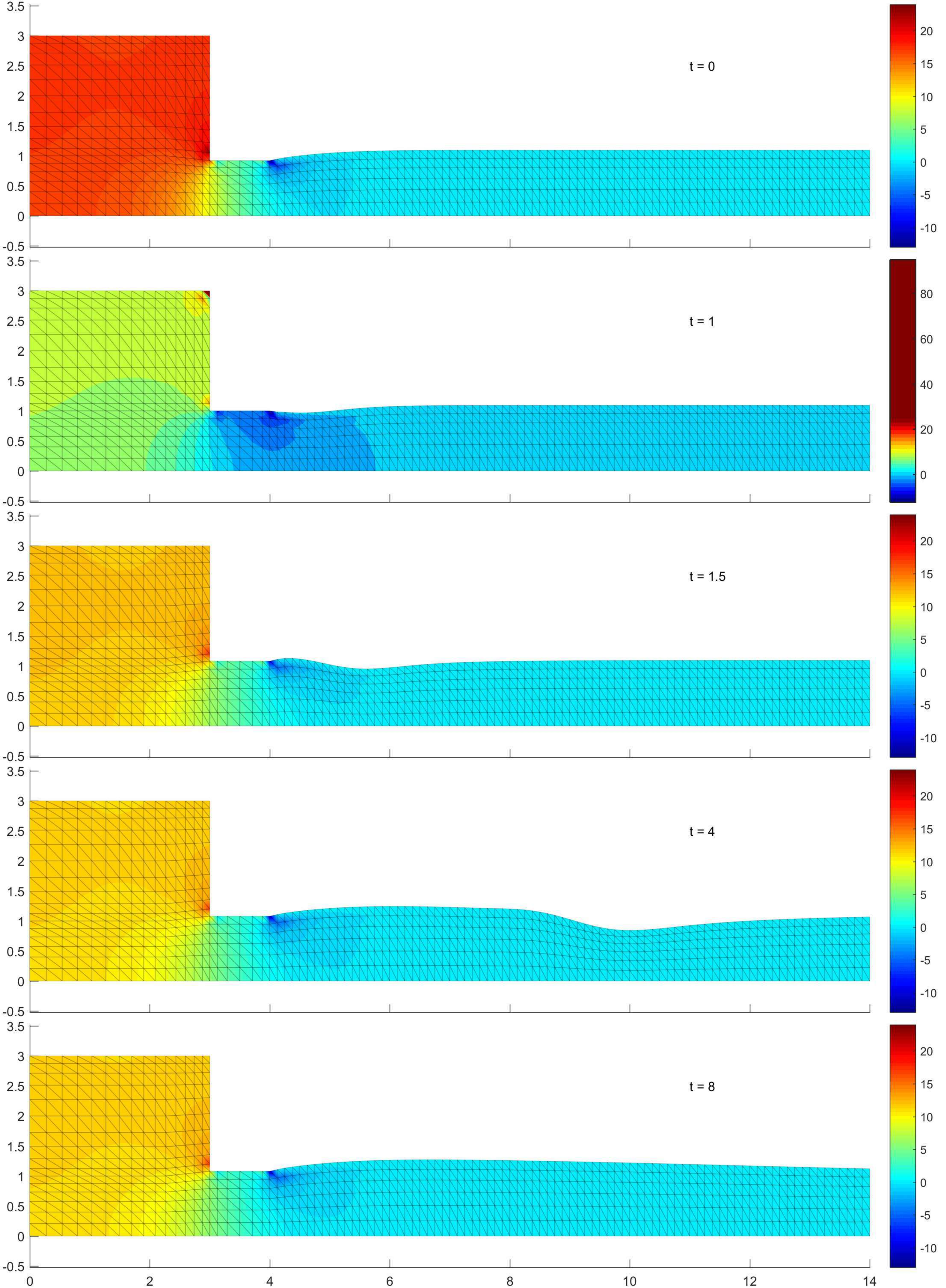}
	\centering
	\caption{$p$ of opening die for $Re=1$.}
	\label{fig:oRey1p}
\end{figure}
\chapter{$Re = 10$}
\label{sec:Re10}

\begin{figure}
	\includegraphics[width=.9\linewidth]{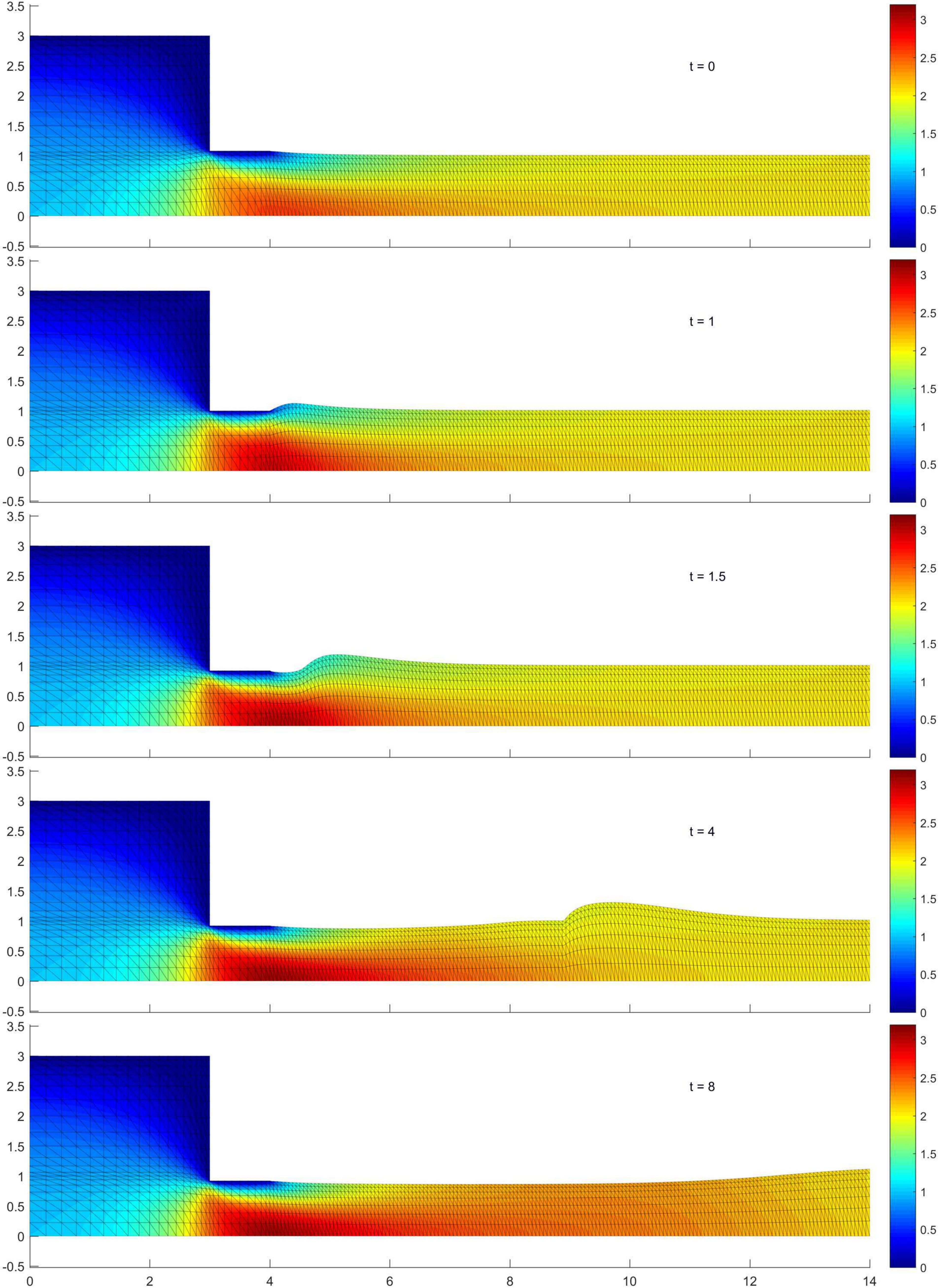}
	\centering
	\caption{$u_x$ of closing die for $Re=10$.}
	\label{fig:cRey10u}
\end{figure}

\begin{figure}
	\includegraphics[width=.9\linewidth]{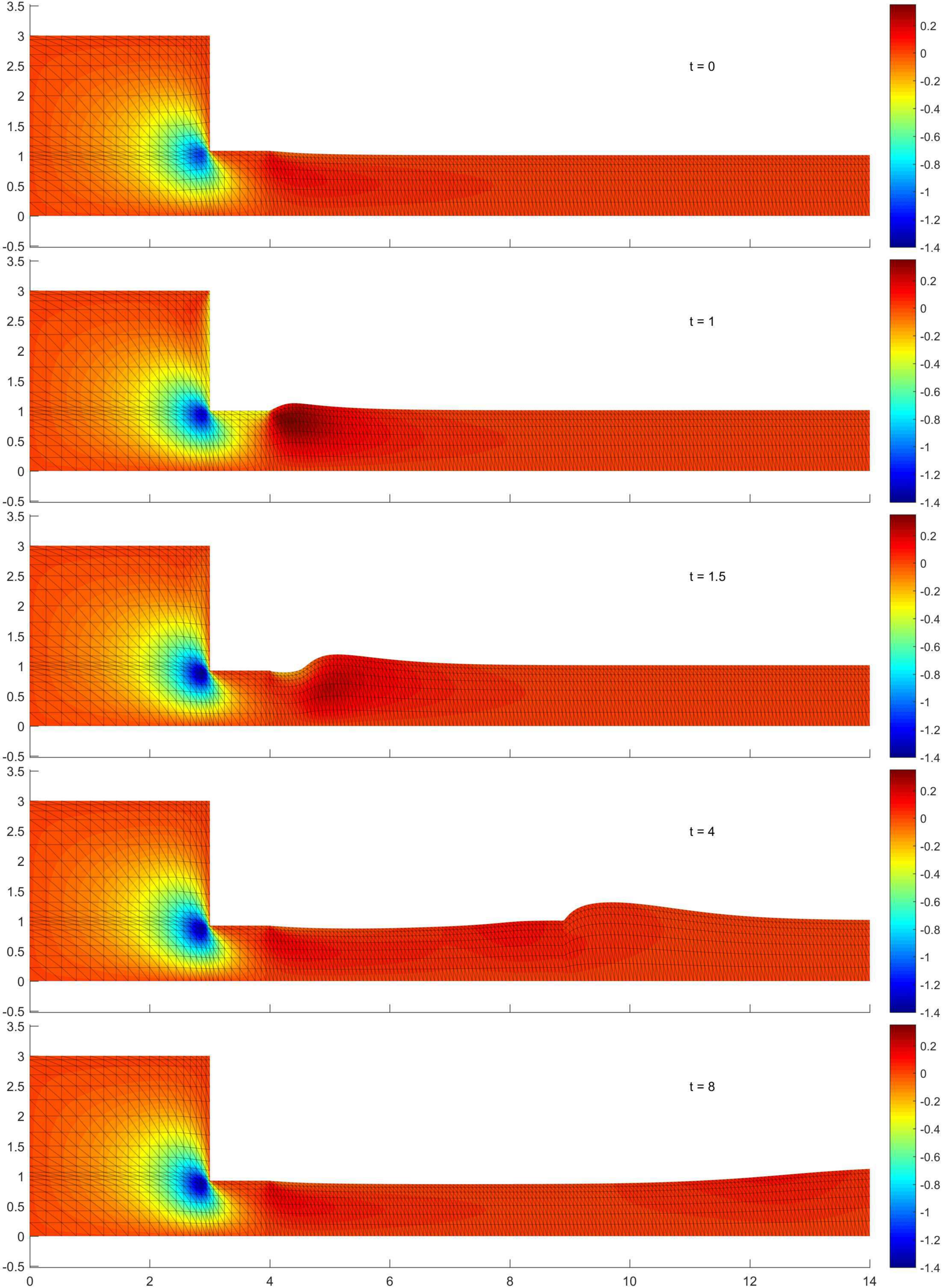}
	\centering
	\caption{$u_y$ of closing die for $Re=10$.}
	\label{fig:cRey10v}
\end{figure}

\begin{figure}
	\includegraphics[width=.9\linewidth]{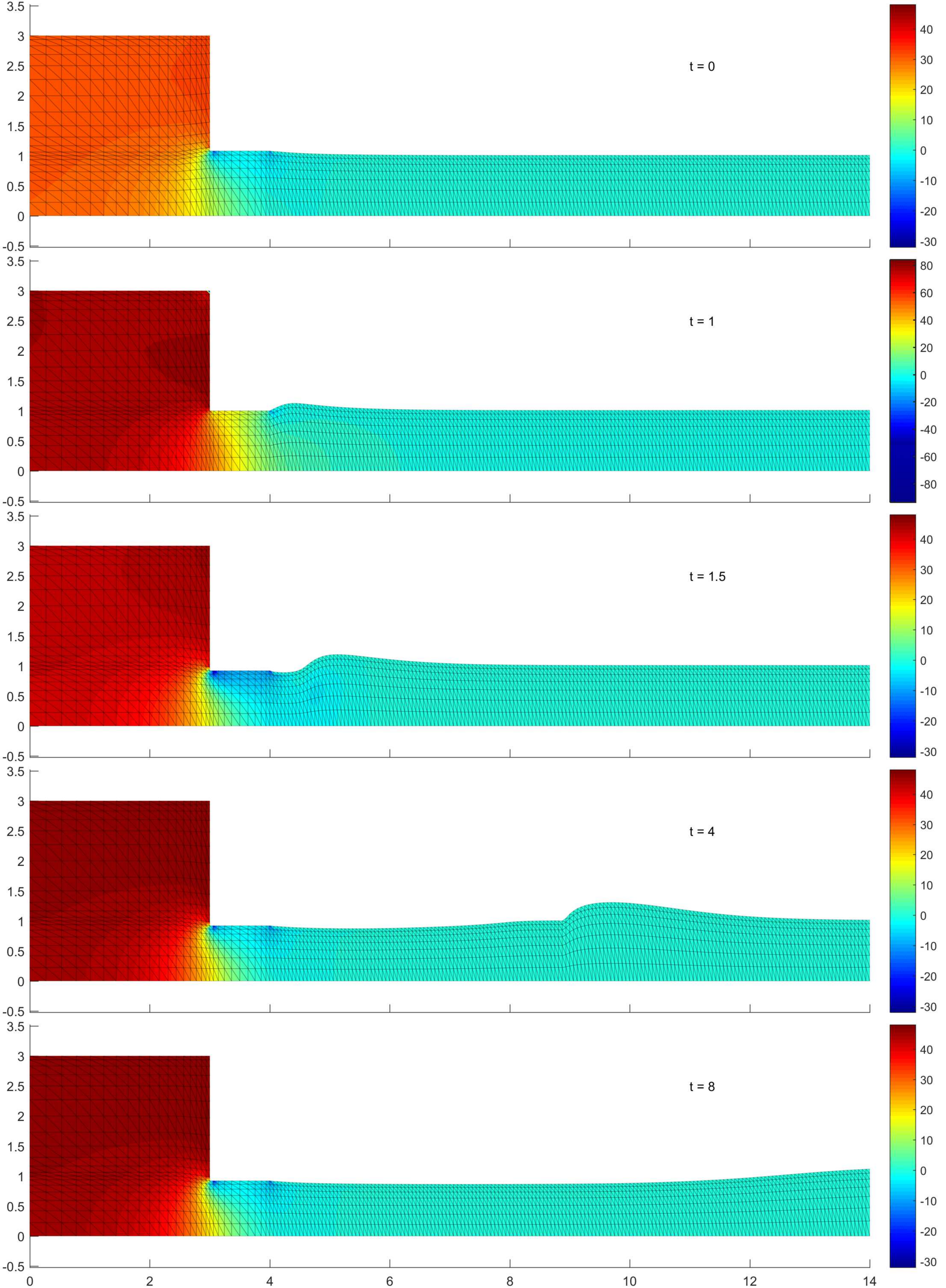}
	\centering
	\caption{$p$ of closing die for $Re=10$.}
	\label{fig:cRey10p}
\end{figure}

\begin{figure}
	\includegraphics[width=.9\linewidth]{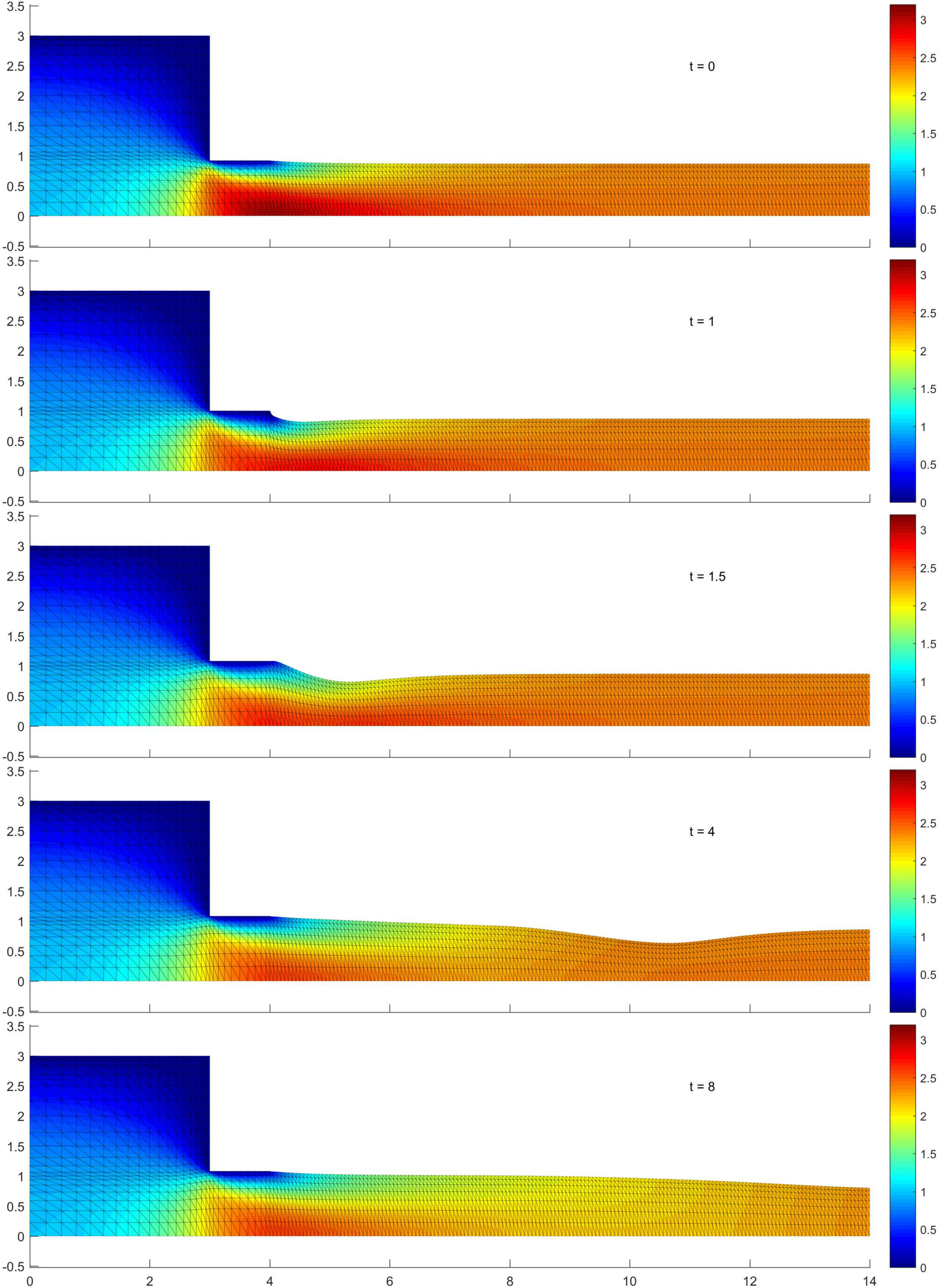}
	\centering
	\caption{$u_x$ of opening die for $Re=10$.}
	\label{fig:oRey10u}
\end{figure}

\begin{figure}
	\includegraphics[width=.9\linewidth]{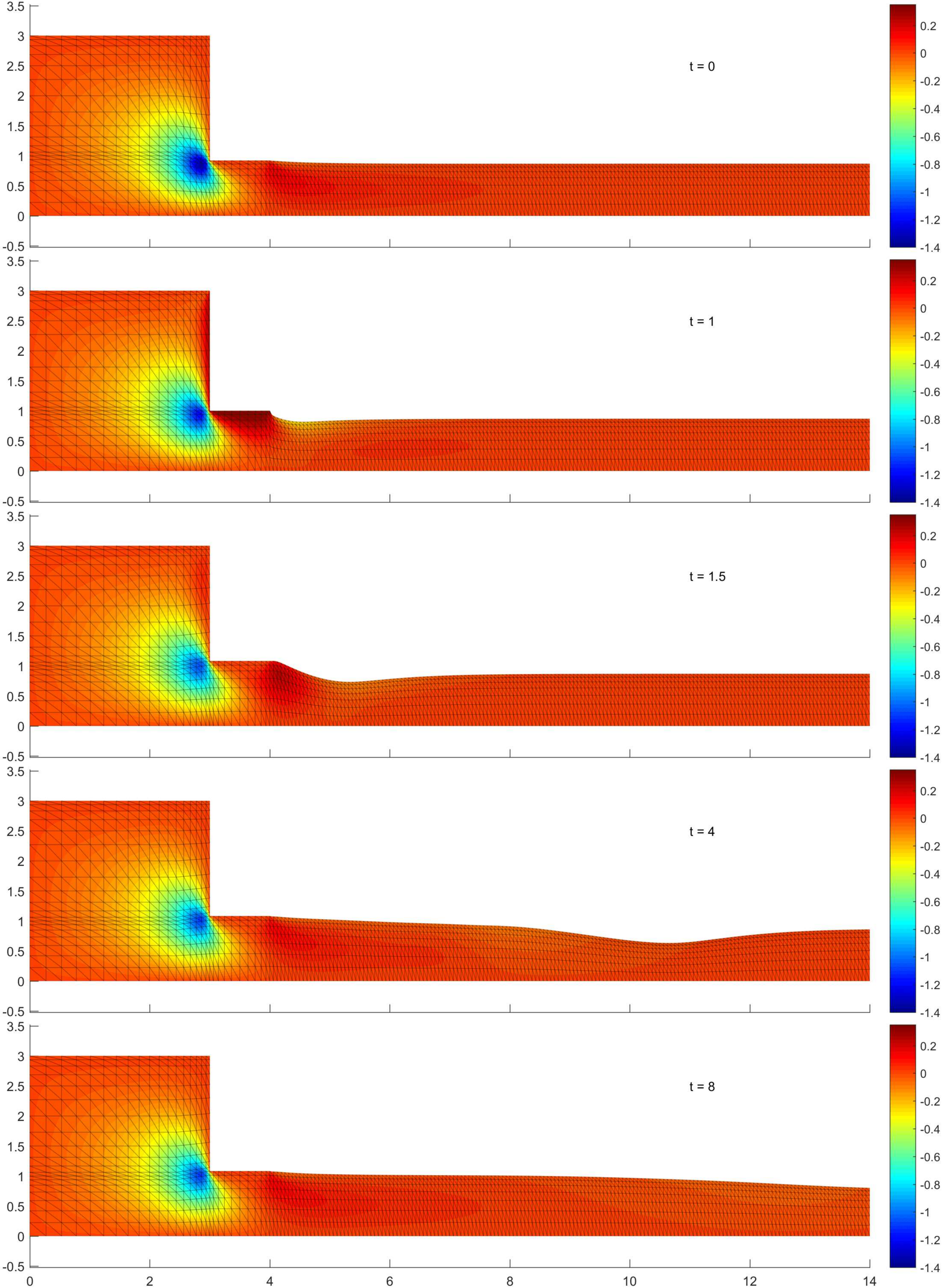}
	\centering
	\caption{$u_y$ of opening die for $Re=10$.}
	\label{fig:oRey10v}
\end{figure}

\begin{figure}
	\includegraphics[width=.9\linewidth]{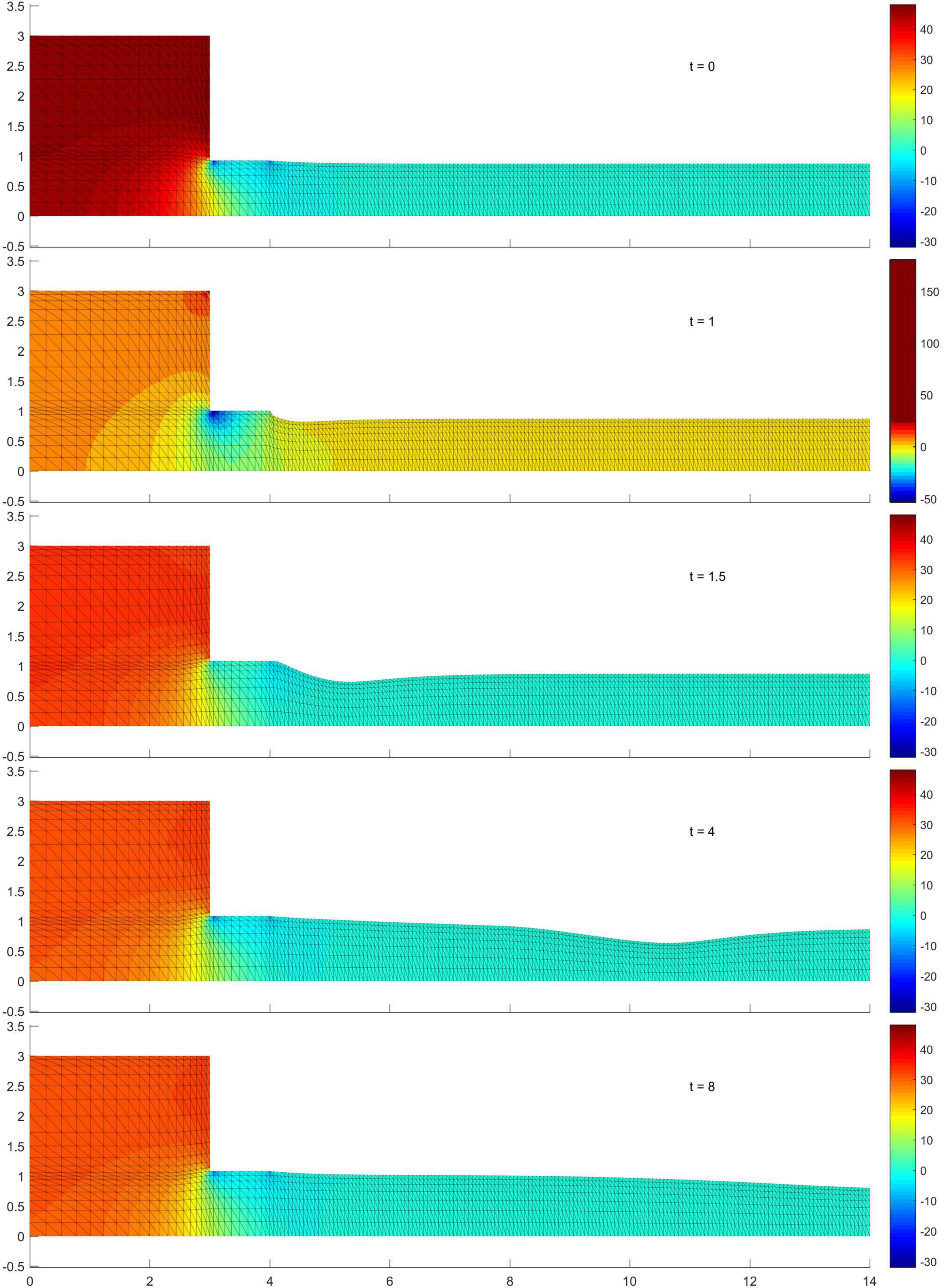}
	\centering
	\caption{$p$ of opening die for $Re=10$.}
	\label{fig:oRey10p}
\end{figure}
\end{appendices}

\end{document}